\documentclass[aps,prl,showpacs,twocolumn,superscriptaddress,10]{revtex4-1}
\pdfpageattr {/Group << /S /Transparency /I true /CS /DeviceRGB>>}

\usepackage{graphicx}
\usepackage{amsmath}
\usepackage{bm}
\usepackage{hyperref}
\hypersetup{%
pdftitle={Universal linear and nonlinear electrodynamics of the Dirac fluid}, %
pdfauthor={Zhiyuan Sun, D.N. Basov, and M.M. Fogler},%
pdfpagemode={UseNone},%
pdfstartview={FitH},%
breaklinks=true,%
citecolor=blue,%
colorlinks=true,%
linkcolor=blue,%
urlcolor=blue
}

\usepackage[T1]{fontenc}
\usepackage{fourier}
\usepackage{baskervald}

\newcommand{\unit}[1]{\,\mathrm{#1}} % for specifying units in math mode
\newcommand{\equa}[1]{Eq.~\eqref{#1}} % for specifying units in math mode
\DeclareMathOperator{\div1}{div}
\DeclareMathOperator{\curl}{curl}
\DeclareMathOperator{\im}{Im}
\DeclareMathOperator{\re}{Re}

\renewcommand{\vec}{\mathbf}

\graphicspath{{./}}

\begin{document}

\title{Universal linear and nonlinear electrodynamics of the Dirac fluid}

\author{Zhiyuan Sun}
\affiliation{Department of Physics, University of California San Diego, 9500 Gilman Drive, La Jolla, California 92093, USA}

\author{D. N. Basov}
\affiliation{Department of Physics, University of California San Diego, 9500 Gilman Drive, La Jolla, California 92093, USA}
\affiliation{Department of Physics, Columbia University,
538 West 120th Street, New York, New York 10027}

\author{M. M. Fogler}
\affiliation{Department of Physics, University of California San Diego, 9500 Gilman Drive, La Jolla, California 92093, USA}

\date{\today}

\begin{abstract}

A general relation is derived between 
the linear and second-order nonlinear ac conductivities of
an electron system in the hydrodynamic regime
of frequencies below the interparticle scattering rate.
The magnitude and tensorial structure of the
hydrodynamic nonlinear conductivity
are shown to differ
from their counterparts in
the more familiar kinetic regime of higher frequencies.
Due to
universality of the hydrodynamic equations, the obtained formulas are valid for systems with an arbitrary Dirac-like dispersion,
ranging from solid-state electron gases to 
free-space plasmas,
either massive or massless, at any temperature, chemical potential or space dimension.
Predictions for photon drag and second-harmonic generation in graphene are presented as one application of this theory.

\end{abstract}

\maketitle

%%%%%%%%%%%%%%%%%%%%%%%%%%%%%%%%%%%%%%%%%%%%%%%%%%%%%%%%%%%%%%%%%%%%%%%%%%%%%%%%%
%\emph{Introduction}.---
There has been a renewed interest to hydrodynamic phenomena in
electron systems with a Dirac-like energy-momentum dispersion
$\varepsilon_p^2 = (p v)^2 + (m v^2)^2$.
This subject was revived by studies in quantum criticality and
holographic field theory~\cite{Hartnoll2007}
and outspread in research on two-dimensional (2D) conductors,
e.g., graphene where the massless dispersion $m = 0$ is realized~\cite{DeJong1995,
Muller2008a, Muller2009, Briskot2015, Narozhny2015, Principi2015b, Principi2015, Bandurin2016, Crossno2016, Moll2016, Sun2016, Lucas2016, Guo2017}.
An experimental observation of a viscous electron flow
in graphene has been recently reported~\cite{Bandurin2016}.
Although not uncommon in plasmas~\cite{Tsytovich.1970},
this type of transport
is highly unusual in a solids.
It may be possible only in a limited range of temperatures $T$
and chemical potentials $\mu$ in
pristine samples
where the combined rate of electron-impurity (ei) and
electrons-phonon (ep) scattering  $\Gamma_{d} = \Gamma_{ei} + \Gamma_{ep}$
is lower than the momentum-conserving electron-electron (ee) scattering
rate $\Gamma_{ee}$ \cite{Gurzhi1968, Andreev2011}.
The respective mean-free paths must obey the inequality $l_d \gg l_{ee}$.
Under these conditions,
the electron dynamics at frequencies $\omega \ll\Gamma_{ee}$
and momenta $q \ll l_{ee}^{-1}$ is governed by
collective variables
that obey hydrodynamic equations~\cite{Landau.6}.
The frequency range
$\Gamma_{d} < \omega < \Gamma_{ee}$
may be as wide as several THz in graphene (see below).
Therefore exploring electrodynamics of Dirac fluids may be worthwhile.

\begin{figure}[b]
	\includegraphics[width=2.8in]{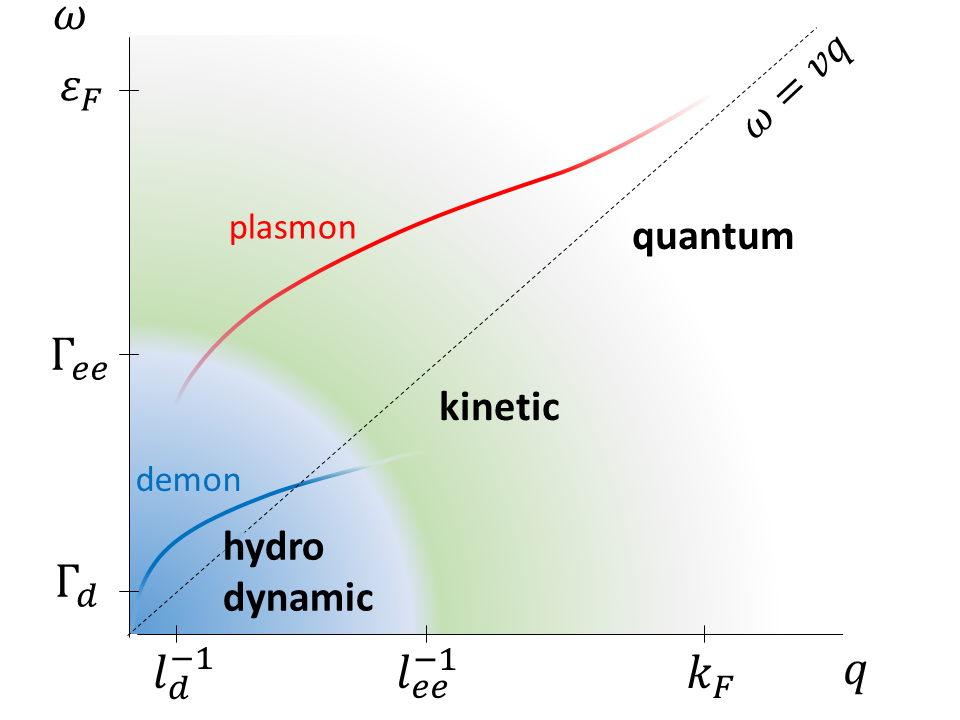} 
	\caption{A sketch of hydrodynamic, kinetic, and quantum domains in the frequency-momentum space. The collective modes of a massless fluid (plasmons and demons) are also shown, see text.}
	\label{fig:Parameter_regime}
\end{figure}

In this Letter we focus on the
second-order ac conductivity, which controls
nonlinear optical phenomena such as sum (difference) frequency generation and also photon drag. In the Supplemental material \cite{SM}, we also discuss the third-order conductivity important for the Kerr effect.
Prior work~\cite{Glazov2011a, Mikhailov2011, Yao2014, Mikhailov2016, Wang2016, Tokman2016, Cheng2016, Manzoni2015, Rostami2016}
has indicated that in graphene such effects may or may not~\cite{Khurgin2014} be stronger
than in typical metals and semiconductors.
We find significant differences of our results from
what one obtains at frequencies $\omega > \Gamma_{ee}$ where the dynamics is
described by the Boltzmann kinetic equation.
The still higher frequency quantum regime (Fig.~\ref{fig:Parameter_regime})
is beyond the scope of our investigation.

%%%%%%%%%%%%%%%%%%%%%%%%%%%%%%%%%%%%%%%%%%%%%%%%%%%%%%%%%%%%%%%%%%%%%%%%%%%%%%%%%
%%%%%%%%%%%%%%%%%%%%%%%%%%%%%%%%%%%%%%%%%%%%%%%%%%%%%%%%%%%%%%%%%%%%%%%%%%%%%%%%%

Recall that the second-order conductivity is a third-rank tensor
$\sigma^{(2)}_{ilm}\left(\vec{q}_1, \omega_1; \vec{q}_2, \omega_2\right)$
that describes the current
of frequency $\omega_3 = \omega_1 + \omega_2$ and momentum
$\vec{q}_3 = \vec{q}_1 + \vec{q}_2$
generated,
to the order ${O}\left(\vec{E}^2\right)$,
in response to an electric field
\begin{equation}
\vec{E}(\vec{r}, t) = \vec{E}(\vec{q}_1,\omega_1) e^{i \vec{q}_1 \vec{r} - i\omega_1 t}
+ \vec{E}(\vec{q}_2, \omega_2) e^{i \vec{q}_2 \vec{r} - i \omega_2 t} + \mathrm{c.c.}
\label{eqn:E}
\end{equation}
By convention, $\sigma^{(2)}_{ilm}$ is symmetrized, i.e.,
invariant under the interchange $(1 \leftrightarrow 2, l \leftrightarrow m)$.
If the system preserves parity,
which we assume to be the case, $\sigma^{(2)}_{ilm}$ must vanish
if both $\vec{q}_\nu$, $\nu = 1, 2$ are zero.
At small $\vec{q}_\nu$,
relevant
for optical/THz experiments, $\sigma^{(2)}_{ilm}$
should scale linearly with $\vec{q}_\nu$.
In comparison,
dissipative effects due to viscosity and heat conduction~\cite{Andreev2011, Forcella2014}, which scale as $|\vec{q}_\nu|^2$,
are subleading: the fluid dynamics
is approximately isentropic (ise)~\cite{Landau.6}.
Below we show that in this regime
the second-order conductivity
has the universal form
\begin{equation}
\sigma^{(2)}_{ilm} =  
\frac{D_h^{(2)}}{\omega_1 \omega_2 \omega_3}
\left( \frac{\omega_3}{\omega_1} q_{1l} \delta_{im} + q_{1i} 
\delta_{lm} \right) +
\left(\begin{smallmatrix}
1 &\leftrightarrow& 2\\ l &\leftrightarrow& m
\end{smallmatrix} \right)\,,
%\frac{D_h^{(2)}}{\omega_1 \omega_2}
%\left( \frac{q_{1l}}{\omega_1} \delta_{im} + \frac{q_{2m}}{\omega_2} \delta_{il} 
%+ \frac{q_{3i}}{\omega_3}\delta_{lm} \right)
\label{eqn:sigma2_1}
\end{equation}
for an arbitrary mass $m$, equilibrium charge density $\rho$,
temperature $T$, and space dimension $d$.
All the material-specific parameters are contained in the second-order spectral weight $D_h^{(2)}$, which we find to be
equal to the derivative
\begin{equation}
D_h^{(2)} = -\frac{1}{4 \pi^2} \left(\frac{\partial (D_h)^2}{\partial \rho}\right)_{\mathrm{ise}}
\label{eqn:D_h2_from_D_h}
\end{equation}
of the squared linear-response (i.e., Drude) spectral weight
\begin{equation}
D_h = \pi\, \frac{e^2 n}{m^{\ast}}\,,
\quad n \equiv \frac{\rho}{e}\,.
\label{eqn:D_h}
\end{equation}
As stated above, these formulas hold for either massless or massive
electrons.
Conventional metals and semiconductors have a parabolic dispersion.
This case is exemplified by the
nonrelativistic limit $|\mu|, T \ll m v^2$ of our equations,
yielding $m^\ast = m$.
This result
can also be understood as the consequence of Galilean invariance,
which demands that the ee interactions affect the linear and nonlinear conductivities only
in higher orders in $\vec{q}_\nu$. 
This is why the effective mass $m^{\ast}$ in Eq.~\eqref{eqn:D_h}
is equal to the bare mass $m$
and the leading $\vec{q}_\nu$-linear terms of
$\sigma^{(2)}_{ilm}$ [Eq.~\eqref{eqn:sigma2_1}] are the same in hydrodynamic~\cite{Tsytovich.1970},  kinetic~\cite{Aliev1992}, and quantum~\cite{Stolz1967} domains.
The equality of $m^*$ and $m$
does not hold if either $|\mu|$ or $T$ are comparable or larger than the energy gap $2 m v^2$, e.g., in the case of graphene.
(For linear conductivity at $T = 0$ this has been discussed
at length~\cite{Kotov2012, Basov2014, Link2016}.)

\begin{figure}[t]
\includegraphics[width=3.4 in]{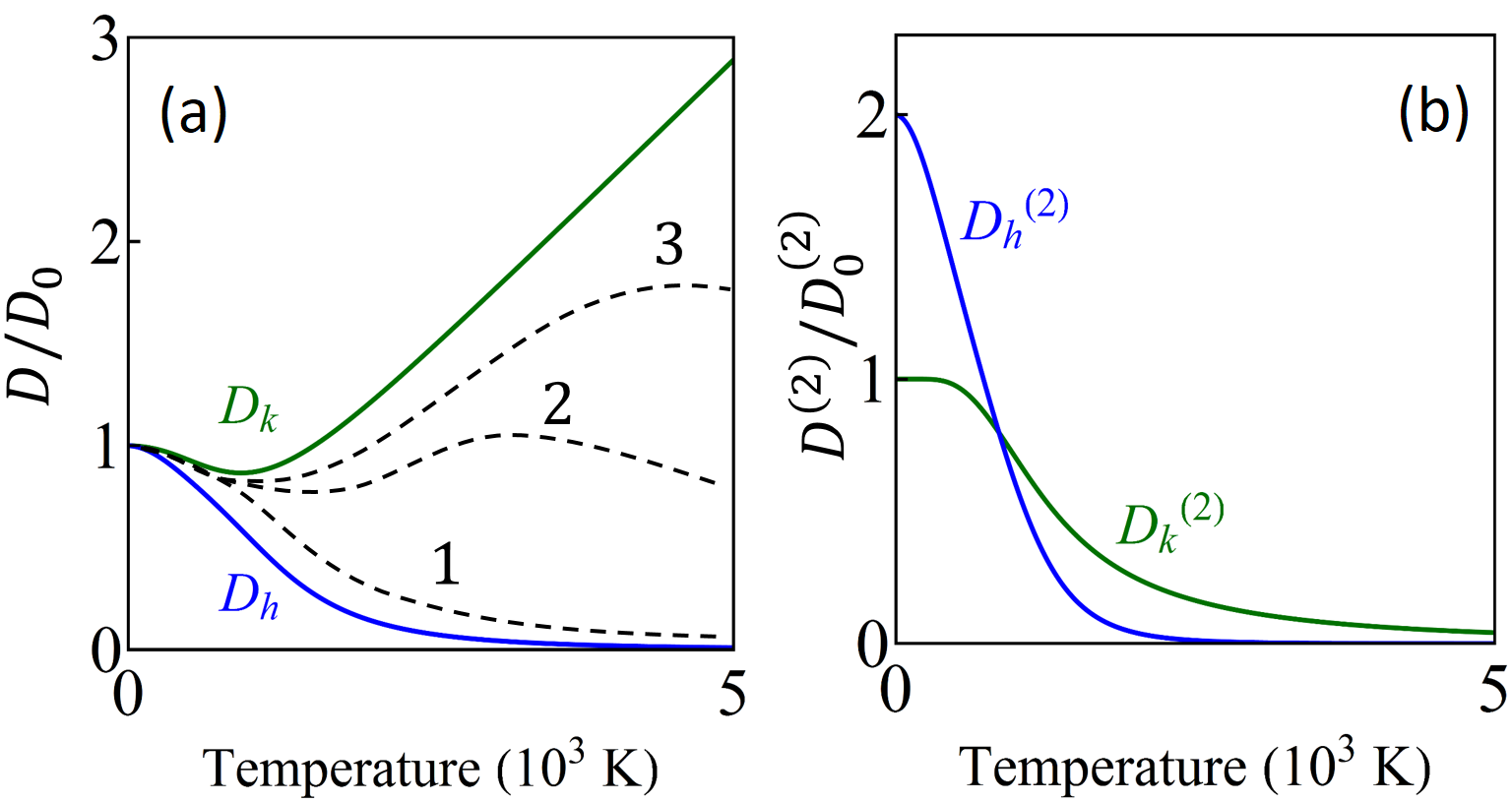} 
\caption{(Color online) (a) Hydrodynamic $D_h$ and kinetic $D_k$ Drude weights of doped graphene as functions of $T$, normalized to their common $T=0$ value. The dashed lines are sketches of the effective Drude weight $\pi \omega \mathrm{Im}[\sigma(\omega)]$ at three different frequencies $\omega \sim \varepsilon_F$
marked $1$--$3$ from low to high.
(b) The second-order spectral weights $D_k^{(2)}$ and $D_h^{(2)}$ of graphene in units of $D_0^{(2)}$, \equa{eqn:D_h2_D0}. The Fermi energy $\varepsilon_F = 0.207 \unit{eV}$ in both panels, corresponding to
$n = 3.14 \times 10^{12} \unit{cm^{-2}}$.
}
\label{fig:drude_temperature}
\end{figure}

In the hydrodynamic regime of graphene, frequent collisions force electrons and holes to move together, causing cancellation of their partial currents.
This enhances $m^{\ast}$ and reduces $D_h$
below its kinetic counterpart $D_k$ at all $T > 0$,
see Fig.~\ref{fig:drude_temperature}(a). Similarly, $D_h^{(2)}$ decreases with $T$ at fixed $\rho$ much faster in the hydrodynamic regime
than in the previously studied kinetic one,
see Fig.~\ref{fig:drude_temperature}(b).

%\emph{Hydrodynamics}.---

Let us now present a qualitative argument for Eq.~\eqref{eqn:sigma2_1}.
Consider the expansion of a given Fourier harmonic of the
electric current $j_i(\vec{q}, \omega) = j_i^{(1)}(\vec{q}, \omega) + j_i^{(2)}(\vec{q}, \omega) + \ldots$
in power series of the driving electric field $\vec{E}$.
The first term is given by
$j_i^{(1)}(\vec{q}, \omega) = \sigma_{ij}(\vec{q}, \omega) F_i^{(1)}(\vec{q}, \omega)$
where
$F_i^{(1)}(\vec{q}, \omega) = E_i(\vec{q}, \omega) + \mathcal{O}(\vec{q})$
is the driving force per unit charge
and $\sigma_{ij}$ is the linear-response conductivity tensor.
It suffices to consider the $\vec{q} \to 0$ limit in which $F_i^{(1)} \to E_i$, $\sigma_{ij} \to \delta_{ij} \sigma$.
The scalar $\sigma$ can be in general separated into the Drude pole and
a nonsingular correction $\sigma_0$ (to be discussed below):
\begin{align}
\sigma(q = 0, \omega) = \frac{1}{\pi}\, \frac{D_h}{-i\omega + \Gamma_d}
 + \sigma_0\,,
 \quad
 \omega \ll \Gamma_{ee}\,.
\label{eqn:sigma_hydro}
\end{align}
Next, to the second order we expect
$j_i^{(2)} = \sigma^{(1)} F_i^{(1)} + \sigma F_i^{(2)}$.
Here
$\sigma^{(1)} = (\partial \sigma / \partial \rho) \rho^{(1)}$
and $\rho^{(1)}$
are the perturbations of the conductivity and charge density.
The latter perturbation can be found from the continuity equation~\eqref{eqn:charge_c},
which gives
$\rho^{(1)}(\vec{q}, \omega) =  \vec{q}\cdot \vec{j}^{(1)}(\vec{q}, \omega) / \omega$.
Calculation of the second-order driving force $F_i^{(2)}$ is the
difficult part of the problem.
We glean the answer from the $\omega_1 \simeq -\omega_2 \gg \omega_3$ case
where it is equal to the sum of the pondermotive
and Abraham forces~\cite{Landau.8}.
The former is of order $(\omega_3)^0$, the latter is the 
leading $(\omega_3)^1$ correction.
Following~\cite{Landau.8}, Sec.~81,
we find the real-space representation of the pondermotive force to be
\begin{equation}
\begin{split}
F^{(2)}_i(\vec{r})
&= \frac{i}{2}\, \partial_i
\left[ \frac{\vec{E}(\vec{r}, \omega_1)}{\omega_2} \frac{\partial \sigma(\omega_2)}{\partial \rho} \vec{E}(\vec{r}, \omega_2)
+ (1 \leftrightarrow 2)
\right]\\
&= -\frac{i}{2}\, \partial_i
\left[ \frac{\vec{E}(\vec{r}, \omega_1)}{\omega_1} \frac{\partial \sigma(\omega_2)}{\partial \rho} \vec{E}(\vec{r}, \omega_2)
+ (1 \leftrightarrow 2)
\right].
\label{eqn:ponderomotive}
\end{split}
\end{equation}
The replacement of $\omega_2$ by $-\omega_1$ in the second line
cannot be strictly justified if $\omega_3 \neq 0$.
However, it is a natural way to ensure
the triangular permutation symmetry of $\sigma^{(2)}_{ilm}$,
which follows from the energy conservation~\cite{Ilinskii1994}
in the dissipationless limit $\Gamma_d, \sigma_0 \to 0$.
Assembling all the terms of $j_i^{(2)}$, we can read off $\sigma^{(2)}_{ilm}$
and see it coincides with Eq.~\eqref{eqn:sigma2_1}.
One can verify that for a nonrelativistic electron gas our formulas agree with those in literature~\cite{Tsytovich.1970, Aliev1992}.
%: Eq.~\eqref{eqn:ponderomotive} reproduces the well-known
%expression
%$F^{(2)}_i(\vec{r}) = -({e^2} / {m \omega^2}) \partial_i |\vec{E}(\vec{r})|^2$; Eq.~\eqref{eqn:sigma2_1} matches Eq.~(11) of \cite{Aliev1992} and Eq.~(2.41) of \cite{Tsytovich.1970}.

The case of a Lorentz-invariant Dirac fluid can be studied rigorously.
Proposed solid-state examples of such fluids~\cite{Hartnoll2007}
actually lack true Lorentz invariance.
Their matter and field components have different limiting velocities, $v$ and $c$.
However, 
if Coulomb interactions are weak, the approximate Lorentz invariance with velocity $v$ holds.
In graphene this is so 
if the dielectric constant $\kappa$ of the environment is large,
so that the interaction constant $e^2 / (\hbar \kappa v)$ is small. 
We will use relativistic hydrodynamics to derive $D_h$ and $D_h^{(2)}$ for this model and verify our key result~\eqref{eqn:sigma2_1}.

Let us introduce two additional quanitites.
One is the flow velocity $\vec{u}$
that defines the electric current $\vec{j} = \rho \vec{u}$.
The other is
the energy density $n_E = \gamma^2 W - P$ related to
the pressure $P = P(\mu, T)$ and enthalpy density $W = W(\mu, T)$ at thermal equilibrium, $u = 0$~\cite{Landau.6}.
Here $\gamma \equiv {1} / \sqrt{1 - (u / v)^2}$ and $n_E$ is referenced to the $\mu = T = u = 0$ state.
Relativistic hydrodynamic equations
admit many equivalent formulations~\cite{Landau.6, Muller2008a, Muller2009, Kovtun2012, Briskot2015}, e.g.,
\begin{subequations}
\label{eqn:hydro}
\begin{align}
	 & \partial_t \rho + \div1 \vec{j} = 0\,,
\label{eqn:charge_c}\\
	 & \partial_t n_{E} + \div1 \left(\gamma^2 W \vec{u}\right)
	  = \vec{j}\cdot \vec{E}\,,
\label{eqn:energy_c}\\
	 & (\gamma^2 m^{\ast}\! n)\, \mathcal{D}_t
	  \vec{u}
	  = \rho \vec{F}_L
	    - \frac{\vec{u}}{v^2} \left(\vec{j}\cdot \vec{E}\right)
	    -\vec{\mathcal{D}} P\,,
\label{eqn:euler}\\
	 & m^{\ast} = \frac{W}{n v^2}\,,
	 \quad \vec{F}_L = \vec{E} + \frac{1}{c}\, \vec{u}\times \vec{B}\,,
	  \quad
	  \curl \vec{E} = -\frac{1}{c} \partial_t \vec{B}\,.
\label{eqn:Lorentz}
\end{align}
\end{subequations}
The first pair is the charge continuity equation
and the energy conservation equation sans the subleading viscous and thermal conductivity terms.
Equation~\eqref{eqn:Lorentz} for the Lorentz force $\vec{F}_L$
includes the force from the ac magnetic field $\vec{B}$
induced by $\vec{E}$. (We assume that no static magnetic field is present.)
This term is important
if $\vec{E}$-field has a transverse component.
Equation~\eqref{eqn:euler} is the relativistic Euler equation
written in ``covariant derivatives''
$\mathcal{D}_i = \partial_i + ({u}_i / {v^2}) \partial_t$,
$\mathcal{D}_t = \partial_t  + \Gamma_{d} + {u}_i \partial_i$ with
the scattering rate $\Gamma_{d}$
accounting for momentum dissipation.
We solve these equations for $\vec{j}$ perturbatively in $\vec{E}$ to get the desired conductivities.

\begin{figure}[b]
\includegraphics[width=3 in]{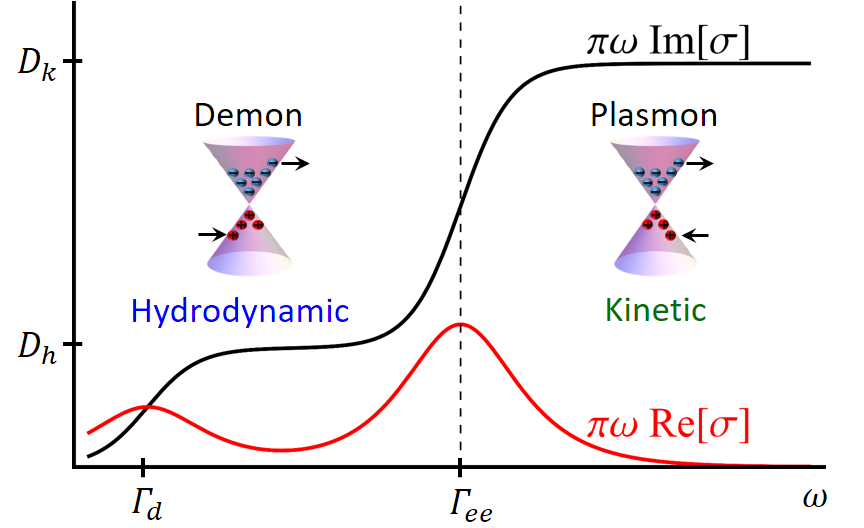} 
\caption{(Color online) Schematic illustration of Eq.~\eqref{eqn:sigma}.
The black curve is the effective Drude weight $\pi \omega \im \sigma$ as a function of $\omega$ at fixed $\rho$ and $T$.
The red curve represents $\pi \omega \re \sigma$
and $\omega$ should be understood as plotted on a logarithmic scale.
The insets depict collective motion of electrons and holes in plasmons and demons.
}
\label{fig:linear_conductivity}
\end{figure}

%\emph{Linear ac conductivity}.---
The linear response has already been treated at length~\cite{Hartnoll2007,
Muller2008a, Muller2009, Kovtun2012, Briskot2015, Narozhny2015, Sun2016, Lucas2016}.
For massless particles, $W = (d + 1) P = \frac{d + 1}{d} n_{E} \propto T^{d + 1}$.
The hydrodynamic Drude weight $D_h = \pi \rho^2 v^2 /W$
[cf.~Eqs.~\eqref{eqn:D_h} and \eqref{eqn:Lorentz}]
decreases as $T^{-d - 1}$ with $T$, i.e., as $T^{-3}$ in graphene.
The usual, kinetic Drude weight $D_k(\mu, T) = (g/2)({e} / {\hbar})^2 T \ln \left[2 \cosh ({\mu} / {2 T}) \right]$
where $g = 4$ is the number of Dirac cones~\cite{Basov2014}
behaves differently.
After some initial drop, $D_k$
\textit{increases} with $T$ because of thermal excitation of carriers,
see Fig.~\ref{fig:drude_temperature}(a).
The question how the opposite trends of $D_h$ and $D_k$ could be reconciled
has not been given proper attention in prior literature.
As a tentative answer, we suggest the interpolation formula:
\begin{align}
\sigma(q = 0, \omega) &= \frac{1}{\pi} \frac{D_h}{-i\omega + \Gamma_d}
+ \frac{1}{\pi}\frac{D_k - D_h}{-i\omega + \Gamma_d + \Gamma_{ee}}\,.
\label{eqn:sigma}
\end{align}
This formula can be derived from the Boltzmann kinetic equation
with the ee scattering rate $\Gamma_{ee}$
added to the collision integral~\cite{SM}.
Matching it with Eq.~\eqref{eqn:sigma_hydro}
at $\omega \ll \Gamma_{ee}$, we
deduce the parameter $\sigma_0 = (D_k - D_h) / (\pi \Gamma_{ee})$
therein (we assume $\Gamma_{d} \ll \Gamma_{ee}$)~\cite{Briskot2015}.
According to Eq.~\eqref{eqn:sigma}, the effective Drude weight $\pi \omega \im\sigma$ as a function of $\omega$ exhibits two plateaus,
see Fig.~\ref{fig:linear_conductivity},
and $\sigma$ as a function of $T$ at fixed $\omega$ may look like as sketched in Fig.~\ref{fig:drude_temperature}(a).
A quantitative theory of these crossover behaviors
is a challenge for future work.
Meanwhile, Fig.~\ref{fig:linear_conductivity} indicates the existence of
two \textit{separate} frequency intervals where $\im \sigma(\omega) \gg \re \sigma(\omega)$.
In these intervals weakly damped collective modes are possible:
sound waves~\cite{Kovtun2012} (or energy waves~\cite{Phan2013} or ``demons''~\cite{Sun2016} ) in the hydrodynamic regime and
plasmons in the kinetic one,
see also Fig.~\ref{fig:Parameter_regime}.

%%

%\emph{Second order nonlinear conductivity}.---
Let us move on to the second-order conductivity,
ignoring the momentum dissipation for now, $\Gamma_d \to 0$.
In the hydrodynamic regime we have two ways to derive $\sigma^{(2)}_{ilm}$.
The quick one is via Eq.~\eqref{eqn:sigma2_1}.
The only unknown parameter is
$D_h^{(2)}$, which we can calculate from Eq.~\eqref{eqn:D_h2_from_D_h}
applied to $D_h = \pi \rho^2 v^2 / W$.
This yields
\begin{align}
 D_h^{(2)} = -\frac{1}{2} \frac{e^3 n}{m^{\ast 2}} (1 - C_{\mathrm{ise}}) \,,
\label{eqn:D_h2} 
\end{align}
where
\begin{align}
C_{\mathrm{ise}} = \frac{n}{W}
 \left( \frac{\partial P}{\partial n} \right)_{\mathrm{ise}}
= \frac{1}{m^{\ast}\! v^2}
\left( \frac{\partial P}{\partial n} \right)_{\mathrm{ise}}
\end{align}
is the dimensionless isentropic bulk modulus.
Note that for massless electrons $C_{\mathrm{ise}} = 1 / d$.
The second derivation we can do is from hydrodynamic Eqs.~\eqref{eqn:hydro},
which is more tedious~\cite{SM}
but gives the same result.
This verifies the validity of our universal formula~\eqref{eqn:sigma2_1}
for Dirac fluids.

Let us examine
the $T$-dependence of the spectral weight $D_h^{(2)}$.
As one can anticipate, $D_h^{(2)}$ rapidly decreases at high $T$,
e.g., $D_h^{(2)} \propto T^{-2d - 2} = T^{-6}$ for graphene.
At $T \to 0$, Eq.~\eqref{eqn:D_h2} predicts
$D_h^{(2)}\to 2 D_0^{(2)} \mathrm{sign}\,n$,
where
\begin{align}
% D_h^{(2)}= -\frac{1}{9} \frac{\rho^3 v^4}{n_{E}^2} ,\quad
D_0^{(2)} = -\frac{g}{32\pi}
\frac{e^3 v^2}{\hbar^2}\,.
\label{eqn:D_h2_D0}
\end{align}
It may seem unusual that $D_h^{(2)}$ becomes doping-independent
in this limit (except for
the overall sign) but this can be rationalized by the dimensional analysis.
Of course, at $T = 0$ the system must be in the kinetic not hydrodynamic regime.
Surprisingly, in the kinetic regime of graphene, $\sigma^{(2)}_{i l m}$
has a different tensorial structure:
\begin{align}
\sigma^{(2)}_{ilm} =\mbox{}& \frac{D^{(2)}_k}{\omega_1 \omega_2 \omega_3}\, \Sigma_{ilmn}(\omega_1,\omega_2) q_{1n}
+
\left(\begin{smallmatrix}
1 &\leftrightarrow& 2\\ l &\leftrightarrow& m
\end{smallmatrix} \right)\,,
\label{eqn:sigma2_kin}\\
\begin{split}
\Sigma_{ilmn} = 
&- \left(1 - 3\, \frac{\omega_3}{\omega_1} \right)
\delta_{im} \delta_{nl}
- \left(1 + \frac{\omega_3}{\omega_1} \right)
 \delta_{il} \delta_{nm}
\\
&\mbox{} + \left(3 - \frac{\omega_3}{\omega_1} \right)
\delta_{in} \delta_{lm}
\,.
\end{split}
\label{eqn:Sigma}
\end{align}
This result can be obtained from either the Boltzmann kinetic equation~\cite{Manzoni2015}
or the semiclassical limit $q \ll k_F$, $\omega \ll \varepsilon_F$  of the quantum random-phase approximation~\cite{Cheng2016, Wang2016, Rostami2016}.
Here $\varepsilon_F$ and $k_F$ are the Fermi energy and momentum. Note that some of the related formulas in prior literature,
e.g., Eq.~(A.8) of \cite{Manzoni2015} and Eq.~(42) of \cite{Rostami2016} are valid only for response to a longitudinal $\vec{E}$-field.
If $\curl \vec{E} \neq 0$,
the correct result is obtained
only if the induced $\vec{B}$-field is included~\cite{Cheng2016, Wang2016,SM}.
When extended further~\cite{SM},
such calculations show that $D^{(2)}_k = D_0^{(2)}$
at $T = 0$ and $D^{(2)}_k \propto T^{-2}$ at high $T$.
Hence, $D_h^{(2)}$ is twice larger than
$D_k^{(2)}$ at $T = 0$ but becomes smaller at high $T$,
see Fig.~\ref{fig:drude_temperature}(b). 

\begin{figure}[t]
\includegraphics[width=3.5 in]{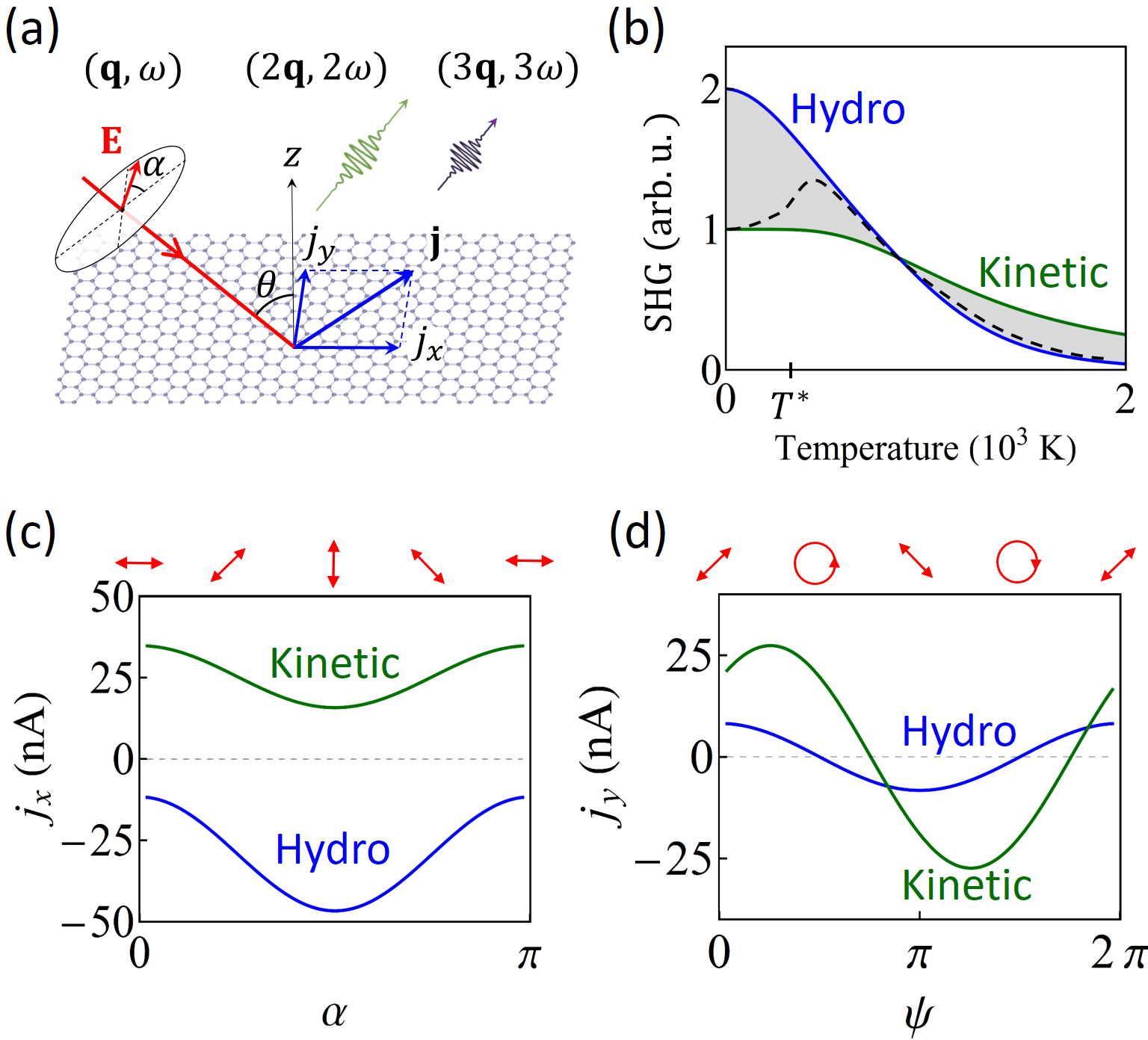} 
\caption{(Color online)
(a) Geometry for measuring PD, second,
and third harmonic generation.
(b) SHG signal as a function of $T$ at fixed $\omega$.
The ``Kinetic'' curve is from Eq.~\eqref{eqn:sigma2_1};
the ``Hydro'' curve is from Eq.~\eqref{eqn:sigma2_kin};
the dashed curve is a sketch of the actual signal.
(c) PD photocurrent $j_x$ in graphene \textit{vs}.
polarization angle $\alpha$ (illustrated by the red arrows).
(d) $j_y$ \textit{vs}. phase delay $\psi$ (degree of circular polarization)
at $\alpha = \pi/4$.
Parameters in (c,d): $T = 0$ for the `Kinetic'' curves,
$T=300\unit{K}$ for the ``Hydro'' curves,
$n = 3.14 \times 10^{12} \unit{cm^{-2}}$,
$\omega = 5 \unit{THz}$, $\Gamma_d = 1 \unit{THz}$,
$\theta = \pi / 4$, $E = 10^3 \unit{V/cm}$.
}
\label{fig:jy_phi}
\end{figure}

A direct experimental probe of
the second-order spectral weight is the second harmonic generation (SHG),
which corresponds to $\omega_2 = \omega_1 = \omega$, $\omega_3 = 2 \omega$,
see Fig.~\ref{fig:jy_phi}(a).
As explained above, the hydrodynamics predicts the SHG signal that is twice larger
at low $T$ and much smaller at high $T$
compared to the standard kinetic theory~\cite{Mikhailov2011, Mikhailov2016},
see Fig.~\ref{fig:jy_phi}(b).
The crossover from the kinetic regime to the hydrodynamic one
would occur at temperature $T^{\ast}$
such that $\Gamma_{ee}\left(T^{\ast}\right) = \omega$.
The measured SHG signal may look
like as sketched by the dashed curve in Fig.~\ref{fig:jy_phi}(b).

Another effect controlled by $\sigma^{(2)}_{ilm}$
is the photon drag (PD),
the generation of a dc current in response to
a monochromatic beam of frequency $\omega$,
see Fig.~\ref{fig:jy_phi}(a).
(A recent work~\cite{Tomadin2013}
studied a similar phenomenon for a surface plasmon playing the role of
the incident beam.)
To the second order in the in-plane field
$\vec{E}(\vec{q}, \omega) = (E_x, E_y)$
the PD is described by
$\sigma^{(2)}_{i l m}$ evaluated at $\omega_2 = -\omega_1 = \omega$,
and $\vec{q}_1 = -\vec{q}_2 = \vec{q}$.
The PD in graphene
has been previously studied in the kinetic regime~\cite{Glazov2014, Jiang2011, Karch2010}.
It was shown that the dc current can be parametrized by three constants $T_1$, $T_2$ and $\tilde{T}_1$,
which multiply the three Stokes parameters
of the incident beam.
Coefficients $T_1$ and $T_2$ quantify the linear PD, $T_2$ and
$\tilde{T}_1$ characterize the circular PD.
Instead of the Stokes parameters,
we find it convenient to use the incident angle $\theta$ and the 
$E_y$--$E_x$ phase delay $\psi$,
so that $E_x=E \cos\alpha \cos\theta$, $E_y=E\sin\alpha e^{i\psi}$.
Note that $\alpha=0$ means p-polarization and $\alpha=\pi/2$ means s-polarization. 
For a beam with the in-plane momentum $\vec{q} = (q_x, 0)$,
the longitudinal and transverse current components are:
\begin{subequations}
\label{eqn:PD}
\begin{align}
\frac{j_x}{C_j} &=
(T_1 + T_2) \cos^2\!\alpha \cos^2\theta + (T_1 - T_2) \sin^2\! \alpha\,,
\label{eqn:j_x}\\
\frac{j_y}{C_j} &=
\cos\theta \sin 2\alpha \left(T_2 \cos\psi - 2\tilde{T}_1 \sin\psi\right)\,,
\label{eqn:j_y}
\end{align}
\end{subequations}
where $C_j = \frac{1}{2} q_x |E|^2$, cf.~Eq.~(10) of \cite{Glazov2014}.
To compute  $T_1$, $T_2$ and $\tilde{T}_1$ for graphene in the hydrodynamic regime, we use the dissipative version of Eq.~\eqref{eqn:sigma2_1}, which corresponds to retaining $\Gamma_d$ in the Euler
equation~\eqref{eqn:euler}.
The resultant expression for $\sigma^{(2)}_{i l m}$ at arbitrary $\omega$
is ponderous~\cite{SM}.
We present only the formulas for the drag coefficients:
\begin{align}
T_1 = -3 T_2\,,
\quad
T_2 = \frac{4 D_h^{(2)}}{\omega\left(\omega^2 + \Gamma_d^2 \right)}\,,
\quad
\tilde{T}_1 = 0\,.
\label{eqn:T_hydro}
\end{align}
They are quite unlike those in the kinetic regime
in which $\tilde{T}_1$ is nonzero, e.g.,
\begin{align}
\tilde{T}_1 = -
\frac{48 D_0^{(2)}\, \Gamma_d}{\left(\omega^2 + \Gamma_d^2 \right)
                \left(\omega^2 + 4\Gamma_d^2 \right)} \,.
\end{align}
This expression, which is
a particular case of a general formula given in~\cite{SM, Glazov2014}, assumes that the scattering rate $\Gamma_d$ is due to short-range scatterers.
The difference between the two regimes
is illustrated in Fig.~\ref{fig:jy_phi}(c,d).

The following estimates suggest that the hydrodynamic regime $\Gamma_d < \omega < \Gamma_{ee}$ could be fairly wide in ultra clean graphene where
electrons are scattered primarily by acoustic phonons,
$\Gamma_d \approx \Gamma_{ep}$.
The electron-phonon scattering rate $\Gamma_{ep}(T_l, T, n)$ \cite{Principi2014} is a function of the lattice temperature $T_l$, electron temperature $T$, and doping $n$.
From~\cite{Ni2016, Ni2017} we estimate $\Gamma_{ep}(150 \unit{K}, 150 \unit{K}, 2\times 10^{12} \unit{cm}^{-2}) \sim 0.3 \unit{THz}$.
On the other hand, $\Gamma_{ee}(T, n)$ is a function of $T$ and $n$.
(In the kinetic regime $\omega \gg \Gamma_{ee}$, it may also depend on frequency.)
Recent dc transport experiments~\cite{Bandurin2016} indicate $\Gamma_{ee}(150 \unit{K}, 10^{12} \unit{cm}^{-2}) \sim 0.5 \unit{THz}$,
so the hydrodynamic region is narrow.
There are two possible schemes to diminish $\Gamma_{ep}$ or
enhance $\Gamma_{ee}$.
The first one is to reduce $n$ to make electron gas non-degenerate,
which should bring $\Gamma_{ee}$ to the theoretical maximum~\cite{Schutt2011}
of $4 (e^2 / \hbar \kappa v)^2 T \sim 10 \unit{THz}$.
The other route is ultrafast pump-probe experiments~\cite{Ni2016}
that can keep the lattice cold, perhaps,
at $T_l \sim 30 \unit{K}$ but heat electrons to $T \sim 3000 \unit{K}$.
 
The universal relation \eqref{eqn:D_h2_from_D_h} between linear and nonlinear ac conductivities is the most important result
of this Letter. Although we have used graphene as the example, this
and our other formulas Eqs.~\eqref{eqn:sigma2_1}, \eqref{eqn:D_h2}, \textit{etc}.,
should apply as well to ultrapure metals and semiconductors~\cite{Moll2016, DeJong1995},
to surface states of topological insulators and Dirac/Weyl semimetals, provided they are in the hydrodynamic regime. 

This work is supported by the DOE under
Grant DE-SC0012592,
by the ONR under Grant N00014-15-1-2671,
by the NSF under Grant ECCS-1640173,
and by the SRC.
D.~N.~B. is an investigator in Quantum Materials funded by
the Gordon and Betty Moore Foundation's EPiQS Initiative
through Grant No. GBMF4533.
We thank G.~Falkovich, M.~Glazov, and G.~Ni for discussions.

%%%%%%%%%%%%%%%%%%%%%%%%%%%%%%%%%%%%%%%%%%%%%
\bibliographystyle{apsrev4-1}
\bibliography{./Library_hydrodynamics}

\begin{thebibliography}{48}%
\makeatletter
\providecommand \@ifxundefined [1]{%
 \@ifx{#1\undefined}
}%
\providecommand \@ifnum [1]{%
 \ifnum #1\expandafter \@firstoftwo
 \else \expandafter \@secondoftwo
 \fi
}%
\providecommand \@ifx [1]{%
 \ifx #1\expandafter \@firstoftwo
 \else \expandafter \@secondoftwo
 \fi
}%
\providecommand \natexlab [1]{#1}%
\providecommand \enquote  [1]{``#1''}%
\providecommand \bibnamefont  [1]{#1}%
\providecommand \bibfnamefont [1]{#1}%
\providecommand \citenamefont [1]{#1}%
\providecommand \href@noop [0]{\@secondoftwo}%
\providecommand \href [0]{\begingroup \@sanitize@url \@href}%
\providecommand \@href[1]{\@@startlink{#1}\@@href}%
\providecommand \@@href[1]{\endgroup#1\@@endlink}%
\providecommand \@sanitize@url [0]{\catcode `\\12\catcode `\$12\catcode
  `\&12\catcode `\#12\catcode `\^12\catcode `\_12\catcode `\%12\relax}%
\providecommand \@@startlink[1]{}%
\providecommand \@@endlink[0]{}%
\providecommand \url  [0]{\begingroup\@sanitize@url \@url }%
\providecommand \@url [1]{\endgroup\@href {#1}{\urlprefix }}%
\providecommand \urlprefix  [0]{URL }%
\providecommand \Eprint [0]{\href }%
\providecommand \doibase [0]{http://dx.doi.org/}%
\providecommand \selectlanguage [0]{\@gobble}%
\providecommand \bibinfo  [0]{\@secondoftwo}%
\providecommand \bibfield  [0]{\@secondoftwo}%
\providecommand \translation [1]{[#1]}%
\providecommand \BibitemOpen [0]{}%
\providecommand \bibitemStop [0]{}%
\providecommand \bibitemNoStop [0]{.\EOS\space}%
\providecommand \EOS [0]{\spacefactor3000\relax}%
\providecommand \BibitemShut  [1]{\csname bibitem#1\endcsname}%
\let\auto@bib@innerbib\@empty
%</preamble>
\bibitem [{\citenamefont {Hartnoll}\ \emph {et~al.}(2007)\citenamefont
  {Hartnoll}, \citenamefont {Kovtun}, \citenamefont {M{\"{u}}ller},\ and\
  \citenamefont {Sachdev}}]{Hartnoll2007}%
  \BibitemOpen
  \bibfield  {author} {\bibinfo {author} {\bibfnamefont {S.~A.}\ \bibnamefont
  {Hartnoll}}, \bibinfo {author} {\bibfnamefont {P.~K.}\ \bibnamefont
  {Kovtun}}, \bibinfo {author} {\bibfnamefont {M.}~\bibnamefont
  {M{\"{u}}ller}}, \ and\ \bibinfo {author} {\bibfnamefont {S.}~\bibnamefont
  {Sachdev}},\ }\href {\doibase 10.1103/PhysRevB.76.144502} {\bibfield
  {journal} {\bibinfo  {journal} {Phys. Rev. B}\ }\textbf {\bibinfo {volume}
  {76}},\ \bibinfo {pages} {144502} (\bibinfo {year} {2007})}\BibitemShut
  {NoStop}%
\bibitem [{\citenamefont {de~Jong}\ and\ \citenamefont
  {Molenkamp}(1995)}]{DeJong1995}%
  \BibitemOpen
  \bibfield  {author} {\bibinfo {author} {\bibfnamefont {M.~J.~M.}\
  \bibnamefont {de~Jong}}\ and\ \bibinfo {author} {\bibfnamefont {L.~W.}\
  \bibnamefont {Molenkamp}},\ }\href {\doibase 10.1103/PhysRevB.51.13389}
  {\bibfield  {journal} {\bibinfo  {journal} {Phys. Rev. B}\ }\textbf {\bibinfo
  {volume} {51}},\ \bibinfo {pages} {13389} (\bibinfo {year}
  {1995})}\BibitemShut {NoStop}%
\bibitem [{\citenamefont {M{\"{u}}ller}\ \emph {et~al.}(2008)\citenamefont
  {M{\"{u}}ller}, \citenamefont {Fritz},\ and\ \citenamefont
  {Sachdev}}]{Muller2008a}%
  \BibitemOpen
  \bibfield  {author} {\bibinfo {author} {\bibfnamefont {M.}~\bibnamefont
  {M{\"{u}}ller}}, \bibinfo {author} {\bibfnamefont {L.}~\bibnamefont {Fritz}},
  \ and\ \bibinfo {author} {\bibfnamefont {S.}~\bibnamefont {Sachdev}},\ }\href
  {\doibase 10.1103/PhysRevB.78.115406} {\bibfield  {journal} {\bibinfo
  {journal} {Phys. Rev. B}\ }\textbf {\bibinfo {volume} {78}},\ \bibinfo
  {pages} {115406} (\bibinfo {year} {2008})}\BibitemShut {NoStop}%
\bibitem [{\citenamefont {M{\"{u}}ller}\ \emph {et~al.}(2009)\citenamefont
  {M{\"{u}}ller}, \citenamefont {Schmalian},\ and\ \citenamefont
  {Fritz}}]{Muller2009}%
  \BibitemOpen
  \bibfield  {author} {\bibinfo {author} {\bibfnamefont {M.}~\bibnamefont
  {M{\"{u}}ller}}, \bibinfo {author} {\bibfnamefont {J.}~\bibnamefont
  {Schmalian}}, \ and\ \bibinfo {author} {\bibfnamefont {L.}~\bibnamefont
  {Fritz}},\ }\href {\doibase 10.1103/PhysRevLett.103.025301} {\bibfield
  {journal} {\bibinfo  {journal} {Phys. Rev. Lett.}\ }\textbf {\bibinfo
  {volume} {103}},\ \bibinfo {pages} {025301} (\bibinfo {year}
  {2009})}\BibitemShut {NoStop}%
\bibitem [{\citenamefont {Briskot}\ \emph {et~al.}(2015)\citenamefont
  {Briskot}, \citenamefont {Sch{\"{u}}tt}, \citenamefont {Gornyi},
  \citenamefont {Titov}, \citenamefont {Narozhny},\ and\ \citenamefont
  {Mirlin}}]{Briskot2015}%
  \BibitemOpen
  \bibfield  {author} {\bibinfo {author} {\bibfnamefont {U.}~\bibnamefont
  {Briskot}}, \bibinfo {author} {\bibfnamefont {M.}~\bibnamefont
  {Sch{\"{u}}tt}}, \bibinfo {author} {\bibfnamefont {I.~V.}\ \bibnamefont
  {Gornyi}}, \bibinfo {author} {\bibfnamefont {M.}~\bibnamefont {Titov}},
  \bibinfo {author} {\bibfnamefont {B.~N.}\ \bibnamefont {Narozhny}}, \ and\
  \bibinfo {author} {\bibfnamefont {A.~D.}\ \bibnamefont {Mirlin}},\ }\href
  {\doibase 10.1103/PhysRevB.92.115426} {\bibfield  {journal} {\bibinfo
  {journal} {Phys. Rev. B}\ }\textbf {\bibinfo {volume} {92}},\ \bibinfo
  {pages} {115426} (\bibinfo {year} {2015})}\BibitemShut {NoStop}%
\bibitem [{\citenamefont {Narozhny}\ \emph {et~al.}(2015)\citenamefont
  {Narozhny}, \citenamefont {Gornyi}, \citenamefont {Titov}, \citenamefont
  {Sch{\"{u}}tt},\ and\ \citenamefont {Mirlin}}]{Narozhny2015}%
  \BibitemOpen
  \bibfield  {author} {\bibinfo {author} {\bibfnamefont {B.~N.}\ \bibnamefont
  {Narozhny}}, \bibinfo {author} {\bibfnamefont {I.~V.}\ \bibnamefont
  {Gornyi}}, \bibinfo {author} {\bibfnamefont {M.}~\bibnamefont {Titov}},
  \bibinfo {author} {\bibfnamefont {M.}~\bibnamefont {Sch{\"{u}}tt}}, \ and\
  \bibinfo {author} {\bibfnamefont {A.~D.}\ \bibnamefont {Mirlin}},\ }\href
  {\doibase 10.1103/PhysRevB.91.035414} {\bibfield  {journal} {\bibinfo
  {journal} {Phys. Rev. B}\ }\textbf {\bibinfo {volume} {91}},\ \bibinfo
  {pages} {035414} (\bibinfo {year} {2015})}\BibitemShut {NoStop}%
\bibitem [{\citenamefont {Principi}\ and\ \citenamefont
  {Vignale}(2015{\natexlab{a}})}]{Principi2015b}%
  \BibitemOpen
  \bibfield  {author} {\bibinfo {author} {\bibfnamefont {A.}~\bibnamefont
  {Principi}}\ and\ \bibinfo {author} {\bibfnamefont {G.}~\bibnamefont
  {Vignale}},\ }\href {\doibase 10.1103/PhysRevB.91.205423} {\bibfield
  {journal} {\bibinfo  {journal} {Phys. Rev. B}\ }\textbf {\bibinfo {volume}
  {91}},\ \bibinfo {pages} {205423} (\bibinfo {year}
  {2015}{\natexlab{a}})}\BibitemShut {NoStop}%
\bibitem [{\citenamefont {Principi}\ and\ \citenamefont
  {Vignale}(2015{\natexlab{b}})}]{Principi2015}%
  \BibitemOpen
  \bibfield  {author} {\bibinfo {author} {\bibfnamefont {A.}~\bibnamefont
  {Principi}}\ and\ \bibinfo {author} {\bibfnamefont {G.}~\bibnamefont
  {Vignale}},\ }\href {\doibase 10.1103/PhysRevLett.115.056603} {\bibfield
  {journal} {\bibinfo  {journal} {Phys. Rev. Lett.}\ }\textbf {\bibinfo
  {volume} {115}},\ \bibinfo {pages} {056603} (\bibinfo {year}
  {2015}{\natexlab{b}})}\BibitemShut {NoStop}%
\bibitem [{\citenamefont {Bandurin}\ \emph {et~al.}(2016)\citenamefont
  {Bandurin}, \citenamefont {Torre}, \citenamefont {Kumar}, \citenamefont {{Ben
  Shalom}}, \citenamefont {Tomadin}, \citenamefont {Principi}, \citenamefont
  {Auton}, \citenamefont {Khestanova}, \citenamefont {Novoselov}, \citenamefont
  {Grigorieva}, \citenamefont {Ponomarenko}, \citenamefont {Geim},\ and\
  \citenamefont {Polini}}]{Bandurin2016}%
  \BibitemOpen
  \bibfield  {author} {\bibinfo {author} {\bibfnamefont {D.~A.}\ \bibnamefont
  {Bandurin}}, \bibinfo {author} {\bibfnamefont {I.}~\bibnamefont {Torre}},
  \bibinfo {author} {\bibfnamefont {R.~K.}\ \bibnamefont {Kumar}}, \bibinfo
  {author} {\bibfnamefont {M.}~\bibnamefont {{Ben Shalom}}}, \bibinfo {author}
  {\bibfnamefont {A.}~\bibnamefont {Tomadin}}, \bibinfo {author} {\bibfnamefont
  {A.}~\bibnamefont {Principi}}, \bibinfo {author} {\bibfnamefont {G.~H.}\
  \bibnamefont {Auton}}, \bibinfo {author} {\bibfnamefont {E.}~\bibnamefont
  {Khestanova}}, \bibinfo {author} {\bibfnamefont {K.~S.}\ \bibnamefont
  {Novoselov}}, \bibinfo {author} {\bibfnamefont {I.~V.}\ \bibnamefont
  {Grigorieva}}, \bibinfo {author} {\bibfnamefont {L.~A.}\ \bibnamefont
  {Ponomarenko}}, \bibinfo {author} {\bibfnamefont {A.~K.}\ \bibnamefont
  {Geim}}, \ and\ \bibinfo {author} {\bibfnamefont {M.}~\bibnamefont
  {Polini}},\ }\href {\doibase 10.1126/science.aad0201} {\bibfield  {journal}
  {\bibinfo  {journal} {Science}\ }\textbf {\bibinfo {volume} {351}},\ \bibinfo
  {pages} {1055} (\bibinfo {year} {2016})}\BibitemShut {NoStop}%
\bibitem [{\citenamefont {Crossno}\ \emph {et~al.}(2016)\citenamefont
  {Crossno}, \citenamefont {Shi}, \citenamefont {Wang}, \citenamefont {Liu},
  \citenamefont {Harzheim}, \citenamefont {Lucas}, \citenamefont {Sachdev},
  \citenamefont {Kim}, \citenamefont {Taniguchi}, \citenamefont {Watanabe},
  \citenamefont {Ohki},\ and\ \citenamefont {Fong}}]{Crossno2016}%
  \BibitemOpen
  \bibfield  {author} {\bibinfo {author} {\bibfnamefont {J.}~\bibnamefont
  {Crossno}}, \bibinfo {author} {\bibfnamefont {J.~K.}\ \bibnamefont {Shi}},
  \bibinfo {author} {\bibfnamefont {K.}~\bibnamefont {Wang}}, \bibinfo {author}
  {\bibfnamefont {X.}~\bibnamefont {Liu}}, \bibinfo {author} {\bibfnamefont
  {A.}~\bibnamefont {Harzheim}}, \bibinfo {author} {\bibfnamefont
  {A.}~\bibnamefont {Lucas}}, \bibinfo {author} {\bibfnamefont
  {S.}~\bibnamefont {Sachdev}}, \bibinfo {author} {\bibfnamefont
  {P.}~\bibnamefont {Kim}}, \bibinfo {author} {\bibfnamefont {T.}~\bibnamefont
  {Taniguchi}}, \bibinfo {author} {\bibfnamefont {K.}~\bibnamefont {Watanabe}},
  \bibinfo {author} {\bibfnamefont {T.~A.}\ \bibnamefont {Ohki}}, \ and\
  \bibinfo {author} {\bibfnamefont {K.~C.}\ \bibnamefont {Fong}},\ }\href
  {\doibase 10.1126/science.aad0343} {\bibfield  {journal} {\bibinfo  {journal}
  {Science}\ }\textbf {\bibinfo {volume} {351}},\ \bibinfo {pages} {1058}
  (\bibinfo {year} {2016})}\BibitemShut {NoStop}%
\bibitem [{\citenamefont {Moll}\ \emph {et~al.}(2016)\citenamefont {Moll},
  \citenamefont {Kushwaha}, \citenamefont {Nandi}, \citenamefont {Schmidt},\
  and\ \citenamefont {Mackenzie}}]{Moll2016}%
  \BibitemOpen
  \bibfield  {author} {\bibinfo {author} {\bibfnamefont {P.~J.~W.}\
  \bibnamefont {Moll}}, \bibinfo {author} {\bibfnamefont {P.}~\bibnamefont
  {Kushwaha}}, \bibinfo {author} {\bibfnamefont {N.}~\bibnamefont {Nandi}},
  \bibinfo {author} {\bibfnamefont {B.}~\bibnamefont {Schmidt}}, \ and\
  \bibinfo {author} {\bibfnamefont {A.~P.}\ \bibnamefont {Mackenzie}},\ }\href
  {\doibase 10.1126/science.aac8385} {\bibfield  {journal} {\bibinfo  {journal}
  {Science}\ }\textbf {\bibinfo {volume} {351}},\ \bibinfo {pages} {1061}
  (\bibinfo {year} {2016})}\BibitemShut {NoStop}%
\bibitem [{\citenamefont {Sun}\ \emph {et~al.}(2016)\citenamefont {Sun},
  \citenamefont {Basov},\ and\ \citenamefont {Fogler}}]{Sun2016}%
  \BibitemOpen
  \bibfield  {author} {\bibinfo {author} {\bibfnamefont {Z.}~\bibnamefont
  {Sun}}, \bibinfo {author} {\bibfnamefont {D.~N.}\ \bibnamefont {Basov}}, \
  and\ \bibinfo {author} {\bibfnamefont {M.~M.}\ \bibnamefont {Fogler}},\
  }\href {\doibase https://doi.org/10.1103/PhysRevLett.117.076805} {\bibfield
  {journal} {\bibinfo  {journal} {Phys. Rev. Lett.}\ }\textbf {\bibinfo
  {volume} {117}},\ \bibinfo {pages} {076805} (\bibinfo {year}
  {2016})}\BibitemShut {NoStop}%
\bibitem [{\citenamefont {Lucas}\ \emph {et~al.}(2016)\citenamefont {Lucas},
  \citenamefont {Crossno}, \citenamefont {Fong}, \citenamefont {Kim},\ and\
  \citenamefont {Sachdev}}]{Lucas2016}%
  \BibitemOpen
  \bibfield  {author} {\bibinfo {author} {\bibfnamefont {A.}~\bibnamefont
  {Lucas}}, \bibinfo {author} {\bibfnamefont {J.}~\bibnamefont {Crossno}},
  \bibinfo {author} {\bibfnamefont {K.~C.}\ \bibnamefont {Fong}}, \bibinfo
  {author} {\bibfnamefont {P.}~\bibnamefont {Kim}}, \ and\ \bibinfo {author}
  {\bibfnamefont {S.}~\bibnamefont {Sachdev}},\ }\href {\doibase
  10.1103/PhysRevB.93.075426} {\bibfield  {journal} {\bibinfo  {journal} {Phys.
  Rev. B}\ }\textbf {\bibinfo {volume} {93}},\ \bibinfo {pages} {075426}
  (\bibinfo {year} {2016})}\BibitemShut {NoStop}%
\bibitem [{\citenamefont {Guo}\ \emph {et~al.}(2017)\citenamefont {Guo},
  \citenamefont {Ilseven}, \citenamefont {Falkovich},\ and\ \citenamefont
  {Levitov}}]{Guo2017}%
  \BibitemOpen
  \bibfield  {author} {\bibinfo {author} {\bibfnamefont {H.}~\bibnamefont
  {Guo}}, \bibinfo {author} {\bibfnamefont {E.}~\bibnamefont {Ilseven}},
  \bibinfo {author} {\bibfnamefont {G.}~\bibnamefont {Falkovich}}, \ and\
  \bibinfo {author} {\bibfnamefont {L.~S.}\ \bibnamefont {Levitov}},\ }\href
  {\doibase 10.1073/pnas.1612181114} {\bibfield  {journal} {\bibinfo  {journal}
  {Proc. Nat. Acad. Sci.}\ }\textbf {\bibinfo {volume} {114}},\ \bibinfo
  {pages} {3068} (\bibinfo {year} {2017})}\BibitemShut {NoStop}%
\bibitem [{\citenamefont {Tsytovich}(1970)}]{Tsytovich.1970}%
  \BibitemOpen
  \bibfield  {author} {\bibinfo {author} {\bibfnamefont {V.~N.}\ \bibnamefont
  {Tsytovich}},\ }\href {\doibase 10.1007/978-1-4684-1788-3} {\emph {\bibinfo
  {title} {{Nonlinear Effects in Plasma}}}}\ (\bibinfo  {publisher}
  {Springer},\ \bibinfo {address} {Boston},\ \bibinfo {year}
  {1970})\BibitemShut {NoStop}%
\bibitem [{\citenamefont {Gurzhi}(1968)}]{Gurzhi1968}%
  \BibitemOpen
  \bibfield  {author} {\bibinfo {author} {\bibfnamefont {R.~N.}\ \bibnamefont
  {Gurzhi}},\ }\href {\doibase 10.1070/PU1968v011n02ABEH003815} {\bibfield
  {journal} {\bibinfo  {journal} {Sov. Phys. Uspekhi}\ }\textbf {\bibinfo
  {volume} {11}},\ \bibinfo {pages} {255} (\bibinfo {year} {1968})}\BibitemShut
  {NoStop}%
\bibitem [{\citenamefont {Andreev}\ \emph {et~al.}(2011)\citenamefont
  {Andreev}, \citenamefont {Kivelson},\ and\ \citenamefont
  {Spivak}}]{Andreev2011}%
  \BibitemOpen
  \bibfield  {author} {\bibinfo {author} {\bibfnamefont {A.~V.}\ \bibnamefont
  {Andreev}}, \bibinfo {author} {\bibfnamefont {S.~A.}\ \bibnamefont
  {Kivelson}}, \ and\ \bibinfo {author} {\bibfnamefont {B.}~\bibnamefont
  {Spivak}},\ }\href {\doibase 10.1103/PhysRevLett.106.256804} {\bibfield
  {journal} {\bibinfo  {journal} {Phys. Rev. Lett.}\ }\textbf {\bibinfo
  {volume} {106}},\ \bibinfo {pages} {256804} (\bibinfo {year}
  {2011})}\BibitemShut {NoStop}%
\bibitem [{\citenamefont {Landau}\ and\ \citenamefont
  {Lifshitz}(1987)}]{Landau.6}%
  \BibitemOpen
  \bibfield  {author} {\bibinfo {author} {\bibfnamefont {L.~D.}\ \bibnamefont
  {Landau}}\ and\ \bibinfo {author} {\bibfnamefont {E.~M.}\ \bibnamefont
  {Lifshitz}},\ }\href@noop {} {\emph {\bibinfo {title} {{Fluid Mechanics}}}},\
  \bibinfo {edition} {2nd}\ ed.\ (\bibinfo  {publisher} {Pergamon Press},\
  \bibinfo {address} {Oxford},\ \bibinfo {year} {1987})\BibitemShut {NoStop}%
\bibitem [{SM()}]{SM}%
  \BibitemOpen
  \href@noop {} {}\bibinfo {note} {See {S}upplemental {M}aterial at [{URL} to
  be inserted by publisher] for technical details.}\BibitemShut {Stop}%
\bibitem [{\citenamefont {Glazov}(2011)}]{Glazov2011a}%
  \BibitemOpen
  \bibfield  {author} {\bibinfo {author} {\bibfnamefont {M.~M.}\ \bibnamefont
  {Glazov}},\ }\href {\doibase 10.1134/S0021364011070046} {\bibfield  {journal}
  {\bibinfo  {journal} {JETP Lett.}\ }\textbf {\bibinfo {volume} {93}},\
  \bibinfo {pages} {366} (\bibinfo {year} {2011})}\BibitemShut {NoStop}%
\bibitem [{\citenamefont {Mikhailov}(2011)}]{Mikhailov2011}%
  \BibitemOpen
  \bibfield  {author} {\bibinfo {author} {\bibfnamefont {S.~A.}\ \bibnamefont
  {Mikhailov}},\ }\href {\doibase 10.1103/PhysRevB.84.045432} {\bibfield
  {journal} {\bibinfo  {journal} {Phys. Rev. B}\ }\textbf {\bibinfo {volume}
  {84}},\ \bibinfo {pages} {045432} (\bibinfo {year} {2011})}\BibitemShut
  {NoStop}%
\bibitem [{\citenamefont {Yao}\ \emph {et~al.}(2014)\citenamefont {Yao},
  \citenamefont {Tokman},\ and\ \citenamefont {Belyanin}}]{Yao2014}%
  \BibitemOpen
  \bibfield  {author} {\bibinfo {author} {\bibfnamefont {X.}~\bibnamefont
  {Yao}}, \bibinfo {author} {\bibfnamefont {M.}~\bibnamefont {Tokman}}, \ and\
  \bibinfo {author} {\bibfnamefont {A.}~\bibnamefont {Belyanin}},\ }\href
  {\doibase 10.1103/PhysRevLett.112.055501} {\bibfield  {journal} {\bibinfo
  {journal} {Phys. Rev. Lett.}\ }\textbf {\bibinfo {volume} {112}},\ \bibinfo
  {pages} {055501} (\bibinfo {year} {2014})}\BibitemShut {NoStop}%
\bibitem [{\citenamefont {Mikhailov}(2016)}]{Mikhailov2016}%
  \BibitemOpen
  \bibfield  {author} {\bibinfo {author} {\bibfnamefont {S.~A.}\ \bibnamefont
  {Mikhailov}},\ }\href {\doibase 10.1103/PhysRevB.93.085403} {\bibfield
  {journal} {\bibinfo  {journal} {Phys. Rev. B}\ }\textbf {\bibinfo {volume}
  {93}},\ \bibinfo {pages} {085403} (\bibinfo {year} {2016})}\BibitemShut
  {NoStop}%
\bibitem [{\citenamefont {Wang}\ \emph {et~al.}(2016)\citenamefont {Wang},
  \citenamefont {Tokman},\ and\ \citenamefont {Belyanin}}]{Wang2016}%
  \BibitemOpen
  \bibfield  {author} {\bibinfo {author} {\bibfnamefont {Y.}~\bibnamefont
  {Wang}}, \bibinfo {author} {\bibfnamefont {M.}~\bibnamefont {Tokman}}, \ and\
  \bibinfo {author} {\bibfnamefont {A.}~\bibnamefont {Belyanin}},\ }\href
  {\doibase 10.1103/PhysRevB.94.195442} {\bibfield  {journal} {\bibinfo
  {journal} {Phys. Rev. B}\ }\textbf {\bibinfo {volume} {94}},\ \bibinfo
  {pages} {195442} (\bibinfo {year} {2016})}\BibitemShut {NoStop}%
\bibitem [{\citenamefont {Tokman}\ \emph {et~al.}(2016)\citenamefont {Tokman},
  \citenamefont {Wang}, \citenamefont {Oladyshkin}, \citenamefont {Kutayiah},\
  and\ \citenamefont {Belyanin}}]{Tokman2016}%
  \BibitemOpen
  \bibfield  {author} {\bibinfo {author} {\bibfnamefont {M.}~\bibnamefont
  {Tokman}}, \bibinfo {author} {\bibfnamefont {Y.}~\bibnamefont {Wang}},
  \bibinfo {author} {\bibfnamefont {I.}~\bibnamefont {Oladyshkin}}, \bibinfo
  {author} {\bibfnamefont {A.~R.}\ \bibnamefont {Kutayiah}}, \ and\ \bibinfo
  {author} {\bibfnamefont {A.}~\bibnamefont {Belyanin}},\ }\href {\doibase
  10.1103/PhysRevB.93.235422} {\bibfield  {journal} {\bibinfo  {journal} {Phys.
  Rev. B}\ }\textbf {\bibinfo {volume} {93}},\ \bibinfo {pages} {235422}
  (\bibinfo {year} {2016})}\BibitemShut {NoStop}%
\bibitem [{\citenamefont {Cheng}\ \emph {et~al.}(2017)\citenamefont {Cheng},
  \citenamefont {Vermeulen},\ and\ \citenamefont {Sipe}}]{Cheng2016}%
  \BibitemOpen
  \bibfield  {author} {\bibinfo {author} {\bibfnamefont {J.~L.}\ \bibnamefont
  {Cheng}}, \bibinfo {author} {\bibfnamefont {N.}~\bibnamefont {Vermeulen}}, \
  and\ \bibinfo {author} {\bibfnamefont {J.~E.}\ \bibnamefont {Sipe}},\ }\href
  {\doibase 10.1038/srep43843} {\bibfield  {journal} {\bibinfo  {journal} {Sci.
  Rep.}\ }\textbf {\bibinfo {volume} {7}},\ \bibinfo {pages} {43843} (\bibinfo
  {year} {2017})}\BibitemShut {NoStop}%
\bibitem [{\citenamefont {Manzoni}\ \emph {et~al.}(2015)\citenamefont
  {Manzoni}, \citenamefont {Silveiro}, \citenamefont {Abajo},\ and\
  \citenamefont {Chang}}]{Manzoni2015}%
  \BibitemOpen
  \bibfield  {author} {\bibinfo {author} {\bibfnamefont {M.~T.}\ \bibnamefont
  {Manzoni}}, \bibinfo {author} {\bibfnamefont {I.}~\bibnamefont {Silveiro}},
  \bibinfo {author} {\bibfnamefont {F.~J. G.~D.}\ \bibnamefont {Abajo}}, \ and\
  \bibinfo {author} {\bibfnamefont {D.~E.}\ \bibnamefont {Chang}},\ }\href
  {\doibase 10.1088/1367-2630/17/8/083031} {\bibfield  {journal} {\bibinfo
  {journal} {New J. Phys.}\ }\textbf {\bibinfo {volume} {17}},\ \bibinfo
  {pages} {83031} (\bibinfo {year} {2015})}\BibitemShut {NoStop}%
\bibitem [{\citenamefont {Rostami}\ \emph {et~al.}(2017)\citenamefont
  {Rostami}, \citenamefont {Katsnelson},\ and\ \citenamefont
  {Polini}}]{Rostami2016}%
  \BibitemOpen
  \bibfield  {author} {\bibinfo {author} {\bibfnamefont {H.}~\bibnamefont
  {Rostami}}, \bibinfo {author} {\bibfnamefont {M.~I.}\ \bibnamefont
  {Katsnelson}}, \ and\ \bibinfo {author} {\bibfnamefont {M.}~\bibnamefont
  {Polini}},\ }\href {\doibase 10.1103/PhysRevB.95.035416} {\bibfield
  {journal} {\bibinfo  {journal} {Phys. Rev. B}\ }\textbf {\bibinfo {volume}
  {95}},\ \bibinfo {pages} {035416} (\bibinfo {year} {2017})}\BibitemShut
  {NoStop}%
\bibitem [{\citenamefont {Khurgin}(2014)}]{Khurgin2014}%
  \BibitemOpen
  \bibfield  {author} {\bibinfo {author} {\bibfnamefont {J.~B.}\ \bibnamefont
  {Khurgin}},\ }\href {\doibase 10.1063/1.4873704} {\bibfield  {journal}
  {\bibinfo  {journal} {Appl. Phys. Lett.}\ }\textbf {\bibinfo {volume}
  {104}},\ \bibinfo {pages} {161116} (\bibinfo {year} {2014})}\BibitemShut
  {NoStop}%
\bibitem [{\citenamefont {Forcella}\ \emph {et~al.}(2014)\citenamefont
  {Forcella}, \citenamefont {Zaanen}, \citenamefont {Valentinis},\ and\
  \citenamefont {van~der Marel}}]{Forcella2014}%
  \BibitemOpen
  \bibfield  {author} {\bibinfo {author} {\bibfnamefont {D.}~\bibnamefont
  {Forcella}}, \bibinfo {author} {\bibfnamefont {J.}~\bibnamefont {Zaanen}},
  \bibinfo {author} {\bibfnamefont {D.}~\bibnamefont {Valentinis}}, \ and\
  \bibinfo {author} {\bibfnamefont {D.}~\bibnamefont {van~der Marel}},\ }\href
  {\doibase 10.1103/physrevb.90.035143} {\bibfield  {journal} {\bibinfo
  {journal} {Phys. Rev. B}\ }\textbf {\bibinfo {volume} {90}},\ \bibinfo
  {pages} {035143} (\bibinfo {year} {2014})}\BibitemShut {NoStop}%
\bibitem [{\citenamefont {Aliev}\ \emph {et~al.}(1992)\citenamefont {Aliev},
  \citenamefont {Bychenkov}, \citenamefont {Jovanovi{\'{c}}},\ and\
  \citenamefont {Frolov}}]{Aliev1992}%
  \BibitemOpen
  \bibfield  {author} {\bibinfo {author} {\bibfnamefont {Y.~M.}\ \bibnamefont
  {Aliev}}, \bibinfo {author} {\bibfnamefont {V.~Y.}\ \bibnamefont
  {Bychenkov}}, \bibinfo {author} {\bibfnamefont {M.~S.}\ \bibnamefont
  {Jovanovi{\'{c}}}}, \ and\ \bibinfo {author} {\bibfnamefont {A.~A.}\
  \bibnamefont {Frolov}},\ }\href {\doibase 10.1017/S0022377800016457}
  {\bibfield  {journal} {\bibinfo  {journal} {J. Plasma Phys.}\ }\textbf
  {\bibinfo {volume} {48}},\ \bibinfo {pages} {167} (\bibinfo {year}
  {1992})}\BibitemShut {NoStop}%
\bibitem [{\citenamefont {Stolz}(1967)}]{Stolz1967}%
  \BibitemOpen
  \bibfield  {author} {\bibinfo {author} {\bibfnamefont {H.}~\bibnamefont
  {Stolz}},\ }\href {\doibase 10.1002/pssb.19670210105} {\bibfield  {journal}
  {\bibinfo  {journal} {Phys. Status Solidi}\ }\textbf {\bibinfo {volume}
  {21}},\ \bibinfo {pages} {77} (\bibinfo {year} {1967})}\BibitemShut {NoStop}%
\bibitem [{\citenamefont {Kotov}\ \emph {et~al.}(2012)\citenamefont {Kotov},
  \citenamefont {Uchoa}, \citenamefont {Pereira}, \citenamefont {Guinea},\ and\
  \citenamefont {Castro~Neto}}]{Kotov2012}%
  \BibitemOpen
  \bibfield  {author} {\bibinfo {author} {\bibfnamefont {V.~N.}\ \bibnamefont
  {Kotov}}, \bibinfo {author} {\bibfnamefont {B.}~\bibnamefont {Uchoa}},
  \bibinfo {author} {\bibfnamefont {V.~M.}\ \bibnamefont {Pereira}}, \bibinfo
  {author} {\bibfnamefont {F.}~\bibnamefont {Guinea}}, \ and\ \bibinfo {author}
  {\bibfnamefont {A.~H.}\ \bibnamefont {Castro~Neto}},\ }\href {\doibase
  10.1103/revmodphys.84.1067} {\bibfield  {journal} {\bibinfo  {journal} {Rev.
  Mod. Phys.}\ }\textbf {\bibinfo {volume} {84}},\ \bibinfo {pages} {1067}
  (\bibinfo {year} {2012})}\BibitemShut {NoStop}%
\bibitem [{\citenamefont {Basov}\ \emph {et~al.}(2014)\citenamefont {Basov},
  \citenamefont {Fogler}, \citenamefont {Lanzara}, \citenamefont {Wang},\ and\
  \citenamefont {Zhang}}]{Basov2014}%
  \BibitemOpen
  \bibfield  {author} {\bibinfo {author} {\bibfnamefont {D.~N.}\ \bibnamefont
  {Basov}}, \bibinfo {author} {\bibfnamefont {M.~M.}\ \bibnamefont {Fogler}},
  \bibinfo {author} {\bibfnamefont {A.}~\bibnamefont {Lanzara}}, \bibinfo
  {author} {\bibfnamefont {F.}~\bibnamefont {Wang}}, \ and\ \bibinfo {author}
  {\bibfnamefont {Y.}~\bibnamefont {Zhang}},\ }\href {\doibase
  10.1103/RevModPhys.86.959} {\bibfield  {journal} {\bibinfo  {journal} {Rev.
  Mod. Phys.}\ }\textbf {\bibinfo {volume} {86}},\ \bibinfo {pages} {959}
  (\bibinfo {year} {2014})}\BibitemShut {NoStop}%
\bibitem [{\citenamefont {Link}\ \emph {et~al.}(2016)\citenamefont {Link},
  \citenamefont {Orth}, \citenamefont {Sheehy},\ and\ \citenamefont
  {Schmalian}}]{Link2016}%
  \BibitemOpen
  \bibfield  {author} {\bibinfo {author} {\bibfnamefont {J.~M.}\ \bibnamefont
  {Link}}, \bibinfo {author} {\bibfnamefont {P.~P.}\ \bibnamefont {Orth}},
  \bibinfo {author} {\bibfnamefont {D.~E.}\ \bibnamefont {Sheehy}}, \ and\
  \bibinfo {author} {\bibfnamefont {J.}~\bibnamefont {Schmalian}},\ }\href
  {\doibase 10.1103/physrevb.93.235447} {\bibfield  {journal} {\bibinfo
  {journal} {Phys. Rev. B}\ }\textbf {\bibinfo {volume} {93}},\ \bibinfo
  {pages} {235447} (\bibinfo {year} {2016})}\BibitemShut {NoStop}%
\bibitem [{\citenamefont {Landau}\ and\ \citenamefont
  {Lifshitz}(1984)}]{Landau.8}%
  \BibitemOpen
  \bibfield  {author} {\bibinfo {author} {\bibfnamefont {L.~D.}\ \bibnamefont
  {Landau}}\ and\ \bibinfo {author} {\bibfnamefont {E.~M.}\ \bibnamefont
  {Lifshitz}},\ }\href@noop {} {\emph {\bibinfo {title} {{Electrodynamics of
  Continuous Media}}}}\ (\bibinfo  {publisher} {Pergamon},\ \bibinfo {address}
  {Oxford},\ \bibinfo {year} {1984})\BibitemShut {NoStop}%
\bibitem [{\citenamefont {Il'inskii}\ and\ \citenamefont
  {Keldysh}(1994)}]{Ilinskii1994}%
  \BibitemOpen
  \bibfield  {author} {\bibinfo {author} {\bibfnamefont {Y.~A.}\ \bibnamefont
  {Il'inskii}}\ and\ \bibinfo {author} {\bibfnamefont {L.~V.}\ \bibnamefont
  {Keldysh}},\ }\href {\doibase 10.1007/978-1-4899-1570-2} {\emph {\bibinfo
  {title} {{Electromagnetic Response of Material Media}}}}\ (\bibinfo
  {publisher} {Springer},\ \bibinfo {address} {Boston},\ \bibinfo {year}
  {1994})\BibitemShut {NoStop}%
\bibitem [{\citenamefont {Kovtun}(2012)}]{Kovtun2012}%
  \BibitemOpen
  \bibfield  {author} {\bibinfo {author} {\bibfnamefont {P.}~\bibnamefont
  {Kovtun}},\ }\href {\doibase 10.1088/1751-8113/45/47/473001} {\bibfield
  {journal} {\bibinfo  {journal} {J. Phys. A Math. Theor.}\ }\textbf {\bibinfo
  {volume} {45}},\ \bibinfo {pages} {473001} (\bibinfo {year}
  {2012})}\BibitemShut {NoStop}%
\bibitem [{\citenamefont {Phan}\ \emph {et~al.}()\citenamefont {Phan},
  \citenamefont {Song},\ and\ \citenamefont {Levitov}}]{Phan2013}%
  \BibitemOpen
  \bibfield  {author} {\bibinfo {author} {\bibfnamefont {T.~V.}\ \bibnamefont
  {Phan}}, \bibinfo {author} {\bibfnamefont {J.~C.~W.}\ \bibnamefont {Song}}, \
  and\ \bibinfo {author} {\bibfnamefont {L.~S.}\ \bibnamefont {Levitov}},\
  }\href {http://arxiv.org/abs/1306.4972} {\enquote {\bibinfo {title}
  {{Ballistic Heat Transfer and Energy Waves in an Electron System}},}\
  }\bibinfo {note} {(unpublished)},\ \Eprint {http://arxiv.org/abs/1306.4972}
  {arXiv:1306.4972} \BibitemShut {NoStop}%
\bibitem [{\citenamefont {Tomadin}\ and\ \citenamefont
  {Polini}(2013)}]{Tomadin2013}%
  \BibitemOpen
  \bibfield  {author} {\bibinfo {author} {\bibfnamefont {A.}~\bibnamefont
  {Tomadin}}\ and\ \bibinfo {author} {\bibfnamefont {M.}~\bibnamefont
  {Polini}},\ }\href {\doibase 10.1103/PhysRevB.88.205426} {\bibfield
  {journal} {\bibinfo  {journal} {Phys. Rev. B}\ }\textbf {\bibinfo {volume}
  {88}},\ \bibinfo {pages} {205426} (\bibinfo {year} {2013})}\BibitemShut
  {NoStop}%
\bibitem [{\citenamefont {Glazov}\ and\ \citenamefont
  {Ganichev}(2014)}]{Glazov2014}%
  \BibitemOpen
  \bibfield  {author} {\bibinfo {author} {\bibfnamefont {M.}~\bibnamefont
  {Glazov}}\ and\ \bibinfo {author} {\bibfnamefont {S.}~\bibnamefont
  {Ganichev}},\ }\href {\doibase 10.1016/j.physrep.2013.10.003} {\bibfield
  {journal} {\bibinfo  {journal} {Phys. Rep.}\ }\textbf {\bibinfo {volume}
  {535}},\ \bibinfo {pages} {101} (\bibinfo {year} {2014})}\BibitemShut
  {NoStop}%
\bibitem [{\citenamefont {Jiang}\ \emph {et~al.}(2011)\citenamefont {Jiang},
  \citenamefont {Shalygin}, \citenamefont {Panevin}, \citenamefont {Danilov},
  \citenamefont {Glazov}, \citenamefont {Yakimova}, \citenamefont {Lara-Avila},
  \citenamefont {Kubatkin},\ and\ \citenamefont {Ganichev}}]{Jiang2011}%
  \BibitemOpen
  \bibfield  {author} {\bibinfo {author} {\bibfnamefont {C.}~\bibnamefont
  {Jiang}}, \bibinfo {author} {\bibfnamefont {V.~A.}\ \bibnamefont {Shalygin}},
  \bibinfo {author} {\bibfnamefont {V.~Y.}\ \bibnamefont {Panevin}}, \bibinfo
  {author} {\bibfnamefont {S.~N.}\ \bibnamefont {Danilov}}, \bibinfo {author}
  {\bibfnamefont {M.~M.}\ \bibnamefont {Glazov}}, \bibinfo {author}
  {\bibfnamefont {R.}~\bibnamefont {Yakimova}}, \bibinfo {author}
  {\bibfnamefont {S.}~\bibnamefont {Lara-Avila}}, \bibinfo {author}
  {\bibfnamefont {S.}~\bibnamefont {Kubatkin}}, \ and\ \bibinfo {author}
  {\bibfnamefont {S.~D.}\ \bibnamefont {Ganichev}},\ }\href {\doibase
  10.1103/PhysRevB.84.125429} {\bibfield  {journal} {\bibinfo  {journal} {Phys.
  Rev. B}\ }\textbf {\bibinfo {volume} {84}},\ \bibinfo {pages} {125429}
  (\bibinfo {year} {2011})}\BibitemShut {NoStop}%
\bibitem [{\citenamefont {Karch}\ \emph {et~al.}(2010)\citenamefont {Karch},
  \citenamefont {Olbrich}, \citenamefont {Schmalzbauer}, \citenamefont {Zoth},
  \citenamefont {Brinsteiner}, \citenamefont {Fehrenbacher}, \citenamefont
  {Wurstbauer}, \citenamefont {Glazov}, \citenamefont {Tarasenko},
  \citenamefont {Ivchenko}, \citenamefont {Weiss}, \citenamefont {Eroms},
  \citenamefont {Yakimova}, \citenamefont {Lara-Avila}, \citenamefont
  {Kubatkin},\ and\ \citenamefont {Ganichev}}]{Karch2010}%
  \BibitemOpen
  \bibfield  {author} {\bibinfo {author} {\bibfnamefont {J.}~\bibnamefont
  {Karch}}, \bibinfo {author} {\bibfnamefont {P.}~\bibnamefont {Olbrich}},
  \bibinfo {author} {\bibfnamefont {M.}~\bibnamefont {Schmalzbauer}}, \bibinfo
  {author} {\bibfnamefont {C.}~\bibnamefont {Zoth}}, \bibinfo {author}
  {\bibfnamefont {C.}~\bibnamefont {Brinsteiner}}, \bibinfo {author}
  {\bibfnamefont {M.}~\bibnamefont {Fehrenbacher}}, \bibinfo {author}
  {\bibfnamefont {U.}~\bibnamefont {Wurstbauer}}, \bibinfo {author}
  {\bibfnamefont {M.~M.}\ \bibnamefont {Glazov}}, \bibinfo {author}
  {\bibfnamefont {S.~A.}\ \bibnamefont {Tarasenko}}, \bibinfo {author}
  {\bibfnamefont {E.~L.}\ \bibnamefont {Ivchenko}}, \bibinfo {author}
  {\bibfnamefont {D.}~\bibnamefont {Weiss}}, \bibinfo {author} {\bibfnamefont
  {J.}~\bibnamefont {Eroms}}, \bibinfo {author} {\bibfnamefont
  {R.}~\bibnamefont {Yakimova}}, \bibinfo {author} {\bibfnamefont
  {S.}~\bibnamefont {Lara-Avila}}, \bibinfo {author} {\bibfnamefont
  {S.}~\bibnamefont {Kubatkin}}, \ and\ \bibinfo {author} {\bibfnamefont
  {S.~D.}\ \bibnamefont {Ganichev}},\ }\href {\doibase
  10.1103/PhysRevLett.105.227402} {\bibfield  {journal} {\bibinfo  {journal}
  {Phys. Rev. Lett.}\ }\textbf {\bibinfo {volume} {105}},\ \bibinfo {pages}
  {227402} (\bibinfo {year} {2010})}\BibitemShut {NoStop}%
\bibitem [{\citenamefont {Principi}\ \emph {et~al.}(2014)\citenamefont
  {Principi}, \citenamefont {Carrega}, \citenamefont {Lundeberg}, \citenamefont
  {Woessner}, \citenamefont {Koppens}, \citenamefont {Vignale},\ and\
  \citenamefont {Polini}}]{Principi2014}%
  \BibitemOpen
  \bibfield  {author} {\bibinfo {author} {\bibfnamefont {A.}~\bibnamefont
  {Principi}}, \bibinfo {author} {\bibfnamefont {M.}~\bibnamefont {Carrega}},
  \bibinfo {author} {\bibfnamefont {M.~B.}\ \bibnamefont {Lundeberg}}, \bibinfo
  {author} {\bibfnamefont {A.}~\bibnamefont {Woessner}}, \bibinfo {author}
  {\bibfnamefont {F.~H.~L.}\ \bibnamefont {Koppens}}, \bibinfo {author}
  {\bibfnamefont {G.}~\bibnamefont {Vignale}}, \ and\ \bibinfo {author}
  {\bibfnamefont {M.}~\bibnamefont {Polini}},\ }\href {\doibase
  10.1103/PhysRevB.90.165408} {\bibfield  {journal} {\bibinfo  {journal} {Phys.
  Rev. B}\ }\textbf {\bibinfo {volume} {90}},\ \bibinfo {pages} {165408}
  (\bibinfo {year} {2014})}\BibitemShut {NoStop}%
\bibitem [{\citenamefont {Ni}\ \emph {et~al.}(2016)\citenamefont {Ni},
  \citenamefont {Wang}, \citenamefont {Goldflam}, \citenamefont {Wagner},
  \citenamefont {Fei}, \citenamefont {McLeod}, \citenamefont {Liu},
  \citenamefont {Keilmann}, \citenamefont {{\"{O}}zyilmaz}, \citenamefont
  {{Castro Neto}}, \citenamefont {Hone}, \citenamefont {Fogler},\ and\
  \citenamefont {Basov}}]{Ni2016}%
  \BibitemOpen
  \bibfield  {author} {\bibinfo {author} {\bibfnamefont {G.~X.}\ \bibnamefont
  {Ni}}, \bibinfo {author} {\bibfnamefont {L.}~\bibnamefont {Wang}}, \bibinfo
  {author} {\bibfnamefont {M.~D.}\ \bibnamefont {Goldflam}}, \bibinfo {author}
  {\bibfnamefont {M.}~\bibnamefont {Wagner}}, \bibinfo {author} {\bibfnamefont
  {Z.}~\bibnamefont {Fei}}, \bibinfo {author} {\bibfnamefont {A.~S.}\
  \bibnamefont {McLeod}}, \bibinfo {author} {\bibfnamefont {M.~K.}\
  \bibnamefont {Liu}}, \bibinfo {author} {\bibfnamefont {F.}~\bibnamefont
  {Keilmann}}, \bibinfo {author} {\bibfnamefont {B.}~\bibnamefont
  {{\"{O}}zyilmaz}}, \bibinfo {author} {\bibfnamefont {A.~H.}\ \bibnamefont
  {{Castro Neto}}}, \bibinfo {author} {\bibfnamefont {J.}~\bibnamefont {Hone}},
  \bibinfo {author} {\bibfnamefont {M.~M.}\ \bibnamefont {Fogler}}, \ and\
  \bibinfo {author} {\bibfnamefont {D.~N.}\ \bibnamefont {Basov}},\ }\href
  {\doibase 10.1038/nphoton.2016.45} {\bibfield  {journal} {\bibinfo  {journal}
  {Nature Photon.}\ }\textbf {\bibinfo {volume} {10}},\ \bibinfo {pages} {244}
  (\bibinfo {year} {2016})}\BibitemShut {NoStop}%
\bibitem [{\citenamefont {Ni}\ \emph {et~al.}()\citenamefont {Ni},
  \citenamefont {McLeod}, \citenamefont {Wang}, \citenamefont {Xiong},
  \citenamefont {Charnukha}, \citenamefont {Post}, \citenamefont {Keilmann},
  \citenamefont {Hone}, \citenamefont {Dean}, \citenamefont {Fogler},\ and\
  \citenamefont {Basov}}]{Ni2017}%
  \BibitemOpen
  \bibfield  {author} {\bibinfo {author} {\bibfnamefont {G.~X.}\ \bibnamefont
  {Ni}}, \bibinfo {author} {\bibfnamefont {A.~S.}\ \bibnamefont {McLeod}},
  \bibinfo {author} {\bibfnamefont {L.}~\bibnamefont {Wang}}, \bibinfo {author}
  {\bibfnamefont {L.}~\bibnamefont {Xiong}}, \bibinfo {author} {\bibfnamefont
  {A.}~\bibnamefont {Charnukha}}, \bibinfo {author} {\bibfnamefont
  {K.}~\bibnamefont {Post}}, \bibinfo {author} {\bibfnamefont {F.}~\bibnamefont
  {Keilmann}}, \bibinfo {author} {\bibfnamefont {J.}~\bibnamefont {Hone}},
  \bibinfo {author} {\bibfnamefont {C.~R.}\ \bibnamefont {Dean}}, \bibinfo
  {author} {\bibfnamefont {M.~M.}\ \bibnamefont {Fogler}}, \ and\ \bibinfo
  {author} {\bibfnamefont {D.~N.}\ \bibnamefont {Basov}},\ }\href@noop {}
  {\enquote {\bibinfo {title} {Ballistic plasmon polaritons in high mobility
  electron liquid of graphene},}\ }\bibinfo {note} {{i}n
  preparation}\BibitemShut {NoStop}%
\bibitem [{\citenamefont {Sch{\"{u}}tt}\ \emph {et~al.}(2011)\citenamefont
  {Sch{\"{u}}tt}, \citenamefont {Ostrovsky}, \citenamefont {Gornyi},\ and\
  \citenamefont {Mirlin}}]{Schutt2011}%
  \BibitemOpen
  \bibfield  {author} {\bibinfo {author} {\bibfnamefont {M.}~\bibnamefont
  {Sch{\"{u}}tt}}, \bibinfo {author} {\bibfnamefont {P.~M.}\ \bibnamefont
  {Ostrovsky}}, \bibinfo {author} {\bibfnamefont {I.~V.}\ \bibnamefont
  {Gornyi}}, \ and\ \bibinfo {author} {\bibfnamefont {A.~D.}\ \bibnamefont
  {Mirlin}},\ }\href {\doibase 10.1103/PhysRevB.83.155441} {\bibfield
  {journal} {\bibinfo  {journal} {Phys. Rev. B}\ }\textbf {\bibinfo {volume}
  {83}},\ \bibinfo {pages} {155441} (\bibinfo {year} {2011})}\BibitemShut
  {NoStop}%
\bibitem [{\citenamefont {Basov}\ and\ \citenamefont
  {Timusk}(2005)}]{Basov2005}%
  \BibitemOpen
  \bibfield  {author} {\bibinfo {author} {\bibfnamefont {D.~N.}\ \bibnamefont
  {Basov}}\ and\ \bibinfo {author} {\bibfnamefont {T.}~\bibnamefont {Timusk}},\
  }\href {\doibase 10.1103/RevModPhys.77.721} {\bibfield  {journal} {\bibinfo
  {journal} {Rev. Mod. Phys.}\ }\textbf {\bibinfo {volume} {77}},\ \bibinfo
  {pages} {721} (\bibinfo {year} {2005})}\BibitemShut {NoStop}%
\end{thebibliography}%

%%%%%%%%%% Merge with supplemental materials %%%%%%%%%%
\pagebreak
\begin{widetext}
	\begin{center}
		\textbf{\large Supplementary material for
			``Linear and nonlinear electrodynamics of a Dirac fluid''}
	\end{center}
\end{widetext}
%%%%%%%%%% Merge with supplemental materials %%%%%%%%%%
%%%%%%%%%% Prefix a "S" to all equations, figures, tables and reset the counter %%%%%%%%%%
\setcounter{equation}{0}
\setcounter{figure}{0}
\setcounter{table}{0}
\setcounter{page}{1}
\makeatletter
\renewcommand{\theequation}{S\arabic{equation}}
\renewcommand{\thefigure}{S\arabic{figure}}
%\renewcommand{\bibnumfmt}[1]{[S#1]}
%\renewcommand{\citenumfont}[1]{S#1}
%%%%%%%%%% Prefix a "S" to all equations, figures, tables and reset the counter %%%%%%%%%%

\section{Linear ac conductivity} 

\subsection{Drude weight and demons in the hydrodynamic regime}

As shown in literature~\cite{Hartnoll2007,
	Muller2008a, Muller2009, Kovtun2012, Briskot2015, Narozhny2015, Sun2016, Lucas2016}, the linear-response ac conductivity of a Dirac fluid at $q = 0$ is given by
\begin{equation}
	\sigma(0, \omega) = \frac{D_h/\pi}{-i\omega + \Gamma_d} + \sigma_0\,,
	\label{eq:sigma_hydro}
\end{equation}
which is Eq.~(\textcolor{blue}{5}) of the main text.
The hydrodynamic Drude weight that enters Eq.~\eqref{eq:sigma_hydro} is
\begin{equation}
	D_h = \frac{\pi e^2\! n}{m^{\ast}}  \,.
	\label{eqn:D_h}
\end{equation}
At zero temperature the hydrodynamic mass $m^\ast$ is
no different from the Fermi-liquid effective mass $m^{\ast} = \hbar k_F / v_F$, where $v_F$ is the Fermi velocity.
Hence, $D_h$ is equal to the conventional (kinetic) Drude weight $D_k$.
For example, for parabolic band, $m^\ast$ is simply the band mass $m$.
For graphene with weak ee interactions,
\begin{equation}
	D_h(T = 0) = \frac{g}{4} \frac{e^2 k_F v}{\hbar}  \,,
\end{equation}
where $g = 4$ is the total spin-valley degeneracy~\cite{Basov2014}.

As usual, at finite $q$, the conductivity becomes a tensor
\begin{equation}
	\sigma_{i j}\left(\vec{q}, \omega\right)
	= \sigma_L(q, \omega) \,\frac{q_i q_j}{q^2} + \sigma_T(q, \omega) \left(\delta_{i j} - \frac{q_i q_j}{q^2}\right).
\end{equation}
Neglecting two subleading dissipative effects [$\sigma_{0}$
in Eq.~\eqref{eq:sigma_hydro} and
viscous damping], 
the longitudinal conductivity $\sigma_L$ is given by~\cite{Kovtun2012, Briskot2015}
\begin{align}
	\sigma_{L}(q, \omega)
	= \frac{D_h/\pi}{-i\omega + \Gamma_d -\frac{v^2_d q^2 }{\omega}}
	\,.
	\label{eqn:sigma_L}
\end{align}
The longitudinal conductivity enters the equation for the dispersion of longitudinal collective modes.
In 2D case, this equation reads~\cite{Basov2014}
\begin{equation}
	q = \frac{i\kappa \omega}{2\pi \sigma_{L}(q, \omega)}\,.
	\label{eqn:q_demon}
\end{equation}
The longitudinal mode in the hydrodynamic regime has been variously referred to as
the sound~\cite{Kovtun2012}, the energy wave~\cite{Phan2013, Briskot2015}, and finally, the demon~\cite{Sun2016}, which is our preference here.
Equations~\eqref{eqn:sigma_L} and \eqref{eqn:q_demon}
imply that away from charge neutrality, $n \neq 0$,
the dispersion of the demon varies from $\omega \propto \sqrt{q}$
at low $q$ to $\omega \simeq v_d q$ at large $q$,
see Fig.~1 of the main text.
For neutral fluid, $D_h \to 0$,
the demon dispersion is acoustic starting from $q = 0$.
The (asymptotic) speed of the demon
is given by $v_d = v \sqrt{C_{\mathrm{ise}}}$ where
$C_{\mathrm{ise}}$
is the dimensionless isentropic bulk modulus
[Eq.~(\textcolor{blue}{10}) of the main text or
Eq.~\eqref{eqn:C_ise} below].
For a degenerate Fermi gas
\begin{equation}
	C_{\mathrm{ise}} 
	= \frac{n}{W} \frac{\hbar k_F v_F}{d}\,,
\end{equation}
where $W \simeq n \varepsilon_F$ is the enthalpy density
and $d$ is the space dimension; therefore,
\begin{equation}
	C_{\mathrm{ise}} = \frac{1}{d}
	\frac{\hbar k_F v_F}{\varepsilon_F} \,.
	\label{eq:C_T=0}
\end{equation}
For Dirac dispersion $\varepsilon_p^2 = (p v)^2 + (m v^2)^2$, the relation $\hbar k_F = \varepsilon_F v_F /v^2$ holds; thus,
\begin{equation}
	C_{\mathrm{ise}}(T = 0) = \frac{1}{d}\, \frac{v^2_F}{v^2}
\end{equation}
and $v_d = v_F / \sqrt{d}$,
same as the speed of the first sound in a neutral Fermi liquid.
For graphene,
\begin{equation}
	C_{\mathrm{ise}} = \frac12
	\label{eq:C_graphene}
\end{equation}
at \textit{any} $T$ (see below). Therefore,
$v_d = v / \sqrt{2}$~\cite{Kovtun2012,
	Phan2013, Briskot2015, Sun2016}.

\subsection{Interpolation formula for the ac conductivity}

In this Section we derive Eq.~(\textcolor{blue}{8}) of the main text,
which smoothly
connects the hydrodynamic and kinetic regimes of the linear-response theory.
We start with the formula for the current
\begin{align}
	\vec{j} = e \sum \vec{v}_{\vec{p}}
	f_{\vec{p}}(t)
	\label{eqn:j_from_f}
\end{align}
in terms of the quasiparticle distribution function $f_{\vec{p}}$
and velocity $\vec{v}_{\vec{p}} =  \partial \varepsilon_{p} / \partial \vec{p}$
as a function of momentum $\vec{p}$.
For simplicity of notations, all the other quantum numbers such as spin, valley, and band index are omitted.
We use $\sum \ldots$ to denote
the summation over these quantum numbers
combined with the integration $\int d^2 p / (2\pi)^2\ldots$
over momentum.

\begin{figure}[t]
	\includegraphics[width= 2.1in]{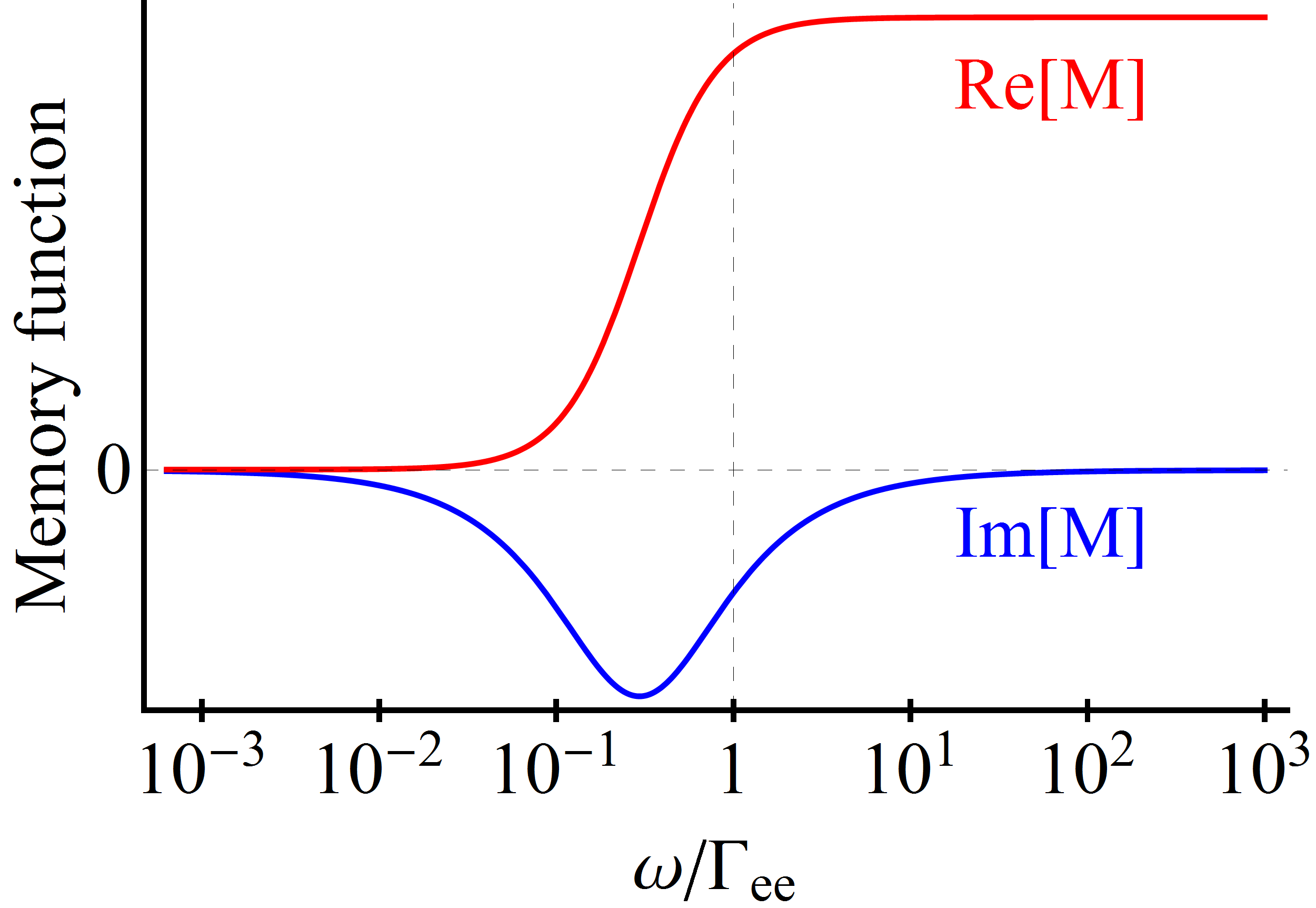} 
	\includegraphics[width= 2.1in]{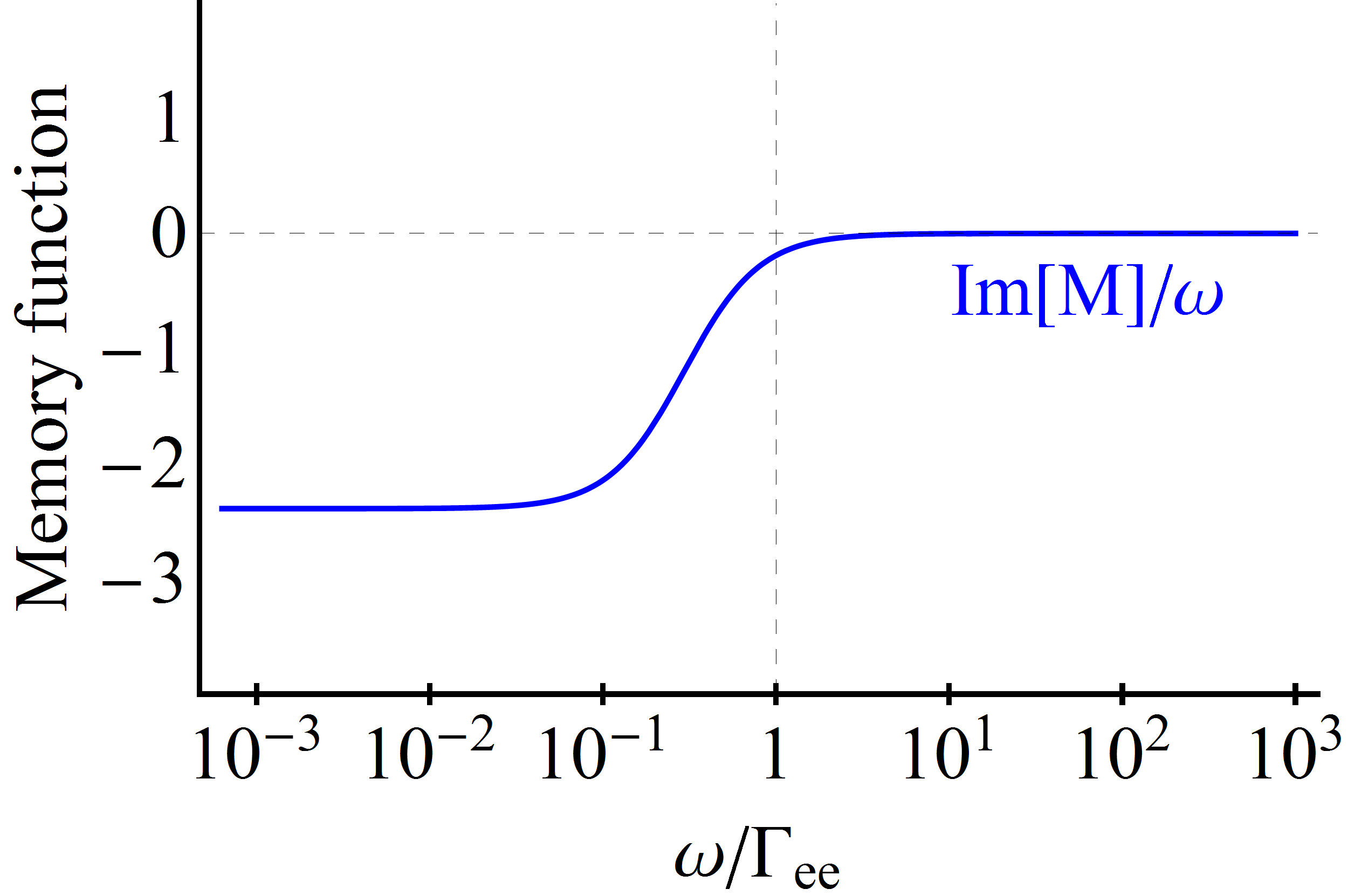} 
	\caption{(Top) Real and imaginary parts of the memory function $M(\omega)$
		in Eq.~\eqref{eqn:memory_function}.
		(Bottom) Mass renormalization factor $\im M(\omega)/\omega \equiv -\lambda$.
		The wide dynamic range of $\omega$
		is used to illustrate the features more clearly.
	}
	\label{fig:memory}
\end{figure}

Let us assume that the electric field in the system
is position-independent and directed along $x$, i.e.,
$\vec{E} = \hat{\mathbf{x}} E(t)$.
We want to compute the current $\vec{j}$ to the first order in $E(t)$.
The result is different in the two regimes
because the deviation $f_{\vec{p}} - f_{\vec{p}}^{(0)}$ of the distribution function from 
the equilibrium value $f_{\vec{p}}^{(0)}
= \left[e^{\left(\varepsilon_{p} - \mu\right) / T} + 1\right]^{-1}$ has different forms.
It is proportional to $g_p = p_x \partial_\varepsilon f_{\vec{p}}^{(0)}$
in the hydrodynamic limit but
to $g_v = v_x \partial_\varepsilon f_{\vec{p}}^{(0)}$ in the kinetic limit.
To obtain the desired interpolation,
we postulate that in general, $f_{\vec{p}} - f_{\vec{p}}^{(0)}$ is a certain linear combination
\begin{align}
	f_{\vec{p}} - f_{\vec{p}}^{(0)} = a_v(t) g_v + a_p(t) g_p\,.
	\label{eqn:distribution_decompose}
\end{align}
To find the coefficients $a_v$ and $a_p$ we consider the Boltzmann kinetic equation
\begin{align}
	\left(\partial_t + e \vec{E} \partial_{\vec{p}} \right)f_{\vec{p}}
	= -\partial_t a_v g_v + \partial_t a_p g_p + e E g_v
	= \hat{I}\left[f_{\vec{p}}\right] \,.
	\label{eqn:boltzmann0}
\end{align}
We further assume that the linearized collision operator $\hat{I}$
acts within the space of functions given by
Eq.~\eqref{eqn:distribution_decompose} and is
characterized by two parameters:
$\Gamma_d$, the scattering rate due to disorder and phonons, and $\Gamma_{ee}$, the electron-electron (ee) scattering rate.
Mode $g_v$ is damped by both types of scattering but
$g_p$ is immune to the ee one,
which implies
\begin{align}
	\hat{I}[g_p] = -\Gamma_d g_p\,,
	\quad 
	\hat{I}[g_v] = -(\Gamma_d + \Gamma_{ee}) g_v
	+ a_{pv} \Gamma_{ee} g_p\,.
\end{align}
The condition that ee scattering conserves momentum
fixes the coefficient $a_{pv} = D_h / (\pi e^2 n) = 1/m^{\ast}$
[Eq.~\eqref{eqn:D_h}],
leading us to
\begin{equation}
	\partial_t a_v + e E = - (\Gamma_d + \Gamma_{ee}) a_v\,,
	\:\:\:\:
	\partial_t a_p = -\Gamma_d a_p + \frac{\Gamma_{ee}}{m^{\ast}} a_v\,.
\end{equation}
For $E(t) \propto e^{-i \omega t}$, the solution is
\begin{align}
	a_v = \frac{e E}{-i\omega + \Gamma_d + \Gamma_{ee}}\,,
	\quad
	a_p = \frac{\Gamma_{ee}/ m^{\ast}}{-i\omega + \Gamma_d} \,
	a_v\,.
	\label{eqn:a_v_and_a_p}
\end{align}
Combining Eqs.~\eqref{eqn:j_from_f}, \eqref{eqn:distribution_decompose},
and \eqref{eqn:a_v_and_a_p}, we get the linear conductivity 
\begin{align}
	\sigma(\omega)
	= \frac{j(\omega)}{E(\omega)} 
	= \frac{1}{\pi}\frac{D_k - D_h}{-i\omega + \Gamma_d + \Gamma_{ee}} + \frac{1}{\pi} \frac{D_h}{-i\omega + \Gamma_d} \,,
	\label{eqn:sigma}
\end{align}
which is Eq.~(\textcolor{blue}{8}) of the main text.
Note that the obtained $\sigma(\omega)$ can be recast
in the form of an extended Drude model~\cite{Basov2005}:
\begin{align}
	\sigma(\omega) = \frac{1}{\pi}\frac{D_k}{-i\omega +  M(\omega)}\,,
	\quad
	M(\omega) \equiv \frac{1}{\tau(\omega)} - i\omega \lambda(\omega) 
	\,.
	\label{eqn:memory_function}
\end{align}
The complex memory function $M(\omega)$ appearing in this equation
is illustrated by Fig.~\ref{fig:memory}.
Both the effective scattering rate $1 / \tau = \re M(\omega)$
and the mass renormalization factor
$\lambda = -\im M(\omega) / \omega$ show step-like crossovers
at the boundary $\omega \sim \Gamma_{ee}$ of the hydrodynamic and kinetic
regimes.

%%%%%%%%%%%%%%%%%%%%%%%%%%%%%%%%%%%%%%%%%%%%%%%%%%%%%%%%%%%%%%%%%%%%%%%%%%%%
\section{Second-order conductivity: general}

The second-order nonlinear conductivity $\sigma_{ilm}^{(2)}$
determines the second-order current
\begin{equation}
	\begin{split}
		j_i^{(2)}\left(\vec{q}, \omega\right) &=
		\int \frac{d\omega^\prime d^2 {q}^\prime}{(2\pi)^3}
		\sigma_{ilm}^{(2)} \left(\vec{q} - \vec{q}^{\prime}, \omega - \omega^{\prime};
		\vec{q}^{\prime}, \omega^{\prime}\right)
		\\
		&\times E_l\left(\vec{q} - \vec{q}^{\prime}, \omega - \omega^{\prime}) E_m(\vec{q}^{\prime}, \omega^{\prime}\right)
		\label{eqn:nonlinear_conductivity}
	\end{split}
\end{equation}
in response to the total electric field $\vec{E}$ in the system.
By convention,
$\sigma_{ilm}^{(2)}\left(\vec{q_1}, \omega_1;\vec{q_2}, \omega_2\right)$
is chosen to be symmetrized, i.e.,
invariant under the interchange $(1 \leftrightarrow 2, l \leftrightarrow m)$.
Expanded to the linear order in
momenta, the second-order conductivity must have the form
\begin{align}
	&\sigma^{(2)}_{ilm} = \Sigma_{ilmn} (\omega_1,\omega_2) q_{1n}
	+
	\left(\begin{smallmatrix}
		1 &\leftrightarrow& 2\\ l &\leftrightarrow& m
	\end{smallmatrix} \right)\,,
	\label{eqn:Sigma}
\end{align}
where $\Sigma_{ilmn}$ is some isotropic rank-$4$ tensor.
Any such tensor is
a linear combination of the following three:
\begin{align}
	B^1_{ilmn} = \delta_{il} \delta_{nm}\,,
	\:\:\:
	B^2_{ilmn} = \delta_{im} \delta_{nl}\,,
	\:\:\:
	B^3_{ilmn} = \delta_{in} \delta_{lm}\,.
	\label{eqn:B}
\end{align}
In other words,
$\sigma^{(2)}_{i l m}$ is fully characterized by three functions $G_1$, $G_2$, and $G_3$ such that
\begin{align}
	\Sigma_{ilmn} (\omega_1,\omega_2) = \sum_{a=1}^{3} G_a(\omega_1,\omega_2) B^a_{ilmn} 
	\,.
	\label{eqn:Sigma_from_G}
\end{align}
Below we derive $\sigma^{(2)}_{i l m}$ and show
it has a different form in the
hydrodynamic and the kinetic regimes.

%%%%%%%%%%%%%%%%%%%%%%%%%%%%%%%%%%%%%%%%%%%%%%%%%%%%%%%%%%%%%%%%%%%%%%%%%%%%
\section{Second-order conductivity in the hydrodynamic regime}

To derive $\sigma_{ilm}^{(2)}$ in the hydrodynamic regime
we solve the equations
\begin{subequations}
	\label{eqn:hydro}
	\begin{align}
		&
		\partial_t n + \partial_i j_i = 0\,,
		\:\:\: j_i = n u_i\,,
		\label{eqn:charge_c}\\
		&
		\left(\partial_t + \Gamma_{E}\right) n_{E} + \partial_i \left(\gamma^2 W u_i\right)
		= j_m E_m\,,
		\:\:\: n_E = \gamma^2 W - P\,,
		\label{eqn:energy_c}\\
		&
		\begin{aligned}
			\left(\partial_t + \Gamma_{d} + u_k\partial_k\right) u_i
			&= \frac{1}{\gamma^2 W}
			\Biggl(-\partial_i P - u_i \partial_t P + n E_i
			\\
			\mbox{} &+ \frac{n}{c}\,
			\epsilon_{ikl} u_k B_l - u_i j_m E_m
			\Biggr)\,.
		\end{aligned}
		\label{eqn:euler}
	\end{align}
\end{subequations}
These equations are the same as Eqs.~(\textcolor{blue}{7}) of the main text, except we 
added phenomenological energy dissipation rate
$\Gamma_{E}$ in Eq.~\eqref{eqn:energy_c} and
chose the units $e = v = 1$ to lighten the notations.
Hence, the Lorentz factor in Eq.~\eqref{eqn:euler} is now $\gamma = 1 / \sqrt{1 - u^2}$.
% and
%the ``covariant derivatives'' are
%$\mathcal{D}_i = \partial_i + {u}_i \partial_t$,
%$\mathcal{D}_t = \partial_t  + \Gamma_{d} + {u}_i \partial_i$.
The derivation of Eqs.~\eqref{eqn:hydro} can be found in literature~\cite{Landau.8,
	Muller2008a, Muller2009, Kovtun2012, Briskot2015}.
The definitions of pressure $P$, energy density $n_E$, and
enthalpy density $W$ deserve a comment.
Whereas the current $\vec{j}$ is proportional to the actual charge density $n$, the pressure $P = P(n_0, n_{E0})$ is
the equilibrium thermodynamic parameter, which is a function
of the proper density $n_0 = n / \gamma$
and the proper energy density  $n_{E0}$.
The actual energy density is $n_E$ [Eq.~\eqref{eqn:energy_c}]
and the enthalpy density is $W = n_{E0} + P$.
Another key thermodynamic parameter is the dimensionless
isentropic (ise) bulk modulus $C_{\mathrm{ise}}$.
It is defined by Eq.~(\textcolor{blue}{10}) of the main text:
\begin{align}
	\begin{split}
		C_{\mathrm{ise}} = \frac{n_0}{W} \left( \frac{\partial P}{\partial n_0} \right)_{s_n}
		= \frac{n_0}{W} \left(\frac{\partial P}{\partial n_0} \right)_{n_{E0}}
		+ \left( \frac{\partial P}{\partial n_{E0}} \right)_{n_0}.
	\end{split}
	\label{eqn:C_ise}
\end{align}
The second equation in Eq.~\eqref{eqn:C_ise} follows from
the thermodynamic relation
\[
T d s_n = n_0^{-1} d n_{E0} - n_0^{-2} W d n_0
\]
for the quantity $s_n = s / n_0$, with $s = s(n_0, n_{E0})$ being the entropy density.

%Note that $C_ise$ has another form in terms of pressure as a function of $T$ and $\mu$
%%%
%\begin{align}
%C_ise &= 
%\frac{1}{T\partial_T P + \mu\partial_\mu P}
%\begin{pmatrix}
%	\partial_T P &
%	\partial_\mu P
%\end{pmatrix}
%\begin{pmatrix}
%	\partial^2_T P & \partial^2_{\mu T} P \\
%	\partial^2_{\mu T} P
%	& \partial^2_\mu P
%\end{pmatrix}^{-1}
%\begin{pmatrix}
%	\partial_T P \\
%	\partial_\mu P
%\end{pmatrix} 
%\,.
%\label{eqn:Ca}
%\end{align}
%%%

\begin{widetext}
	Suppose $\vec{E}(\vec{r}, t) \propto e^{i \vec{q} \vec{r} - i\omega t}$
	and define $\omega^+ = \omega + i \Gamma_{d}$.
	To the first order in $\vec{E}$ we obtain,
	for $\Gamma_{E} = 0$:
	\begin{equation}
		\begin{split}
			n^{(1)} &= n_{0}^{(1)} = \frac{n_0}{\omega} q_m u^{(1)}_m  \,, \quad 
			n_E^{(1)} = n_{E0}^{(1)} = \frac{W}{\omega} q_m u^{(1)}_m \,, \quad
			E_m^{(1)} = -i\omega^+ \frac{W}{n_0} u^{(1)}_m \,,
			\\
			W^{(1)} &= \frac{\partial W}{\partial {n_0}} n_{0}^{(1)} + \frac{\partial W}{\partial {n_{E0}}} n_{E0}^{(1)} 
			= \left( \frac{n_0}{W} \frac{\partial W}{\partial {n_0}} + \frac{\partial W}{\partial {n_{E0}}} \right) \frac{W}{\omega} q_m u_m 
			= \left(1 + C_{\mathrm{ise}} \right) \frac{W}{\omega} q_m u_m \,,
			\\
			P^{(1)} &= \frac{\partial P}{\partial {n_0}} n_{0}^{(1)} + \frac{\partial P}{\partial {n_{E0}}} n_{E0}^{(1)} 
			= \left( \frac{n_0}{W} \frac{\partial P}{\partial {n_0}} + \frac{\partial P}{\partial {n_{E0}}} \right) \frac{W}{\omega} q_m u_m
			= C_{\mathrm{ise}} \frac{W}{\omega} q_m u_m \,,
			\\
			\left(\frac{n}{W} \right)^{(1)} &= \frac{n^{(1)}}{W} - n_0 \frac{W^{(1)}}{W^2} 
			= - C_{\mathrm{ise}} \frac{n_0}{W} \frac{1}{\omega} q_m u_m \,.
			\label{eqn:first}
		\end{split}
	\end{equation}
	Now let us assume that the electric field consists of two plane waves:
	\begin{equation}
		\vec{E}(\vec{r}, t) = \vec{E}_1 e^{i \vec{q}_1 \vec{r} - i\omega_1 t}
		+ \vec{E}_2 e^{i \vec{q}_2 \vec{r} - i \omega_2 t} + \mathrm{c.c.}
		\label{eqn:E}
	\end{equation}
	To the second order in $\vec{E}$, various quantities of interest
	develop Fourier amplitudes of frequency and momenta $(\vec{q}_3,\, \omega_3) = (\vec{q}_1+\vec{q}_2,\, \omega_1 + \omega_2)$. These amplitudes are given by
	\begin{equation}
		\begin{split}
			n^{(2)} &= O\left(q_\nu^2 u\right)\,,
			\quad
			n_0^{(2)} = n^{(2)} -  \frac{1}{2} n_0 u^2 = -\frac{1}{2}n_0 u^{(1)}_{1i} u^{(1)}_{2i} + (1 \leftrightarrow 2) + O\left(q_\nu^2\right)\,,
			\\
			n_E^{(2)} &= \frac{\omega_2^+}{\omega_3} W u^{(1)}_{1i} u^{(1)}_{2i} + (1 \leftrightarrow 2) + O\left(q_\nu^2\right)\,,
			\quad
			n_{E0}^{(2)} = n^{(2)}_E - W u^2
			= \frac{i\Gamma_{d} - \omega_1}{\omega_3} W u^{(1)}_{1i} u^{(1)}_{2i}
			+ (1 \leftrightarrow 2) + O\left(q_\nu^2\right)\,,
			\\
			P^{(2)} &=  n^{(2)}_0 \frac{\partial P}{\partial n_0}
			+ n^{(2)}_{E0} \frac{\partial P}{\partial n_{E0}}
			+ \frac{1}{2} \left\{\left[{n_{0}^{(1)}}\right]^2 \frac{\partial^2 P}{\partial {n_0}^2}
			+ 2 n_{0}^{(1)} n_{E0}^{(1)} \frac{\partial^2 P}{\partial {n_0}\partial {n_{E0}}}
			+ \left[{n_{E0}^{(1)}}\right]^2 \frac{\partial^2 P}{\partial {n_{E0}}^2} \right\}
			\\
			&= -\frac{1}{2} n_0
			u^{(1)}_{1i} u^{(1)}_{2i}  \frac{\partial P}{\partial n_0}
			+ \frac{i\Gamma_{d} - \omega_1}{\omega_3}  u^{(1)}_{1i} u^{(1)}_{2i}
			W \frac{\partial P}{\partial n_{E0}}
			+ (1 \leftrightarrow 2)  + O\left(q_\nu^2\right)
			= \left(-C_{\mathrm{ise}}
			+ \frac{2 i \Gamma_{d}}{\omega_3}
			\frac{\partial P}{\partial n_{E0}}\right)
			W u^{(1)}_{1i} u^{(1)}_{2i} + O\left(q_\nu^2\right) \,.
			\label{eqn:second}
		\end{split}
	\end{equation}
	From Eq.~\eqref{eqn:euler} and
	the Faraday law $\epsilon_{ikl} \partial_k E_l = -c^{-1} \partial_t B_i$
	we obtain
	\begin{align}
		(\partial_t + \Gamma_{d} + u_k\partial_k) u_i
		&= \frac{n}{\gamma^2 W} \hat{A}_{ij}
		\left(-i\bm{\partial},\, i\partial_t\right) E_j
		- \frac{1}{\gamma^2 W} (\partial_i P + u_i \partial_t P)
		- \frac{1}{\gamma^2 W} u_i j_m E_m \,,
		\label{eqn:euler_2}
		\\
		\hat{A}_{ij}\left(\vec{k}, \omega\right) &= \left(
		1 - \frac{k_m u_m}{\omega}\right) \delta_{i j} + \frac{k_i u_j}{\omega}\,.
		\label{eqn:alpha_h}
	\end{align}
	Keeping only the terms linear in $\vec{q}_\nu$, we find
	\begin{equation}
		\begin{split}
			-i \omega_3^+ u_{i}^{(2)} &= - i{q_{2k}} u_{1k}^{(1)} u^{(1)}_{2i}
			+ \frac{n_0}{W} \left( -\frac{q_{2m}}{\omega_2} u_{1m}^{(1)} E_{2i}  + \frac{q_{2i}}{\omega_2} u_{1m}^{(1)} E_{2m} \right)
			+ \left(\frac{n}{W}\right)_1^{(1)} E_{2i}^{(1)}
			- \frac{1}{W} \left(\frac{i}{2} q_{3i} P^{(2)} - i\omega_1 P_1^{(1)} u_{2i}^{(1)}\right) + (1 \leftrightarrow 2)
			\\
			&= -i q_{2i} u_{1m}^{(1)} u^{(1)}_{2m}
			- \frac{\Gamma_{d}}{\omega_2} q_{2m}u_{1m}^{(1)} u^{(1)}_{2i}
			+ \frac{\Gamma_{d}}{\omega_2} q_{2i} u_{1m}^{(1)} u^{(1)}_{2m}
			\\
			&\mbox{} + i C_{\mathrm{ise}} \frac{\omega_2^+}{\omega_1}
			q_{1m}u_{1m}^{(1)} u^{(1)}_{2i}
			+ i C_{\mathrm{ise}} q_{2m} u_{1i}^{(1)} u^{(1)}_{2m}
			+ \frac{i}{2} C_{\mathrm{ise}} q_{3i} u_{1m}^{(1)} u^{(1)}_{2m}
			+ \frac{\partial P}{\partial n_{E0}}
			\frac{\Gamma_{d}}{\omega_3} q_{3i} u_{1m}^{(1)} u^{(1)}_{2m}
			+ (1 \leftrightarrow 2)\,. 
			\label{eqn:utwo}
		\end{split}
	\end{equation}
	The current to the second order in field is
	\begin{align}
		j_i^{(2)} &= n^{(1)} u^{(1)}_i + n^{(0)} u^{(2)}_i
		\label{eqn:current}\\
		\mbox{} &= \frac{n_0}{\omega_3^+} \bigg( \frac{\omega_3^+}{\omega_1} q_{1m} u^{(1)}_{1m} u^{(1)}_{2i} +   q_{2i} u_{1m}^{(1)} u^{(1)}_{2m}
		- \frac{i\Gamma_{d}}{\omega_2} q_{2m}u_{1m}^{(1)} u^{(1)}_{2i}  +  \frac{i\Gamma_{d}}{\omega_2} q_{2i} u_{1m}^{(1)} u^{(1)}_{2m}
		\notag \\
		\mbox{} &- C_{\mathrm{ise}} \frac{\omega_2^+}{\omega_1} q_{1m}u_{1m}^{(1)} u^{(1)}_{2i}  - C_{\mathrm{ise}} q_{2m}u_{1i}^{(1)} u^{(1)}_{2m} - \frac{1}{2} C_{\mathrm{ise}} q_{3i} u_{1m}^{(1)} u^{(1)}_{2m} + \frac{\partial P}{\partial n_{E0}}
		\frac{i\Gamma_{d}}{\omega_3} q_{3i} u_{1m}^{(1)} u^{(1)}_{2m} \bigg)
		+ (1 \leftrightarrow 2)
		\notag\\ 
		\mbox{} &= \frac{n_0}{\omega_3^+} \bigg( \frac{\omega_3^+}{\omega_1}q_{1\beta} \delta_{i\nu} +\frac{\omega_3^+}{\omega_2}q_{2\nu} \delta_{i\beta}+   q_{3i} \delta_{\beta\nu} - C_{\mathrm{ise}} \frac{\omega_2^+}{\omega_1} q_{1\beta}\delta_{i\nu} - C_{\mathrm{ise}} \frac{\omega_1^+}{\omega_2} q_{2\nu}\delta_{i\beta}  
		- C_{\mathrm{ise}} q_{2\nu}\delta_{i\beta} - C_{\mathrm{ise}} q_{1\beta}\delta_{i\nu}- C_{\mathrm{ise}} q_{3i} \delta_{\nu\beta}
		\notag \\
		\mbox{} & -\frac{i\Gamma_{d}}{\omega_2} q_{2\beta}\delta_{i\nu}
		- \frac{i\Gamma_{d}}{\omega_1} q_{1\nu}\delta_{i\beta}
		+ \frac{i\Gamma_{d}}{\omega_2} q_{2i} \delta_{\beta\nu}
		+ \frac{i\Gamma_{d}}{\omega_1} q_{1i} \delta_{\beta\nu}
		+ \frac{\partial P}{\partial n_{E0}}\frac{2i\Gamma_{d}}{\omega_3} q_{3i}
		\delta_{\nu\beta} \bigg) u_{1\beta}^{(1)} u^{(1)}_{2\nu}
		\notag \\
		\mbox{} &= \frac{n_0}{\omega_3^+} \Bigg[(1 - C_{\mathrm{ise}}) \left(\frac{\omega_3^+}{\omega_1}q_{1\beta} \delta_{i\nu} +\frac{\omega_3^+}{\omega_2}q_{2\nu} \delta_{i\beta} + q_{3i} \delta_{\beta\nu} \right)
		\notag\\
		\mbox{} &+ {i\Gamma_{d}}
		\left(
		-\frac{q_{2\beta}}{\omega_2} \delta_{i\nu}
		-\frac{q_{1\nu}}{\omega_1}\delta_{i\beta}
		+\frac{q_{2i}}{\omega_2} \delta_{\beta\nu}
		+\frac{q_{1i}}{\omega_1}  \delta_{\beta\nu}
		+\frac{\partial P}{\partial n_{E0}}\frac{2 q_{3i}}{\omega_3}  \delta_{\nu\beta} \right)
		\Bigg] u_{1\beta}^{(1)} u^{(1)}_{2\nu}\,.
		\label{eqn:current_2}
	\end{align}
	Hence, the second-order conductivity is
	\begin{align}
		\sigma^{(2)}_{ilm} &= \frac{D^{(2)}_h}{\omega_1^+ \omega_2^+ \omega_3^+}
		\left\{
		\frac{\omega_3^+}{\omega_1} q_{1l}\delta_{im}
		+ q_{1i} \delta_{lm}
		+ \frac{i\Gamma_{d}}{1 - C_{\mathrm{ise}}} 
		\left[
		\left(
		\frac{q_{1i}}{\omega_1}
		+ \frac{\partial P}{\partial n_{E}} \frac{2 q_{1i}}{\omega_3 + i\Gamma_{E}}
		\right) \delta_{lm}
		- \frac{q_{1m}}{\omega_1}\delta_{il} 
		\right]
		\right\}
		+
		\left(\begin{smallmatrix}
			1 &\leftrightarrow& 2\\ l &\leftrightarrow& m
		\end{smallmatrix} \right)\,.
		\label{eqn:sigma2}
	\end{align}
\end{widetext}
Here we added the neglected earlier energy dissipation rate $\Gamma_{E}$ in one of the terms in Eq.~\eqref{eqn:sigma2}.
In principle, $\Gamma_{E}$ should appear in more than one place.
However, we assume that $\Gamma_{E}$ is very small and its sole
role is to resolve the indeterminacy of the ratio
$q_3 / \omega_3$ in the context of the photon drag problem where $q_3, \omega_3 \to 0$.
The second-order spectral weight appearing in Eq.~\eqref{eqn:sigma2} is
\begin{equation}
	D^{(2)}_h = -\frac{e^3 n^3 v^4}{2 W^2} (1 - C_{\mathrm{ise}}) = -\frac{1}{2} \frac{e^3 n}{m^{\ast 2}} (1 - C_{\mathrm{ise}})\,,
\end{equation}
where we restored physical units and replaced $n_0$ by $n$,
$n_{E0}$ by $n_E$ to simplify notations.
Let us discuss the value of $D^{(2)}_h$ in representative cases,
assuming ee interaction corrections to pressure and enthalpy density are negligible.
The result for particles with a parabolic dispersion can be obtained
taking the nonrelativistic limit, in which
$P \simeq ({2}/{d}) (n_{E} - n m v^2)$ and
$W \simeq n m v^2$.
This gives
\begin{align}
	D^{(2)}_h = -\frac{1}{2} \frac{e^3 n}{m^2}\,,
	\quad
	\frac{\partial P}{\partial n_{E}} = \frac{2}{d}\,,
	\quad
	C_{\mathrm{ise}} \ll 1\,.
\end{align}
In the massless case, one finds $P = n_{E}/d$ and $W = (1+1/d) n_{E}$,
so that
\begin{align}
	D^{(2)}_h = -\frac{d(d - 1)}{2(d + 1)^2}
	\frac{e^3 n^3 v^4}{n_{E}^2}\,,
	\quad
	\frac{\partial P}{\partial n_{E}} = \frac{1}{d}\,,
	\quad
	C_{\mathrm{ise}} = \frac{1}{d}\,.
\end{align}
Taking $d=2$ for graphene, we get
\begin{align}
	D^{(2)}_h &= -\frac{e^3 n^3 v^4}{9 n_{E}^2} \,,
	\label{eqn:D_h2_graphene}\\
	D^{(2)}_h (T = 0) &= -\frac{g}{16 \pi}
	\frac{e^3 v^2}{\hbar^2} \equiv 2 D^{(2)}_0 \,.
	\label{eqn:D_0}
\end{align}
Functions $G_1$, $G_2$, and $G_3$ [Eq.~\eqref{eqn:Sigma_from_G}]
corresponding to Eq.~\eqref{eqn:sigma2} are
\begin{equation}
	\begin{split}
		G_1 &= \frac{D^{(2)}_h}{\omega_1^+ \omega_2^+ \omega_3^+}\left(-\frac{1}{1-C_{\mathrm{ise}}} \frac{i\Gamma_d}{\omega_1}\right) \,, \\
		G_2 &= \frac{D^{(2)}_h}{\omega_1^+ \omega_2^+ \omega_3^+}\left(\frac{\omega_3^+}{\omega_1}\right) \,,
		\\
		G_3 &= \frac{D^{(2)}_h}{\omega_1^+ \omega_2^+ \omega_3^+} \left[1+\frac{i\Gamma_d}{1-C_{\mathrm{ise}}} \left(\frac{1}{\omega_1} +\frac{\partial P}{\partial {n_E}} \frac{2}{\omega_3 + i \Gamma_E} \right) \right] \,.
	\end{split}
	\label{eqn:G_hydro_general}
\end{equation}
In the collisionless limit $\Gamma_{d}\to 0$,
these formulas simplify to
\begin{align}
	(G_1, G_2, G_3)= D^{(2)}_h \left( 0,\, \frac{1}{\omega_1^2 \omega_2},\, \frac{1}{\omega_1\omega_2\omega_3} \right)
	\,,
	\label{eqn:B}
\end{align}
while Eq.~\eqref{eqn:sigma2} reduces to
\begin{align}
	\sigma^{(2)}_{ilm}
	= \frac{D^{(2)}_h}{\omega_1 \omega_2 \omega_3} \bigg( \frac{\omega_3}{\omega_1} q_{1l}\delta_{im} + \frac{\omega_3}{\omega_2} q_{2m}\delta_{il}  +  q_{3i} \delta_{lm} \bigg)\,,
	\label{eqn:sigma2_no_damping}
\end{align}
which is equivalent to Eq.~(\textcolor{blue}{2}) of the main text.

%%%%%%%%%%%%%%%%%%%%%%%%%%%%%%%%%%%%%%%%%%%%%%%%%%%%%%%%%%%%%%%%%%%%%%%%%%%%
\section{Second-order conductivity in the kinetic regime}

The kinetic regime corresponds to the frequency range
$\Gamma_{ee} \ll \omega \ll \varepsilon_F$.
The linear and nonlinear conductivities
in this regime can be computed by solving the Boltzmann kinetic
equation
\begin{align}
	\left[\partial_t + \vec{v}\cdot \bm{\partial} + \left(\vec{E} + \frac{\vec{v}}{c}\times \vec{B}\right)
	\bm{\partial}_{\vec{p}} \right] f = \hat{I}\left[f\right]\,.
	\label{eqn:boltzmann1}
\end{align}
In this section,
we again set $e = 1$ and suppress the subscripts ${\vec{p}}$
in $\vec{v}_{\vec{p}}$, $f_{\vec{p}}$.
One should not confuse the quasiparticle velocity $\vec{v}$
at finite $\vec{p}$, a vector, with $v$, the limiting velocity at
$p = \infty$, a scalar.

The magnetic-field term in Eq.~\eqref{eqn:boltzmann1} can be expressed
with the help of the kernel
\begin{equation}
	{A}_{ij}\left(\vec{k}, \omega\right)
	= \left(1 - \frac{k_m v_m}{\omega} \right)\delta_{i j}
	+ \frac{k_i v_j}{\omega} \,,
	\label{eqn:alpha_k}
\end{equation}
similar to Eq.~\eqref{eqn:alpha_h},
leading to
\begin{equation}
	\left[\partial_t + v_i\partial_i + {A}_{ij}\left(-i\bm{\partial}, i\partial_t\right)
	E_j \partial_{p_i}  \right] f = \hat{I}\left[f\right]\,.
	\label{eqn:boltzmann2}
\end{equation}
The second-order current we want to compute is 
\begin{align}
	j_i \left(\vec{q}_3, \omega_3\right) = \sum v_i f^{(2)}\left(\vec{q}_3, \omega_3\right) \,.
	\label{eqn:j2}
\end{align}
To do this we need to specify the collision integral
$\hat{I}\left[f\right]$.

%%%%%%%%%%%%%%%%%%%%%%%%%%%%%%%%%%%%%%%%%%%%%%%%%%%%%%%%%%%%%%%%%%%%%%%%%%%%
\subsection{Nonconserving relaxation-time approximation}

It is useful to consider first
the approximation
$\hat{I}\left[f\right] = -\Gamma \left(f - f^{(0)}\right)$,
with $\Gamma$ being an energy-independent relaxation rate.
This is probably the simplest model
one can study.
However, one should keep in mind that this approximation is flawed because
it may not conserve the particle number.

Assuming the electric field is composed of two plane waves
[Eq.~\eqref{eqn:E}],
we expand $f$ to the first and second order in field:
\begin{align}
	f^{(1)} \left(\vec{q}_\nu, \omega_\nu\right) &= \frac{-i E_{2m} v_m}{\omega_\nu^+ - v_k q_{\nu\,k}} \partial_{\varepsilon}f^{(0)}  \,,
	\quad \nu = 1, 2\,,
	\\
	f^{(2)} \left(\vec{q}_3, \omega_3\right) &= \frac{-i E_{1l}
		A_{a l}\left(\vec{q}_1, \omega_1\right)}{\omega_3^+ - v_j q_{3j}} \partial_{p_a}
	f^{(1)} \left(\vec{q}_2, \omega_2\right)
	%  \frac{-i e E_{2j} v_j}{\omega_2^+ - v_m q_{2m}}
	% \partial_{\varepsilon}f^{(0)}
	% \notag\\ &
	+ (1 \leftrightarrow 2) \,.
	\label{eqn:f2_single}
\end{align}
Hence, the second-order conductivity is
\begin{widetext}
	\begin{align}
		&\sigma^{(2)}_{ilm} = -\sum
		A_{a l}\left(\vec{q}_1, \omega_1\right)
		\frac{v_i}{\omega_3^+ - v_j q_{3j}}
		\partial_{p_a} \frac{v_m}{\omega_2^+ - v_k q_{2k}} \partial_{\varepsilon}f^{(0)} 
		+
		\left(\begin{smallmatrix}
			1 &\leftrightarrow& 2\\ l &\leftrightarrow& m
		\end{smallmatrix} \right)\,.
		\label{eqn:current_int}
	\end{align}
	The evaluation of this expression for Dirac electrons in graphene
	is tedious but straightforward. The final result is
	\begin{align}
		\sigma^{(2)}_{ilm}
		%	 &=
		%\frac{2D_k^{(2)}}{\omega_3^+ \omega_1^+ \omega_2^+}
		% \bigg[  (q_1 - q_2)_l\delta_{im} + (q_2 - q_1)_m\delta_{il}
		%+ q_{3i} \delta_{lm} + \frac{\omega_1^+}{2\omega_2^+} \left(
		% 3q_{2m} \delta_{il} - q_{2l}\delta_{im} -q_{2i}\delta_{lm} \right)
		% + \frac{\omega_2^+}{2\omega_1^+} \left( 3q_{1l} \delta_{im}
		% - q_{1m}\delta_{il} -q_{1i}\delta_{lm} \right)  \notag\\  
		%&+ i\Gamma 
		%\left( \frac{1}{\omega_3^+} (q_{3m} \delta_{il} + q_{3l} \delta_{im}
		%-q_{3i} \delta_{lm}) + \frac{2}{\omega_1} (q_{1i} \delta_{lm}
		% - q_{1m}\delta_{il}  ) +\frac{2}{\omega_2} (q_{2i} \delta_{lm}
		% - q_{2l}\delta_{im}  ) 
		%\right)
		%\bigg]  
		%\,. \notag \\
		&= \frac{D_k^{(2)}}{\omega_1^+ \omega_2^+ \omega_3^+} 
		\Bigg[   \left(
		\frac{\omega_3}{\omega_3^+} q_{3} - \frac{\omega_1^+}{2\omega_2^+} q_{2} -\frac{\omega_2^+}{2\omega_1^+} q_{1} +\frac{2 i\Gamma}{\omega_1} q_{1} + \frac{2i\Gamma}{\omega_2} q_{2} \right)_i \delta_{lm}
		\notag\\  
		\mbox{} &+ \left(2 q_2 - 2 q_1 - \frac{\omega_2^+}{\omega_1^+} q_1
		+ \frac{3\omega_1^+}{\omega_2^+} q_2
		+ \frac{2i\Gamma}{\omega_3^+} q_3
		- \frac{4i\Gamma}{\omega_1} q_1  \right)_m \delta_{il}
		\Bigg]  
		+
		\left(\begin{smallmatrix}
			1 &\leftrightarrow& 2\\ l &\leftrightarrow& m
		\end{smallmatrix} \right)\,.
		\label{eqn:sigma2_graphene_boltzmann}
	\end{align}
	Its collisionless limit $\Gamma \to 0$ is
	\begin{align}
		\sigma^{(2)}_{ilm}
		%&= \frac{2D_k^{(2)}}{\omega_1 \omega_2 \omega_3} \Bigg[
		%  (q_1 - q_2)_l\delta_{im} + (q_2-q_1)_m\delta_{il} + q_{3i} \delta_{lm}
		% + \frac{\omega_1}{2\omega_2} \left( 3q_{2m} \delta_{il} - q_{2l}\delta_{im}
		% - q_{2i}\delta_{lm} \right) + \frac{\omega_2}{2\omega_1} \left( 3q_{1l}
		% \delta_{im} - q_{1m}\delta_{il} -q_{1i}\delta_{lm} \right) 
		%\Bigg]
		%\notag\\
		&= \frac{D_k^{(2)}}{\omega_1 \omega_2 \omega_3}
		\Bigg[ \left(q_3 -\frac{\omega_1}{2\omega_2}q_2 -\frac{\omega_2}{2\omega_1}q_1 \right)_i \delta_{lm} 
		+ \left(2 q_2 - 2 q_1 - \frac{\omega_2}{\omega_1} q_1 +\frac{3\omega_1}{\omega_2}q_2 \right)_m \delta_{il}
		\Bigg]
		+	\left(\begin{smallmatrix}
			1 &\leftrightarrow& 2\\ l &\leftrightarrow& m
		\end{smallmatrix} \right)\,,
		\label{eqn:sigma2_graphene_boltzmann_2}
	\end{align}
\end{widetext}
which implies [cf.~Eq.~\eqref{eqn:Sigma_from_G}]
\begin{align}
	(G_1,\, G_2,\, G_3)
	= \frac{D^{(2)}_k}{\omega_1 \omega_2 \omega_3}  \left(-2 -\frac{\omega_2}{\omega_1},\,  2 + 3\,\frac{\omega_2}{\omega_1},\,
	2 - \frac{\omega_2}{\omega_1}
	\right)
	\,.
	\label{eqn:G_kinetic_collisionless_nonconserving}
\end{align}
The second-order spectral weight in these equations is
\begin{equation}
	D_k^{(2)} = D_0^{(2)} \left( 2 f_{p=0}^{(0)} - 1 \right)
	= D_0^{(2)} \tanh \left(\frac{\mu}{2 T}\right)\,,
	\label{eqn:graphene_boltzmann_finite_T}
\end{equation}
where [Eq.~\eqref{eqn:D_0}]
\begin{equation}
	D^{(2)}_0 = -\frac{g}{32 \pi}
	\frac{e^3 v^2}{\hbar^2}\,.
	\label{eqn:D_0_again}
\end{equation}
In the limit of $T \to 0$, we have
$D_k^{(2)} = D_0^{(2)}\, \mathrm{sign}(\mu)$.
At $T \gg \mu$, the asymptotic behavior of the chemical potential is $\mu \propto n / T$, see, e.g., Supplemental material of Ref.~\onlinecite{Sun2016}.
Therefore, $D_k^{(2)} \propto n /T^2$ as mentioned in the main text.
However, at high $T$, an interband contribution to $\sigma^{(2)}_{i l m}$, not included in our semiclassical approach, may become important.

%Following the notation of Karch et al \cite{Ganichev2010} and Jiang et al \cite{Jiang2011}, we get the parameters $T_1$ and $T_2$ for the photocurrents driven by linearly polarized light with frequency $\omega$ and in plane momentum $q$
%%%
%\begin{align}
%T_1&=  16 D_0^{(2)} \frac{-\omega^2 -\Gamma^2}{\omega(\omega^2 + \Gamma^2 )^2} \notag \\
%T_2&=  16 D_0^{(2)} \frac{\Gamma^2}{\omega(\omega^2 + \Gamma^2 )^2} 
%\,.
%\end{align}
%%
%The clean limit $\Gamma=0$ gives $T_2=0$, indicating $j_y$ would vanish in the clean limit, which is in contradiction with Eq.~(26b) of Karch et al\cite{Ganichev2010} and Eq.~(14) of Jiang et al\cite{Jiang2011}.

%%%%%%%%%%%%%%%%%%%%%%%%%%%%%%%%%%%%%%%%%%%%%%%%%%%%%%%%%%%%%%%%%%%%%%%%%%%%
\subsection{Multiple relaxation-time approximation}

Let us assume now that the collision operator $\hat{I}\left[f\right]$
is linear and diagonal in the angular momentum basis, so that the Boltzmann equation can be written as 
\begin{align}
	\hat{L} f  = - E_j A_{ij} \partial_{p_i} f \,,
	\quad
	\hat{L} = \hat{L}_0(\omega) + \vec{v}\cdot \bm{\partial}\,,
	\label{eqn:boltzmann}
\end{align}
where $\hat{L}_0$ is the operator 
\begin{align}
	\hat{L}_0 = \sum\limits_{l = -\infty}^\infty
	(\partial_t + \Gamma_l) |l \rangle \langle l|
	\label{eqn:L}
\end{align}
and $\Gamma_l$ is the scattering rate for the angular momentum $l$.
This rate may depend on the quasiparticle energy $\varepsilon$.
The model conserves the number of particles if $\Gamma_0 = 0$.
The action of $\hat{L}_0$
can be written in terms of the complex frequencies
\begin{equation}
	\omega_\nu^{(\mu)} = \omega_\nu + i\Gamma_\mu\,,
	\quad
	\nu = 1,\, 2,\, 3\,.
\end{equation}
Instead of Eq.~\eqref{eqn:f2_single} we now get a more complicated
expression:
\begin{equation}
	f^{(2)} = \hat{L}^{-1} E_l A_{il} \partial_{p_i} \hat{L}^{-1} E_j A_{ij} \partial_{p_i} f^{(0)} \,.
\end{equation}
\begin{widetext}
	\noindent%
	To calculate $\sigma_{ilm}^{(2)} (\vec{q}_1, \omega_1; \vec{q}_2, \omega_2)$ we need the $(\vec{q}_3, \omega_3)$
	Fourier harmonic of $f^{(2)}$:
	\begin{align}
		f^{(2)}
		= \hat{L}^{-1}{\left(\vec{q}_3, \omega_3\right)}
		E_{1l} \left( \partial_{p_l} - \frac{q_{1n}}{\omega_1} v_n \partial_{p_l}
		+\frac{q_{1n}}{\omega_1} v_l \partial_{p_n} \right)
		\hat{L}^{-1}{\left(\vec{q}_2, \omega_2\right)} E_{2m} v_m \partial_\varepsilon f^{(0)}
		+
		\left(\begin{smallmatrix}
			1 &\leftrightarrow& 2\\ l &\leftrightarrow& m
		\end{smallmatrix} \right)\,.
		\label{eqn:f2}
	\end{align}
	[The argument $(\vec{q}_3, \omega_3)$ of $f^{(2)}$ is omitted.] For our purpose of computing the terms linear in gradients the expansion
	\begin{align}
		\hat{L}^{-1} \simeq \hat{R}
		- \hat{R} \left(\vec{v}\cdot \bm{\partial}\right) \hat{R}\,,
		\quad
		\hat{R} = \hat{L}_0^{-1}
		\label{eqn:l}
	\end{align}
	suffices. It yields
	\begin{equation}
		\begin{split}
			f^{(2)} &=
			E_{1l} E_{2m} \Bigg[\hat{R}{(\omega_3)}
			\left( -\frac{1}{\omega_1} q_{1n} v_n \partial_{p_l}
			+ \frac{1}{\omega_1} v_l q_{1n} \partial_{p_n} \right)
			\hat{R}{(\omega_2)} 
			- \hat{R}{(\omega_3)} i q_{3i} v_i \hat{R}{(\omega_3)}
			\partial_{p_l} \hat{R}{(\omega_2)}\\
			\mbox{} &- \hat{R}{(\omega_3)} \partial_{p_l}
			\hat{R}{(\omega_2)} i q_{2i} v_i \hat{R}{(\omega_2)}
			\Bigg] v_m  \partial_\varepsilon f^{(0)}
			+
			\left(\begin{smallmatrix}
				1 &\leftrightarrow& 2\\ l &\leftrightarrow& m
			\end{smallmatrix} \right)\,.
			\label{eqn:f2_again}
		\end{split}
	\end{equation}
	To do the summation over the angular directions,
	we expand all the variables in the angular momentum basis.
	To this end, we do a set of unitary transformations.
	For example, the velocity goes from $\vec{v}
	= \left(v_x, v_y\right)^T = \left|\vec{v}\right| \left(\cos\phi,\, \sin\phi \right)^T$
	to $\left(\tilde{v}_-, \tilde{v}_+\right)^T$:
	\begin{equation}
		\begin{pmatrix}
			\tilde{v}_+ \\
			\tilde{v}_-
		\end{pmatrix} 
		= \mathbf{U}\,
		\begin{pmatrix}
			v_x \\
			v_y
		\end{pmatrix} 
		= \frac{\left|\vec{v}\right|}{\sqrt{2}}
		\begin{pmatrix}
			M_{+} \\
			M_{-}
		\end{pmatrix}\,,
		\quad
		\mathbf{U} = 
		\left\{U_{m j}\right\} = \frac{1}{\sqrt{2}}
		\begin{pmatrix}
			1 & +i\\
			1 & -i
		\end{pmatrix}\,,
		\quad
		\mathbf{U} \mathbf{U}^{\dagger} = \{\delta_{i j}\}\,,
		\quad
		M_m = e^{im \phi}\,,
		\quad
		m = \pm\,.
		\label{eqn:v_ladder}
	\end{equation}
	We do the same transformation for
	the momentum-space derivatives:
	\begin{align}
		\tilde{\partial}_{p_m} =  U_{m j} \partial_{p_j} = \frac{1}{\sqrt{2}}
		\left(\partial_{p_x} + i m \partial_{p_y}\right)
		= M_m \left(\partial_p + i\frac{m}{p} \partial_\phi \right)\,,
		\quad
		m = \pm\,.
		\label{eqn:p_ladder}
	\end{align}
	To the electric fields and spatial momenta we
	apply a conjugate transformation,
	%%
	%\begin{align}
	$
	\tilde{E}_m = E_j U^{\dagger}_{j m}
	$,
	%\quad
	$
	\tilde{q}_m = q_j U^{\dagger}_{j m}
	$,
	%\end{align}
	in order to leave the scalar products 
	$E_i v_i$,  $E_i \partial_{p_i}$, $q_i v_i$ invariant.
	The net effect on Eq.~\eqref{eqn:f2_again} is simply to add tildes for every variable.
	%%%
	%\begin{equation}
	%\begin{split}
	%f^{(2)} &=
	% \tilde{E}_{1l} \tilde{E}_{2m} \Biggl[\hat{R}{(\omega_3)}  \left(-\frac{1}{\omega_1} \tilde{q}_{1n} \tilde{v}_n \tilde{\partial}_{p_l} + \frac{1}{\omega_1} \tilde{v}_l \tilde{q}_{1n} \tilde{\partial}_{p_n} \right) \hat{R}{(\omega_2)}
	%- \hat{R}{(\omega_3)} i \tilde{q}_{3i} \tilde{v}_i \hat{R}{(\omega_3)} \tilde{\partial}_{p_l} \hat{R}{(\omega_2)}\\
	%&- \hat{R}{(\omega_3)} \tilde{\partial}_{p_l} \hat{R}{(\omega_2)} i \tilde{q}_{2i} \tilde{v}_i \hat{R}{(\omega_2)} \Biggr]
	%\tilde{v}_m {\partial}_\varepsilon f^{(0)}
	%+
	%\left(\begin{smallmatrix}
	%1 &\leftrightarrow& 2\\ l &\leftrightarrow& m
	%\end{smallmatrix} \right)\,.
	%\label{eqn:f2_tilde}
	%\end{split}
	%\end{equation}
	%%%
	The second-order current becomes
	\begin{equation}
		\begin{split}
			\tilde{j}_i &= \sum \tilde{v}_i f^{(2)}
			= \tilde{E}_{1l} \tilde{E}_{2m} \sum \tilde{v}_i   \Bigg[\hat{R}{(\omega_3)}  \left( -\frac{1}{\omega_1} \tilde{q}_{1n} \tilde{v}_n \tilde{\partial}_{p_l} +\frac{1}{\omega_1} \tilde{v}_l \tilde{q}_{1n} \tilde{\partial}_{p_n} \right) \hat{R}{(\omega_2)}   - \hat{R}{(\omega_3)} i \tilde{q}_{3n} \tilde{v}_n \hat{R}{(\omega_3)} \tilde{\partial}_{p_l} \hat{R}{(\omega_2)}
			\\
			&- \hat{R}{(\omega_3)} \tilde{\partial}_{p_l} \hat{R}{(\omega_2)} i \tilde{q}_{2n} \tilde{v}_n \hat{R}{(\omega_2)} \Bigg] \tilde{v}_m
			{\partial}_\varepsilon f^{(0)}
			+
			\left(\begin{smallmatrix}
				1 &\leftrightarrow& 2\\ l &\leftrightarrow& m
			\end{smallmatrix} \right)
			\\
			&= (\partial_\phi \, \mathrm{-part}) \:\:+\:\: (\partial_p \, \mathrm{-part})
			\\
			&=  \tilde{E}_{1l} \tilde{E}_{2m} 
			\Bigg[ 
			\frac{ \tilde{q}_{1n}}{\omega_1}
			\sum \tilde{v}_i \hat{R}{(\omega_3)}  \left( -\tilde{v}_n \frac{il}{p}\tilde{M}_{l} + \tilde{v}_l \frac{in}{p} \tilde{M}_{n} \right) \partial_\phi \hat{R}{(\omega_2)} \tilde{v}_m {\partial}_\varepsilon f^{(0)}
			\\
			& - i \tilde{q}_{3n} \sum \tilde{v}_i \hat{R}{(\omega_3)} \tilde{v}_n \hat{R}{(\omega_3)}  \frac{il}{p} \tilde{M}_{l} \partial_\phi \hat{R}{(\omega_2)} \tilde{v}_m {\partial}_\varepsilon f^{(0)} 
			- i \tilde{q}_{2n} \sum \tilde{v}_i \hat{R}{(\omega_3)} \frac{il}{p} \tilde{M}_{l} \partial_\phi \hat{R}{(\omega_2)} \tilde{v}_n \hat{R}{(\omega_2)} \tilde{v}_m {\partial}_\varepsilon f^{(0)} 
			\\
			& - i \tilde{q}_{3n} \sum \tilde{v}_i \hat{R}{(\omega_3)} \tilde{v}_n \hat{R}{(\omega_3)} \tilde{M}_{l} \partial_p \hat{R}{(\omega_2)} \tilde{v}_m {\partial}_\varepsilon f^{(0)} 
			- i \tilde{q}_{2n} \sum \tilde{v}_i \hat{R}{(\omega_3)} \tilde{M}_{l} \partial_p \hat{R}{(\omega_2)} \tilde{v}_n \hat{R}{(\omega_2)} \tilde{v}_m {\partial}_\varepsilon f^{(0)} 
			\Bigg]
			+
			\left(\begin{smallmatrix}
				1 &\leftrightarrow& 2\\ l &\leftrightarrow& m
			\end{smallmatrix} \right)\,.
		\end{split}
		\label{eqn:f2_yet_again}
	\end{equation}
	Only the terms of zero net angular momentum, i.e., $i + n + l + m = 0$ survive after the summation. Since $i$, $n$, $l$, $m$ have values $\pm 1$, they have to appear in opposite-sign pairs.
	This constraint can be implemented with the help
	of the transformed rank-2 and rank-4 isotropic tensors
	\begin{align}
		\tilde{\delta}_{ij} \equiv U_{i m} \delta_{m n} U_{j n}\,,
		\quad
		\left\{\tilde{\delta}_{ij}\right\} =
		\mathbf{U}\mathbf{U}^T =
		\begin{pmatrix}
			0 & 1  \\
			1 & 0
		\end{pmatrix} \,,
		\label{eqn:delta_transformed}
		\qquad
		\tilde{\Delta}_{inlm} = \tilde{\delta}_{in} \tilde{\delta}_{lm}
		+ \tilde{\delta}_{il} \tilde{\delta}_{nm}
		+ \tilde{\delta}_{ln} \tilde{\delta}_{im}\,.
	\end{align}
	The subsequent calculations are done for $T = 0$
	where $\partial_{\varepsilon} f^{(0)}
	= -\delta\left(\varepsilon - \varepsilon_F\right)$.
	We obtain
	\begin{equation}
		\begin{split}
			\tilde{j}_i 
			&=  \frac{v_F^2}{16\pi}
			\frac{\tilde{E}_{1l} \tilde{E}_{2m}}{\omega_3^{(1)}}
			\Bigg\{ 
			\left(\frac{\tilde{q}_{1n}}{\omega_1}\right) 
			\frac{1}{\omega_2^{(1)}} m(l-n) \tilde{\Delta}_{inlm}
			- \frac{1}{\omega_2^{(1)}} \Bigg( \tilde{q}_{3n}
			\frac{m l}{\omega_3^{(l+m)}}
			+ \tilde{q}_{2n} \frac{l(m+n)}{\omega_2^{(n+m)}}  \Bigg) \tilde{\Delta}_{inlm}
			- \frac{1}{\omega_2^{(1)}} \Bigg( \frac{\tilde{q}_{3n}}{\omega_3^{(l+m)}}
			+ \frac{\tilde{q}_{2n}}{\omega_2^{(n+m)}} \Bigg)  \tilde{\Delta}_{inlm}
			\\
			&+ \frac{1}{\omega_2^{(1)}} \Bigg[
			\frac{\tilde{q}_{3n} \varepsilon_F}{\omega_3^{(l+m)}} \left(\frac{\partial_\varepsilon i\Gamma_1}{\omega_3^{(1)}} + \frac{\partial_\varepsilon i\Gamma_{l+m}}{\omega_3^{(l+m)}} \right) 
			+ \frac{\tilde{q}_{2n} \varepsilon_F}{\omega_2^{(n+m)}} \frac{\partial_\varepsilon i\Gamma_1}{\omega_3^{(1)}}   \Bigg]
			\tilde{\Delta}_{inlm}
			\Bigg\}
			\\
			&=  \frac{v_F^2}{16\pi}
			\frac{\tilde{E}_{1l} \tilde{E}_{2m}}{\omega_2^{(1)} \omega_3^{(1)}}
			\tilde{\Delta}_{inlm}
			\Bigg\{
			\left(\frac{\tilde{q}_{1n}}{\omega_1}\right) m(l-n) - \left( \tilde{q}_{3n} \frac{ml}{\omega_3^{(l+m)}}
			+ \tilde{q}_{2n} \frac{l(m+n)}{\omega_2^{(n+m)}}  \right)
			- \left(\frac{\tilde{q}_{3n}}{\omega_3^{(l+m)}}
			+ \frac{\tilde{q}_{2n}}{\omega_2^{(n+m)}} \right)
			\\
			\mbox{} &+ \Bigg[\frac{\tilde{q}_{3n} \varepsilon_F}{\omega_3^{(l+m)}} \Bigg(\frac{\partial_\varepsilon i\Gamma_1}{\omega_3^{(1)}} + \frac{\partial_\varepsilon i\Gamma_{l+m}}{\omega_3^{(l+m)}} \Bigg) 
			+ \frac{\tilde{q}_{2n} \varepsilon_F}{\omega_2^{(n+m)}} \frac{\partial_\varepsilon i\Gamma_1}{\omega_3^{(1)}}   \Bigg]
			\Bigg\}
			\\
			&= -\frac{v_F^2}{16\pi}
			\frac{\tilde{E}_{1l} \tilde{E}_{2m}}{\omega_2^{(1)} \omega_3^{(1)}}
			\tilde{\Delta}_{inlm}
			\Bigg\{
			\frac{\tilde{q}_{3n}}{\omega_3^{(l+m)}} \Bigg[1 + m l - \varepsilon_F \Bigg(\frac{\partial_\varepsilon i\Gamma_1}{\omega_3^{(1)}} + \frac{\partial_\varepsilon i\Gamma_{l+m}}{\omega_3^{(l+m)}} \Bigg)  \Bigg]
			\\
			& + \frac{\tilde{q}_{1n}}{\omega_1} m(n - l) 
			+ \frac{\tilde{q}_{2n}}{\omega_2^{(n+m)}} \left(l(m+n) + 1 - \varepsilon_F \frac{\partial_\varepsilon i\Gamma_1}{\omega_3^{(1)}}  \right)
			\Bigg\}\,.
			\\
			&= -\frac{v_F^2}{16\pi}
			\frac{\tilde{E}_{1l} \tilde{E}_{2m}}{\omega_2^{(1)} \omega_3^{(1)}}
			\Bigg\{ 
			\tilde{q}_{3n} \Bigg[\frac{2}{\omega_3^{(2)}} (-\tilde{\delta}_{lm} \tilde{\delta}_{in} + \tilde{\delta}_{li} \tilde{\delta}_{nm} +  \tilde{\delta}_{ln} \tilde{\delta}_{im})
			\\
			& -\frac{\varepsilon_F}{\omega_3^{(2)}} \left(\frac{\partial_\varepsilon i\Gamma_1}{\omega_3^{(1)}} + \frac{\partial_\varepsilon i\Gamma_{2}}{\omega_3^{(2)}}  \right)\left(-\tilde{\delta}_{lm} \tilde{\delta}_{in} + \tilde{\delta}_{li} \tilde{\delta}_{nm} +  \tilde{\delta}_{ln} \tilde{\delta}_{im}\right) - \frac{2\varepsilon_F}{\omega_3^{(0)}}  \left(\frac{\partial_\varepsilon i\Gamma_1}{\omega_3^{(1)}} + \frac{\partial_\varepsilon i\Gamma_{0}}{\omega_3^{(0)}}  \right) \tilde{\delta}_{lm} \tilde{\delta}_{in}  \Bigg]
			\\
			& + \tilde{q}_{1n} \frac{4}{\omega_1} \left(\tilde{\delta}_{lm} \tilde{\delta}_{in}-\tilde{\delta}_{nm} \tilde{\delta}_{il}\right) 
			+ \tilde{q}_{2n} \Bigg[  \frac{2}{\omega_2^{(0)}} \tilde{\delta}_{il} \tilde{\delta}_{nm} - \frac{1}{\omega_2^{(2)}} \Bigg(\tilde{\delta}_{lm} \tilde{\delta}_{in} - \tilde{\delta}_{li} \tilde{\delta}_{nm} + \tilde{\delta}_{ln} \tilde{\delta}_{im}\Bigg)
			\\
			\mbox{}	&- \frac{\varepsilon_F \partial_\varepsilon i\Gamma_1}{\omega_3^{(1)}} \Bigg(\frac{2}{\omega_2^{(0)}}\tilde{\delta}_{li} \tilde{\delta}_{nm}+ \frac{1}{\omega_2^{(2)}} \left(\tilde{\delta}_{lm} \tilde{\delta}_{in} - \tilde{\delta}_{li} \tilde{\delta}_{nm} +  \tilde{\delta}_{ln} \tilde{\delta}_{im}\right)  \Bigg) \Bigg]
			\Bigg\}
			+
			\left(\begin{smallmatrix}
				1 &\leftrightarrow& 2\\ l &\leftrightarrow& m
			\end{smallmatrix} \right)\,.
		\end{split}
	\end{equation}
	To convert back to the $(x,\,y)$ coordinates, one simply needs
	to drop the tildes everywhere.
	Therefore, the second-order nonlinear optical conductivity is
	\begin{equation}
		\begin{split}
			\sigma^{(2)}_{ilm} 
			&= \frac{2D_0^{(2)}}{\omega_2^{(1)} \omega_3^{(1)}}
			\Bigg\{ 
			q_{3n}  \Bigg[ \frac{2}{\omega_3^{(2)}} \left(-\delta_{lm} \delta_{in} + \delta_{li} \delta_{nm} +  \delta_{ln} \delta_{im}\right)
			\\
			& - \frac{\varepsilon_F}{\omega_3^{(2)}} \Bigg(\frac{\partial_\varepsilon i\Gamma_1}{\omega_3^{(1)}} + \frac{\partial_\varepsilon i\Gamma_{2}}{\omega_3^{(2)}}  \Bigg) \left(-\delta_{lm} \delta_{in} + \delta_{li} \delta_{nm} +  \delta_{ln} \delta_{im}\right) - \frac{2\varepsilon_F}{\omega_3^{(0)}}  \Bigg(\frac{\partial_\varepsilon i\Gamma_1}{\omega_3^{(1)}} + \frac{\partial_\varepsilon i\Gamma_{0}}{\omega_3^{(0)}}  \Bigg) \delta_{lm} \delta_{in}  \Bigg]
			\\
			& + 4q_{1n} \frac{4}{\omega_1} \left(\delta_{lm} \delta_{in}-\delta_{nm} \delta_{il}\right) 
			+ q_{2n} \Bigg[  \frac{2}{\omega_2^{(0)}} \delta_{il} \delta_{nm}   - \frac{1}{\omega_2^{(2)}} \left(\delta_{lm} \delta_{in} - \delta_{li} \delta_{nm} +  \delta_{ln} \delta_{im}\right)
			\\
			\mbox{}	&- \frac{\varepsilon_F \partial_\varepsilon i\Gamma_1}{\omega_3^{(1)}} \Bigg(\frac{2}{\omega_2^{(0)}}\delta_{li} \delta_{nm}+ \frac{1}{\omega_2^{(2)}}\left(\delta_{lm} \delta_{in} - \delta_{li} \delta_{nm} +  \delta_{ln} \delta_{im}\right)  \Bigg)  \Bigg]	\Bigg\}
			+
			\left(\begin{smallmatrix}
				1 &\leftrightarrow& 2\\ l &\leftrightarrow& m
			\end{smallmatrix} \right)\,.
		\end{split}
	\end{equation}
	Completing the symmetrization step $\left(\begin{smallmatrix}
	1 &\leftrightarrow& 2\\ l &\leftrightarrow& m
	\end{smallmatrix} \right)$, we get the following:
	\begin{equation}
		\begin{split}
			\sigma^{(2)}_{ilm} 
			&= \frac{2D_0^{(2)}}{\omega_1^{(1)}  \omega_2^{(1)} \omega_3^{(1)}}
			\Bigg\{ 
			q_{3n}  \Bigg[ \frac{\omega_1^{(1)}+\omega_2^{(1)}}{\omega_3^{(2)}} \left(-\delta_{lm} \delta_{in} + \delta_{li} \delta_{nm} +  \delta_{ln} \delta_{im}\right)
			\\
			& - \frac{\omega_1^{(1)}
				+\omega_2^{(1)}}{2\omega_3^{(2)}} i\varepsilon_F \Bigg(\frac{\partial_\varepsilon \Gamma_1}{\omega_3^{(1)}}
			+ \frac{\partial_\varepsilon \Gamma_{2}}{\omega_3^{(2)}} \Bigg)
			\left(-\delta_{lm} \delta_{in} + \delta_{li} \delta_{nm} +  \delta_{ln} \delta_{im}\right) - \frac{\omega_1^{(1)}+\omega_2^{(1)}}{\omega_3^{(0)}} i\varepsilon_F  \Bigg(\frac{\partial_\varepsilon \Gamma_1}{\omega_3^{(1)}} + \frac{\partial_\varepsilon \Gamma_{0}}{\omega_3^{(0)}} \Bigg) \delta_{lm} \delta_{in}  \Bigg]
			\\
			& + q_{1n} \frac{2\omega_1^{(1)}}{\omega_1} \left(\delta_{lm} \delta_{in}-\delta_{nm} \delta_{il}\right)  + q_{2n} \frac{2\omega_2^{(1)}}{\omega_2} \left(\delta_{lm} \delta_{in}-\delta_{nl} \delta_{im}\right) 
			\\
			& + q_{2n} \Bigg[  \frac{\omega_1^{(1)}}{\omega_2^{(0)}} \delta_{il} \delta_{nm}   - \frac{\omega_1^{(1)}}{2\omega_2^{(2)}} (\delta_{lm} \delta_{in} - \delta_{li} \delta_{nm} +  \delta_{ln} \delta_{im})  - i\varepsilon_F \frac{\partial_\varepsilon \Gamma_1}{\omega_3^{(1)}} \Bigg(
			\frac{\omega_1^{(1)}}{\omega_2^{(0)}}
			\delta_{li} \delta_{nm}+ \frac{\omega_1^{(1)}}{2\omega_2^{(2)}} \left(\delta_{lm} \delta_{in} - \delta_{li} \delta_{nm} +  \delta_{ln} \delta_{im}\right)  \Bigg)  \Bigg]
			\\
			& + q_{1n} \Bigg[  \frac{\omega_2^{(1)}}{\omega_1^{(0)}} \delta_{im} \delta_{nl}   - \frac{\omega_2^{(1)}}{2\omega_1^{(2)}} (\delta_{lm} \delta_{in} - \delta_{mi} \delta_{nl} +  \delta_{mn} \delta_{il})  - i\varepsilon_F \frac{\partial_\varepsilon \Gamma_1}{\omega_3^{(1)}} \Bigg(\frac{\omega_2^{(1)}}{\omega_1^{(0)}}\delta_{mi} \delta_{nl}+ \frac{\omega_2^{(1)}}{2\omega_1^{(2)}}(\delta_{lm} \delta_{in} - \delta_{mi} \delta_{nl} +  \delta_{mn} \delta_{il})  \Bigg)  \Bigg]
			\Bigg\}  \,.
		\end{split}
	\end{equation}
	The corresponding functions $G_1$, $G_2$, and $G_3$ [Eq.~\eqref{eqn:Sigma_from_G}] are
	\begin{align}
		G_1 &= \frac{2D_0^{(2)}}{\omega_1^{(1)}  \omega_2^{(1)} \omega_3^{(1)}} \left\{ \frac{\omega_1^{(1)}+\omega_2^{(1)}}{2\omega_3^{(2)}} \left[2 - i\varepsilon_F \left(\frac{\partial_\varepsilon \Gamma_1}{\omega_3^{(1)}} + \frac{\partial_\varepsilon \Gamma_{2}}{\omega_3^{(2)}}  \right)\right] -\frac{2\omega_1^{(1)}}{\omega_1} - \frac{\omega_2^{(1)}}{2\omega_1^{(2)}}
		\left[1 + i\varepsilon_F \frac{\partial_\varepsilon \Gamma_1}{\omega_3^{(1)}}\right] \right\} \,,
		\\
		G_2 &=  \frac{2D_0^{(2)}}{\omega_1^{(1)}  \omega_2^{(1)} \omega_3^{(1)}} \left\{ \frac{\omega_1^{(1)}+\omega_2^{(1)}}{2\omega_3^{(2)}} \left[2-i\varepsilon_F \left(\frac{\partial_\varepsilon \Gamma_1}{\omega_3^{(1)}} + \frac{\partial_\varepsilon \Gamma_{2}}{\omega_3^{(2)}}  \right)\right] + \frac{\omega_2^{(1)}}{2\omega_1^{(2)}}\left(1 + i\varepsilon_F \frac{\partial_\varepsilon \Gamma_1}{\omega_3^{(1)}}\right) + \frac{\omega_2^{(1)}}{\omega_1^{(0)}}\left(1 - i\varepsilon_F \frac{\partial_\varepsilon \Gamma_1}{\omega_3^{(1)}}\right) \right\} \,,
		\\
		G_3 &= \frac{2D_0^{(2)}}{\omega_1^{(1)} \omega_2^{(1)} \omega_3^{(1)}} \left\{ - \frac{\omega_1^{(1)}+\omega_2^{(1)}}{2\omega_3^{(2)}}
		\left[2 - i\varepsilon_F \left(\frac{\partial_\varepsilon \Gamma_1}{\omega_3^{(1)}} + \frac{\partial_\varepsilon \Gamma_{2}}{\omega_3^{(2)}}  \right)\right] - \frac{\omega_2^{(1)}}{2\omega_1^{(2)}}\left(1 + i\varepsilon_F \frac{\partial_\varepsilon \Gamma_1}{\omega_3^{(1)}}\right) +\frac{2\omega_1^{(1)}}{\omega_1} - \frac{\omega_1^{(1)} + \omega_2^{(1)}}{\omega_3^{(0)}} i\varepsilon_F  \left(\frac{\partial_\varepsilon \Gamma_1}{\omega_3^{(1)}} + \frac{\partial_\varepsilon \Gamma_{0}}{\omega_3^{(0)}}  \right)  \right\} \,.
	\end{align}
	Setting the particle number relaxation rate $\Gamma_0$ to zero,
	which is the physical case, we get
	\begin{align}
		G_1 &= \frac{2D_0^{(2)}}{\omega_1^{(1)}  \omega_2^{(1)} \omega_3^{(1)}} \left\{ \frac{\omega_1^{(1)}+\omega_2^{(1)}}{2\omega_3^{(2)}} \left[2 - i\varepsilon_F \left(\frac{\partial_\varepsilon \Gamma_1}{\omega_3^{(1)}} + \frac{\partial_\varepsilon \Gamma_{2}}{\omega_3^{(2)}}  \right)\right] -\frac{2\omega_1^{(1)}}{\omega_1} - \frac{\omega_2^{(1)}}{2\omega_1^{(2)}}
		\left[1 + i\varepsilon_F \frac{\partial_\varepsilon \Gamma_1}{\omega_3^{(1)}}\right] \right\} \,,
		\label{eqn:G_1}\\
		G_2 &=  \frac{2D_0^{(2)}}{\omega_1^{(1)}  \omega_2^{(1)} \omega_3^{(1)}} \left\{ \frac{\omega_1^{(1)}+\omega_2^{(1)}}{2\omega_3^{(2)}} \left[2-i\varepsilon_F \left(\frac{\partial_\varepsilon \Gamma_1}{\omega_3^{(1)}} + \frac{\partial_\varepsilon \Gamma_{2}}{\omega_3^{(2)}}  \right)\right] + \frac{\omega_2^{(1)}}{2\omega_1^{(2)}}\left(1 + i\varepsilon_F \frac{\partial_\varepsilon \Gamma_1}{\omega_3^{(1)}}\right) + \frac{\omega_2^{(1)}}{\omega_1}\left(1 - i\varepsilon_F \frac{\partial_\varepsilon \Gamma_1}{\omega_3^{(1)}}\right) \right\} \,,
		\label{eqn:G_2}\\
		G_3 &= \frac{2D_0^{(2)}}{\omega_1^{(1)} \omega_2^{(1)} \omega_3^{(1)}} \left\{ - \frac{\omega_1^{(1)}+\omega_2^{(1)}}{2\omega_3^{(2)}}
		\left[2 - i\varepsilon_F \left(\frac{\partial_\varepsilon \Gamma_1}{\omega_3^{(1)}} + \frac{\partial_\varepsilon \Gamma_{2}}{\omega_3^{(2)}}  \right)\right] - \frac{\omega_2^{(1)}}{2\omega_1^{(2)}}\left(1 + i\varepsilon_F \frac{\partial_\varepsilon \Gamma_1}{\omega_3^{(1)}}\right) +\frac{2\omega_1^{(1)}}{\omega_1} - \frac{\omega_1^{(1)} + \omega_2^{(1)}}{\omega_3} i\varepsilon_F
		\frac{\partial_\varepsilon \Gamma_1}{\omega_3^{(1)}}  \right\} \,.
		\label{eqn:G_3}
	\end{align}
	In the collisionless limit, $\Gamma_l \to 0$,
	these formulas reduce to
	Eq.~\eqref{eqn:G_kinetic_collisionless_nonconserving}.
	
	%%%%%%%%%%%%%%%%%%%%%%%%%%%%%%%%%%%%%%%%%%%%%%%%%%%%%%%%%%%%%%%%%%%%%%%%%%%%
	\section{Third-order conductivity in the hydrodynamic regime} 
	
	The third-order ac conductivity $\sigma_{ilmn}^{(3)} \left(\vec{q}_1, \omega_1;
	\vec{q}_2,\omega_2;
	\vec{q}_3,\omega_3\right)$ is defined as
	\begin{equation}
		\begin{split}
			j_i^{(3)}(\vec{q}, \omega)
			&= \int \prod\limits_{j = 1}^3
			\frac{d\omega_j d^2 {q}_j}{(2\pi)^3}
			\delta \left(\vec{q}_1 + \vec{q}_2 + \vec{q}_3 - \vec{q}\right)
			\delta\left(\omega_1 + \omega_2 + \omega_3 - \omega\right)
			\\
			\mbox{} &\times \sigma_{ilmn}^{(3)} \left(\vec{q}_1, \omega_1;
			\vec{q}_2,\omega_2;
			\vec{q}_3,\omega_3\right) E_l\left(\vec{q}_1, \omega_1\right) E_m\left(\vec{q}_2, \omega_2\right)
			E_n\left(\vec{q}_3, \omega_3\right)\,.
		\end{split}
		\label{eqn:nonlinear_conductivity}
	\end{equation}
	Unlike the second-order conductivity,
	$\sigma^{(3)}_{i l m n}$ can approach a nonzero value
	at $q = 0$ in inversion-symmetric systems.
	We will compute this value
	and disregard $\mathcal{O}(q^2)$ nonlocal corrections.
	The calculation is simplified by the observation that Eqs.~\eqref{eqn:first} and \eqref{eqn:second}
	yield $n^{(1)} = n^{(2)} = n_E^{(1)} = W^{(1)} = P^{(1)} = \mathcal{O}(q) \to 0$ in this approximation.
	An alternative way to get the same result
	is to neglect spatial gradients in Eqs.~\eqref{eqn:euler}, \eqref{eqn:energy_c} and \eqref{eqn:charge_c}, after which the hydrodynamic equations reduce to
	\begin{equation}
		(\partial_t + \Gamma_d) u_i = \frac{1}{\gamma^2 W} \left( n E_i - u_i \partial_t P - u_i j_j E_j
		\right) \,,
		%	& \partial_t  u_i =  \frac{1}{\gamma^2 W}
		% \left(- u_i \partial_t P + n E_i - u_i\vec{j} \cdot \vec{E}
		% \right)   \,,
		%\label{eqn:euler2} \\
		\qquad
		\partial_t n_{E}  =  j_j E_j  = n\, u_j E_j \,,
		% \label{eqn:energy_c2}\\
		\qquad
		\partial_t n   =0 \,.
		\label{eqn:charge_c2}
	\end{equation}
	%%
	%And the dissipative case is
	%%%
	%\begin{align}
	%	(\partial_t + \Gamma) u_i = \frac{1}{\gamma^2 W} \left( n E_i - u_i \partial_t P - u_i j_j E_j
	%	\right) \,,
	%\label{eqn:u3}
	%\end{align}
	%%%
	The last equation entails $n = n^{(0)}$, and so $j_i^{(3)} = n^{(0)} u^{(3)}_i$. The third-order velocity can be found from
	\begin{align}
		\left(\partial_t + \Gamma_d\right) u^{(3)}_i = -\frac{n}{\left(\gamma^2 W\right)^2} \left(\gamma^2 W\right)^{(2)} E_i - \frac{1}{\gamma^2 W} u^{(1)}_i \partial_t P^{(2)} - \frac{n}{\gamma^2 W} u^{(1)}_i u^{(1)}_j E_j \,,
		\label{eqn:u3}
	\end{align}
	Since $\gamma^2 W = n_E + P$, we have 
	\begin{align}
		\left(\gamma^2 W\right)^{(2)} = n^{(2)}_E + P^{(2)}
		= W \frac{\omega_2^+}{\omega}  u^{(1)}_{1i} u^{(1)}_{2i}
		+ \left[ \frac{\partial P}{\partial n_0} \left(-\frac{1}{2} n\right)
		+ \frac{\partial P}{\partial n_{E0}}  W \left(\frac{\omega_2^+}{\omega} - 1\right) \right] u^{(1)}_{1i} u^{(1)}_{2i}
		+  \mathrm{perm}.,
	\end{align}
	where ``$\mathrm{perm}.$'' stands for permutations
	among subscripts $1$, $2$, and $3$, corresponding to
	frequencies $\omega_1$, $\omega_2$, and $\omega_3$,
	respectively.
	The equation for the Fourier amplitude $u^{(3)}(\omega)$ of the combined
	frequency
	$\omega = \omega_1 + \omega_2 + \omega_3$ becomes
	\begin{equation}
		\begin{split}
			\left(-i\omega + \Gamma_d\right) u^{(3)}_i &=
			-\frac{n}{\left(\gamma^2 W\right)^2} E_i \left\{ W \frac{\omega_2^+}{\omega_1 + \omega_2}  u^{(1)}_{1i} u^{(1)}_{2i}  + \left[ \frac{\partial P}{\partial n_0} \left(-\frac{1}{2} n\right) + \frac{\partial P}{\partial n_{E0}}  W \left(\frac{\omega_2^+}{\omega_1 + \omega_2} - 1\right) \right] u^{(1)}_{1i} u^{(1)}_{2i} \right\}
			\\
			\mbox{} &- \frac{1}{\gamma^2 W} u^{(1)}_{3i} \partial_t \left\{ \left[ \frac{\partial P}{\partial n_0}\left(-\frac{1}{2} n\right) + \frac{\partial P}{\partial n_{E0}}  W \left(\frac{\omega_2^+}{\omega_1 + \omega_2} -1\right) \right] u^{(1)}_{1i} u^{(1)}_{2i} \right\} - \frac{n}{\gamma^2 W} u^{(1)}_i u^{(1)}_j E_j
			\\
			&= -\frac{n}{ W^2} E_i \left\{ W \frac{\omega_2^+}{\omega_1 + \omega_2}  u^{(1)}_{1i} u^{(1)}_{2i}  + \left[ \frac{\partial P}{\partial n_0} \left(-\frac{1}{2} n\right)
			+ \frac{\partial P}{\partial n_{E0}}  W \left(\frac{\omega_2^+}{\omega_1 + \omega_2} - 1\right) \right] u^{(1)}_{1i} u^{(1)}_{2i} \right\}
			\\
			\mbox{}	&- \frac{1}{W} u^{(1)}_{3i} \left[ \frac{\partial P}{\partial n_0} \left(\frac{i}{2} n\right) \left(\omega_1+\omega_2\right) + \frac{\partial P}{\partial n_{E0}} W (-i) \left(\omega_2^+  - \omega_1 - \omega_2\right) \right] u^{(1)}_{1i} u^{(1)}_{2i}
			- \frac{n}{W} u^{(1)}_i u^{(1)}_j E_j
			\\
			&= \frac{i}{W} \omega_3^+ u^{(1)}_{3i} u^{(1)}_{1j} u^{(1)}_{2j} \left\{ W \frac{\omega_2^+}{\omega_1 + \omega_2}    + \left[ \frac{\partial P}{\partial n_0} \left(-\frac{1}{2} n\right) + \frac{\partial P}{\partial n_{E0}}  W \left(\frac{\omega_2^+}{\omega_1 + \omega_2} - 1\right) \right] \right\} 
			\\
			\mbox{}	&- \frac{1}{W} u^{(1)}_{3i} u^{(1)}_{1j} u^{(1)}_{2j}
			\left\{ \frac{\partial P}{\partial n_0}
			\left(\frac{i}{2} n\right)
			(\omega_1 + \omega_2)
			+ \frac{\partial P}{\partial n_{E0}} W (-i) \left( \omega_2^+  - \omega_1 - \omega_2\right) \right\}
			+ i \omega_2^+  u^{(1)}_{1j} u^{(1)}_{2j} u^{(1)}_{3i}
			\\
			&= i \Bigg[  \frac{\omega_2^+ \omega_3^+  }{\omega_1 + \omega_2}
			+  \frac{\partial P}{\partial n_0}
			\left(-\frac{1}{2} \omega_3^+   \frac{n}{W} \right) + \frac{\partial P}{\partial n_{E0}}
			\left(\frac{\omega_2^+ \omega_3^+ }{\omega_1 + \omega_2} - \omega_3^+ \right)
			\\
			\mbox{} &- \frac{\partial P}{\partial n_0}
			\left( \frac{1}{2} \frac{n}{W}\right) (\omega_1+\omega_2)
			+ \frac{\partial P}{\partial n_{E0}}
			\left( \omega_2^+ - \omega_1 - \omega_2\right)
			+ \omega_2^+ \Bigg] u^{(1)}_{1j} u^{(1)}_{2j} u^{(1)}_{3i}
			\\
			&= i \Bigg[  \frac{\omega_2^+ \omega_3^+  }{\omega_1 + \omega_2}
			+  \omega_2^+
			+ \frac{\partial P}{\partial n_{E0}}
			\left(\frac{\omega_2^+ \omega_3^+  }{\omega_1 + \omega_2} - \omega_3^+ + \omega_2^+  - \omega_1 - \omega_2\right)  -   \frac{\partial P}{\partial n_0}
			\left( \frac{1}{2} \frac{n}{W}\right)
			(\omega_1 + \omega_2 + \omega_3^+)
			\Bigg] u^{(1)}_{1j} u^{(1)}_{2j} u^{(1)}_{3i}\,.
			\label{eqn:u3_eq}
		\end{split}
	\end{equation}
	Therefore,
	\begin{align}
		u^{(3)}_i &= - \frac{1}{\omega^+}
		\left[
		\frac{\omega_2^+ \omega_3^+  }{\omega_1 + \omega_2}  +  \omega_2^+ + \frac{\partial P}{\partial n_{E0}}
		\left(\frac{\omega_2^+ \omega_3^+  }{\omega_1 + \omega_2}  + \omega_2^+  - \omega^+ \right)
		- \frac{\partial P}{\partial n_0}
		\left( \frac{1}{2} \frac{n}{W}\right) \omega^+
		\right] u^{(1)}_{1j} u^{(1)}_{2j} u^{(1)}_{3i}
		\notag \\
		&= \left[
		\frac{\partial P}{\partial n_{E0}} + \frac{\partial P}{\partial n_0}
		\left( \frac{1}{2} \frac{n}{W}\right)
		- \left(\frac{\partial P}{\partial n_{E0}} + 1\right)
		\frac{1}{\omega^+}
		\left(\frac{\omega_2^+ \omega_3^+  }{\omega_1 + \omega_2}  + \omega_2^+  \right)
		\right]
		u^{(1)}_{1j} u^{(1)}_{2j} u^{(1)}_{3i}\,.
		\label{eqn:u3_clean}
	\end{align}
	(For brevity, we omitted ``$\mbox{} + \mathrm{perm}.$'' in the above equations.)
	In the dissipationless limit Eq.~\eqref{eqn:u3_clean} simplifies to [cf.~Eq.~\eqref{eqn:C_ise}]
	\begin{equation}
		u^{(3)}_i = \frac{C_{\mathrm{ise}} - 1}{2}
		u^{(1)}_{1j} u^{(1)}_{2j} u^{(1)}_{3i} 
		+ \mathrm{perm}.
		\label{eqn:u3i}
	\end{equation}
	Therefore, 	
	\begin{align}
		%\sigma^{(3)}_{ilmn} &= \frac{1}{2} \left(C_{\mathrm{ise}} - 1\right)
		%n e \frac{-i}{\omega_1 \omega_2 \omega_3}
		%\left(\frac{e}{m^{\ast}}\right)^3 v^2
		%\frac{2}{3!} \Delta_{ilmn}\,,
		%\\
		\sigma^{(3)}_{ilmn}
		& = \frac{D_h^{(3)}}{\omega_1 \omega_2 \omega_3}
		\left(\delta_{il}\delta_{mn} + \delta_{im}\delta_{ln} + \delta_{in}\delta_{lm}\right)
		%\Delta_{ilmn}
		\,,
		\quad
		D_h^{(3)} = i\frac{1 - C_{\mathrm{ise}}}{3!} \frac{e^4 n }{m^{\ast 3} v^2}
		\,,
		\label{eqn:sigma_3}
	\end{align}
	%%
	%where $\Delta_{ilmn} = \delta_{il}\delta_{mn} + \delta_{im}\delta_{ln}
	% + \delta_{in}\delta_{lm}$.
\end{widetext}
where $e$ and $v$ were restored.

We can compare our formula for the third-order ac conductivity in the hydrodynamic regime with other results in the literature
for the case $\omega_1 = \omega_2 = \omega_3$,
which corresponds to the third harmonic generation.
This effect is controlled by
the conductivity
$\sigma^{(3)}_{ilmn}(0, \omega_1; 0, \omega_1; 0, \omega_1)$.
Applied to graphene at $T = 0$, our result
for $D_h^{(3)}$ is twice larger than the third-order spectral weight from the collisionless Boltzmann transport theory \cite{Mikhailov2016}.
Compared to the linear response, the third-order current is suppressed by the small parameter $\xi = \left( \frac{-eE/\omega}{m^{\ast} v} \right)^2$. At zero temperature, neglecting exchange-correlation corrections, $m^{\ast} v$ is just the Fermi momentum $p_F$,
so that $\xi = ({\delta p} / {p_F})^2$.
The quantity $\delta p = -e E /\omega$ is equal
by the order of magnitude to the change in electron momentum
caused by the electric field during one half cycle of the sum-frequency oscillations, $\delta t \sim \pi / \omega$.
The ratio of $\xi$ factors for
a nonrelativistic and ultrarelativistic Dirac
fluids is $\sim (v_F / v)^2 \ll 1$.
This factor vanishes for a system with a parabolic dispersion
corresponding to $v \to \infty$.
Indeed, for such a system all nonlinearities at zero $q$ should be absent because of
the Galilean invariance.
On the other hand, the linear and second-order conductivities, $\sigma$ and $\sigma^{(2)}_{i l m}$,
do not show this contrasting behavior because they do
not contain $v$ explicitly.

%%%%%%%%%%%%%%%%%%%%%%%%%%%%%%%%%%%%%%%%%%%%%%%%%%%%%%%%%%%%%%%%%%
\section{Applications and summary}

\subsection{Photon drag}

The photon drag effect is the generation of dc current by
a light incident on the sample.
Unlike optical rectification and photogalvanic effect,
the photon drag current
is the result of the transfer of the
linear momentum of photons $\mathbf{q}$ to free carriers~\cite{Glazov2014}.
This is why photon drag can appear only
if $\vec{q}
= \hat{\mathbf{x}} q_x  + \hat{\mathbf{z}} q_z$ is not strictly
normal to the $x$--$y$
plane of the sample,
see Fig.~\ref{fig:jy_phi}(a).
An alternative classical picture of the photon drag is the carrier drift in the
crossed electric and magnetic fields of the electromagnetic wave,
and so the photon drag is also sometimes referred to as
the dynamical Hall effect.
The drag current can have both longitudinal $j_x$ and transverse $j_y$ components.
Let the in-plane component of the electric field be
$\vec{E}=\vec{E}_0 e^{i(\vec{q} \vec{r} -\omega t )} + \mathrm{c.c.}$
where
$\vec{E}_0 =\hat{\mathbf{x}} E_x  + \hat{\mathbf{y}} E_y$.
The polarization of the incident wave in the $x$--$y$ plane
is important.
This polarization can be specified in terms of the Stokes parameters
$s_0 = |E_x|^2 + |E_y|^2$, $s_1 = |E_x|^2 - |E_y|^2$,
$s_2 = E_x E_y^{\ast} + E_y E_x^{\ast}$,
and $s_3 = -i (E_x E_y^{\ast} - E_y E_x^{\ast} )$.
From Eq.~\eqref{eqn:sigma2}, we can calculate the induced dc current components as~\cite{Jiang2011}
\begin{align}
	j_x&= 2 \sigma_{xyy}^{(2)}(\vec{q},\omega;-\vec{q},-\omega) E_y E^{\ast}_y + 2 \sigma_{xxx}^{(2)}(\vec{q},\omega;-\vec{q},-\omega) E_x E^{\ast}_x
	\notag\\
	&= T_1 q_x \frac{1}{2} (|E_x|^2 + |E_y|^2)  + T_2 q_x \frac{1}{2} (|E_x|^2 - |E_y|^2)\,,\notag\\
	j_y&= 2 \sigma_{yxy}^{(2)}(\vec{q},\omega;-\vec{q},-\omega) E_x E^{\ast}_y + 2 \sigma_{yyx}^{(2)}(\vec{q},\omega;-\vec{q},-\omega) E_y E^{\ast}_x
	\notag\\
	&= T_2 q_x \frac{1}{2} (E_x E_y^{\ast} + E_y E_x^{\ast} ) + \tilde{T}_1 q_x (-i) (E_x E_y^{\ast} - E_y E_x^{\ast} )\,.
	%	 \notag \\
	%	&= T_2 q_x \frac{1}{2} (E_x E_y^{\ast} + E_y E_x^{\ast} ) -  \tilde{T}_1 q_x P_{\mathrm{circ}} e_z (|E_x|^2 + |E_y|^2)  \,.
	\label{eqn:dc_current}
\end{align}
The coefficients $T_1$ and $T_2$ are as follows:
\begin{align}
	T_1 = 2 \left(\tilde{G}_1 + \tilde{G}_2 + 2\tilde{G}_3\right)\,,
	\quad
	T_2 = 2 \left(\tilde{G}_1 + \tilde{G}_2 \right) \,,
\end{align}
where
\begin{equation}
	\tilde{G}_a(\omega) \equiv G_a(\omega,-\omega) - G_a(-\omega,\omega)
	\label{eqn:tilde_G}
\end{equation}
and $G_a$ are the functions introduced in
Eq.~\eqref{eqn:Sigma_from_G}.
For $\tilde{T}_1$, we get
\begin{equation}
	\begin{split}
		\tilde{T}_1 
		=  &-i \left[ G_1(\omega,-\omega) + G_1(-\omega,\omega)\right]\\
		\mbox{} &+
		i \left[ G_2(\omega,-\omega) + G_2(-\omega,\omega) \right] \,.
	\end{split}
	\label{eqn:PD_parameters}
\end{equation}
For the hydrodynamic regime,
we take $G_a(\omega,-\omega)$ from Eq.~\eqref{eqn:G_hydro_general} and
obtain
\begin{align}
	T_1 &= \left(-\frac{1}{1 - C_{\mathrm{ise}}} - 1\right) \frac{4 D^{(2)}_h}{\omega(\omega^2 + \Gamma^2 )} ,\\
	%\quad 
	T_2 &= \left(\frac{1}{1 - C_{\mathrm{ise}}} - 1\right) \frac{4 D^{(2)}_h}{\omega(\omega^2 + \Gamma^2 )} ,
	\\
	%\quad 
	\tilde{T}_1 &= 0 \,.
	\label{eqn:PD_parameters_hydro}
\end{align}
For the case of graphene, $C_{\mathrm{ise}} = 1 / 2$, these equations give
\begin{align}
	T_1 = -3 T_2\,,
	\quad 
	T_2 = \frac{4 D^{(2)}_h}{\omega \left(\omega^2 + \Gamma_d^2 \right)}\,,
	\quad 
	\tilde{T}_1 = 0 \,.
	\label{eqn:PD_parameters_hydro_graphene}
\end{align}
which is Eq.~(\textcolor{blue}{15}) of the main text.

In the kinetic regime, the
photon drag coefficients are more complicated.
Equations~\eqref{eqn:G_1}--\eqref{eqn:G_3}
for $G_a$ can be used to compute them
for graphene at zero temperature.
We get the following:
\begin{widetext}
	%%%
	%\begin{equation}
	%\begin{split}
	%	T_2 
	%	&= \frac{4 D_0^{(2)}}{ i\Gamma_1  (-\omega^2 - \Gamma_1^2)} \Bigg[
	%	-\frac{4 i\Gamma_1}{\omega} +
	%	 \frac{1}{2} \left(1+\varepsilon_F \partial_\varepsilon \ln \Gamma_1  \right)\left( -\frac{-\omega+i\Gamma_1}{\omega+i\Gamma_2} + \frac{
	%		\omega+i\Gamma_1}{-\omega+i\Gamma_2} \right)
	%	+ \frac{1}{2} \left(1+\varepsilon_F \partial_\varepsilon \ln \Gamma_1  \right) \left( \frac{-\omega+i\Gamma_1}{\omega+i\Gamma_2} - \frac{
	%		\omega+i\Gamma_1}{-\omega+i\Gamma_2} \right)
	%\\
	%	& +\left(1-\varepsilon_F \partial_\varepsilon \ln \Gamma_1  \right) \left( \frac{-\omega+i\Gamma_1}{\omega+i\Gamma_0} - \frac{
	%		\omega+i\Gamma_1}{-\omega+i\Gamma_0} \right)
	%	\Bigg]
	%\\
	%	&= \frac{4D_0^{(2)}}{ i\Gamma_1 \left(-\omega^2 - \Gamma_1^2\right)} \Bigg[
	%	-\frac{i\Gamma_1}{\omega} +
	%	\left(1 - \varepsilon_F \partial_\varepsilon \ln \Gamma_1  \right)
	%	\frac{-2i\omega \left(\Gamma_1 + \Gamma_0\right)}
	%	{\left(-\omega^2 - \Gamma_0^2\right)} 
	%	\Bigg]
	%\\
	%	&= 8D_0^{(2)} 
	%	\frac{2\left(\omega^2 + \Gamma_0^2\right) -\omega^2 \left(1 - \varepsilon_F \partial_\varepsilon \ln \Gamma_1  \right) \left(1 + \Gamma_0/\Gamma_1\right) }{\omega \left(\omega^2 + \Gamma_0^2\right) \left(\omega^2 + \Gamma_1^2\right)} \,.
	%\end{split}
	%\end{equation}
	%%%
	%%
	\begin{gather}
		T_1 =  8 D_0^{(2)} 
		\frac{\left(1 + \varepsilon_F \partial_\varepsilon \ln \Gamma_1\right) \left[\left(\omega^2 + \Gamma_2^2\right)
			+  \omega^2 \left(1 + \Gamma_2/\Gamma_1\right)
			\right] - 4 \left(\omega^2 + \Gamma_2^2\right)}
		{ \omega \left(\omega^2 + \Gamma_1^2\right)
			\left(\omega^2 + \Gamma_2^2\right)} \,,
		\quad
		T_2 = 8 D_0^{(2)} 
		\frac{1 + \varepsilon_F \partial_\varepsilon \ln \Gamma_1}
		{\omega \left(\omega^2 + \Gamma_1^2\right)}\,,
		\label{eqn:T_2_kin}
		\\
		\tilde{T}_1 =  -4 D_0^{(2)}
		\frac{\left(1 + \Gamma_2 / \Gamma_1 \right) \Gamma_2}
		{\left(\omega^2 + \Gamma_1^2 \right)\left(\omega^2 + \Gamma_2^2 \right)}
		\left(1 + \varepsilon_F \partial_\varepsilon \ln \Gamma_1 \right) \,,
		\label{eqn:T_1_tilde_kin}
	\end{gather}
	%%
	%Next, for $T_1$ we obtain
	%%%
	%\begin{equation}
	%\begin{split}
	%	T_1 &= T_2 + 4\tilde{G}_3
	%	= T_2 + \frac{8 D_0^{(2)}}{ i\Gamma_1 \left(-\omega^2 - \Gamma_1^2\right)} \Bigg[
	%	\left(1 + \varepsilon_F \partial_\varepsilon \ln \Gamma_1  \right) \frac{1}{2}\left( -\frac{-\omega + i\Gamma_1}{\omega + i\Gamma_2}
	%	+ \frac{
	%		\omega + i\Gamma_1}{-\omega+i\Gamma_2} \right)	+ 4 \frac{i\Gamma_1}{\omega}
	%	\Bigg]
	%\\
	%	&= T_2 + \frac{8 D_0^{(2)}}{\left(-\omega^2 - \Gamma_1^2\right)} \Bigg[
	%	\left(1 + \varepsilon_F \partial_\varepsilon \ln \Gamma_1  \right)
	%	\frac{\omega \left(1 + \Gamma_2/\Gamma_1\right)}
	%	{\left(-\omega^2 - \Gamma_2^2\right)} + \frac{4}{\omega}
	%	\Bigg]
	%\\
	%	&= T_2 + 8 D_0^{(2)} 
	%	\frac{\left(1 + \varepsilon_F \partial_\varepsilon \ln \Gamma_1  \right) \omega^2 \left(1 + \Gamma_2/\Gamma_1\right)
	%	 - 4 \left(\omega^2 + \Gamma_2^2\right)}
	%	{\omega \left(\omega^2 + \Gamma_1^2\right) \left(\omega^2 + \Gamma_2^2\right)} \,.
	%\end{split}
	%\end{equation}
	%%%
	%Taking $\Gamma_0 = 0$ again, we obtain
	%%%
	%\begin{equation}
	%	T_1 =  8 D_0^{(2)} 
	%	\frac{\left(1 + \varepsilon_F \partial_\varepsilon \ln \Gamma_1\right) \left[\left(\omega^2 + \Gamma_2^2\right)
	%	 +  \omega^2 \left(1 + \Gamma_2/\Gamma_1\right)
	%	 \right] - 4 \left(\omega^2 + \Gamma_2^2\right)}
	%	 { \omega \left(\omega^2 + \Gamma_1^2\right)
	%	 \left(\omega^2 + \Gamma_2^2\right)} \,.
	%\label{eqn:T_1_kin}
	%\end{equation}
	%%%
\end{widetext}
%%%
%Finally, for $\tilde{T}_1$, we find, after some algebra,
%%%
%\begin{equation}
%\tilde{T}_1 =  -4 D_0^{(2)}
%\frac{\left(1 + \Gamma_2 / \Gamma_1 \right) \Gamma_2}
%     {\left(\omega^2 + \Gamma_1^2 \right)\left(\omega^2 + \Gamma_2^2 \right)}
%     \left(1 + \varepsilon_F \partial_\varepsilon \ln \Gamma_1 \right) \,.
%\label{eqn:T_1_tilde_kin}
%\end{equation}
%%%
in agreement with
Refs.~\onlinecite{Glazov2014, Karch2010, Jiang2011}.
If the dominant electron scattering in graphene is due to short-range impurities,
then
the scattering rates $\Gamma_1$ and  $\Gamma_2$ for the $p$- and
$d$-wave angular deformations of the Fermi surface obey the
relations
\begin{equation}
	\Gamma_2 = 2 \Gamma_1\,,
	\quad
	\varepsilon
	\partial_\varepsilon \ln \Gamma_1 = 1\,.
	\label{eqn:short-range}
\end{equation}
When substituted into the general formulas above, followed by
the notation change $\Gamma_1 \to \Gamma_d$,
these relations lead to
Eq.~(\textcolor{blue}{16}) of the main text.

%Take the clean limit $\Gamma=0$, the angle dependence of the currents are
%%%
%\begin{align}
%j_x
%&= -4 S \frac{k_x}{\omega^3} |E|^2  \left((\cos^2\theta -2)\sin^2\alpha +  2 \right)  \,, \notag \\
%j_y
%&=  2 S \frac{k_x}{\omega^3} |E|^2 \cos\theta \sin(2\alpha) \cos \phi  \,
%\label{eqn:dc_current_2}
%\end{align}
%%
Instead of the Stokes parameters, 
we can use two angles $\psi$ and $\alpha$ such that
$E_x = E \cos\alpha \cos\theta$, $E_y = E\cos\alpha e^{i\psi}$.
Note that $\alpha=0$ means p-polarization and $\alpha=\pi/2$ means s-polarization,
see Fig.~\ref{fig:jy_phi}(a).
The formulas for $j_x$ and $j_y$ become 
\begin{align}
	j_x
	&= \frac{1}{2} q_x |E|^2  \left[ (T_1 + T_2) \cos^2\! \alpha
	\cos^2\!\theta  + (T_1 - T_2) \sin^2\! \alpha \right] \,, \notag \\
	j_y
	&= \frac{1}{2} q_x |E|^2 \left[ \cos\theta \sin 2\alpha
	\left( T_2 \cos\psi - 2 \tilde{T}_1 \sin \psi \right)\right]  \,.
	\notag
	\label{eqn:dc_current}
\end{align}
In the hydrodynamic regime where $\tilde{T}_1 = 0$,
the transverse current $j_y$ has no component proportional to $\sin \psi$.
However, $j_y$ does have such a component in the kinetic regime,
as illustrated by Fig.~\ref{fig:jy_phi}(d).
This distinction may be used to identify
the two regimes in experiments.

\begin{figure}[b]
	\includegraphics[width=3.5 in]{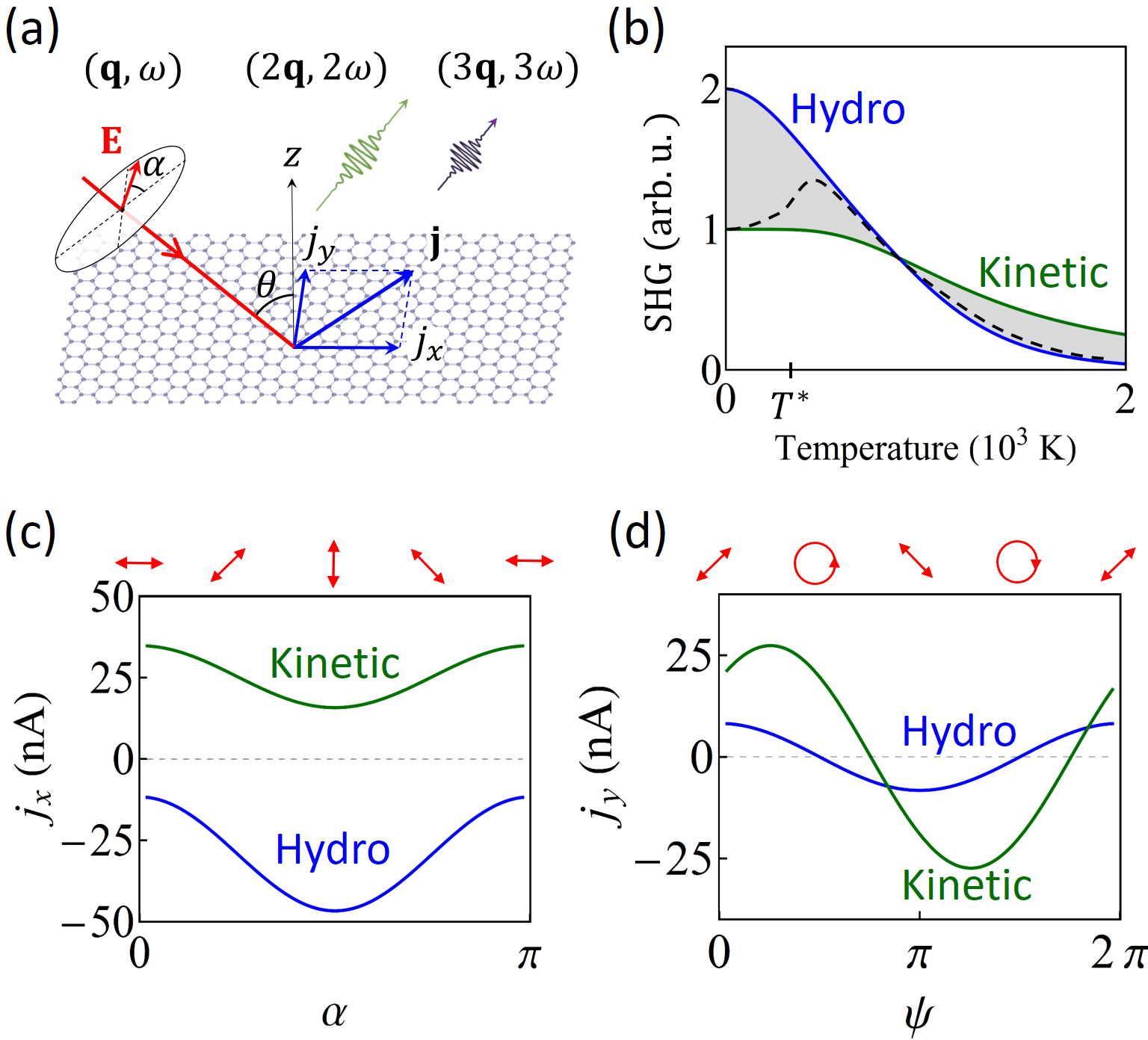} 
	\caption{[Same as Fig.~4 of the main text.] 
		(a) Geometry for measuring photon drag, second,
		and third harmonic generation.
		(b) SHG signal as a function of $T$ at fixed $\omega$.
		The ``Kinetic'' curve is the the kinetic regime;
		the ``Hydro'' curve is for the hydrodynamic one;
		the dashed curve is a sketch of the actual signal.
		(c) Photon drag photocurrent $j_x$ in graphene \textit{vs}.
		polarization angle $\alpha$ (illustrated by the red arrows).
		(d) $j_y$ \textit{vs}. phase delay $\psi$ (degree of circular polarization)
		at $\alpha = \pi/4$.
		Parameters in (c,d): $T = 0$ for the `Kinetic'' curves,
		$T=300\unit{K}$ for the ``Hydro'' curves,
		$n = 3.14 \times 10^{12} \unit{cm^{-2}}$,
		$\omega = 5 \unit{THz}$, $\Gamma_d = 1 \unit{THz}$,
		$\theta = \pi / 4$, $E = 10^3 \unit{V/cm}$.
	}
	\label{fig:jy_phi}
\end{figure}

%%%%%%%%%%%%%%%%%%%%%%%%%%%%%%%%%%%%%%%%%%%%%%%%%%%%%%%%%%%%%%%%%%%%%%%%%%%%
\subsection{Second harmonic generation}

The second-harmonic generation (SHG) signal is proportional
to the $\left(2\vec{q}, 2\omega\right)$  Fourier harmonic of the
second-order ac current, which is given by~\cite{Glazov2011a}
\begin{align}
	j_x&= \sigma_{xyy}^{(2)}\left(\vec{q},\omega;\vec{q},\omega\right) E^2_y
	+ \sigma_{xxx}^{(2)}\left(\vec{q},\omega;\vec{q},\omega\right) E^2_x
	\, \notag\\
	&= S_1 q_x \left(E_x^2 + E_y^2\right)
	+ S_2 q_x \left(E_x^2 - E_y^2\right)\,,
	\\
	j_y&= \sigma_{yxy}^{(2)}(\omega,q,\omega,q) E_x E_y + \sigma_{yyx}^{(2)}(\omega,q,\omega,q) E_y E_x
	\, \notag\\
	&= 2 S_2 q_x E_x E_y  \,,
	\label{eqn:SHG_current}
\end{align}
where
\begin{align}
	S_1 &= G_1(\omega,\omega) + G_2(\omega,\omega) + 2 G_3(\omega,\omega)
	\,,\\
	S_2 &= G_1(\omega,\omega) + G_2(\omega,\omega)\,.
	\label{eqn:SHG_parameters}
\end{align}
Neglecting the damping, for graphene in the kinetic regime,
we get
\begin{align}
	S_1 = \frac{2 D^{(2)}_k}{\omega^3}\,,
	\quad
	S_2 = \frac{D^{(2)}_k}{\omega^3} \,.
	\label{eqn:SHG_parameters_kinetic}
\end{align}
In the hydrodynamic regime, we find
\begin{align}
	S_1 = \frac{2 D^{(2)}_h}{\omega^3}\,,
	\quad
	S_2 = \frac{D^{(2)}_h}{\omega^3} \,.
	\label{eqn:SHG_parameters_hydrodynamic}
\end{align}
This implies that SHG signal has the same polarization dependence in the hydrodynamic and kinetic regimes but the magnitude of the response is different because it is controlled by
either $D^{(2)}_h$ or $D^{(2)}_k$.
At zero temperature the ratio $D^{(2)}_h /  D^{(2)}_k$ is equal to $2$,
but at high temperature it rapidly decreases,
see Fig.~\ref{fig:jy_phi}(b).
Experimentally, this difference may be observed as the electron
temperature is increased and the system crosses over from the kinetic to the hydrodynamic regime at some $T^\ast = T^\ast(\omega)$.
This crossover temperature
is the solution of the equation $\Gamma_{ee}(T^\ast) = \omega$.
As $T^\ast$ is approached from below,
the SHG signal may increase,
by up to a factor of two from its $T = 0$ value,
as sketched by the dashed line in
Fig.~\ref{fig:jy_phi}(b).
When the temperature is raised beyond $T^*$, the system
enters the hydrodynamic regime where the SHG signal
should drop due to decreasing $D^{(2)}_h$.
The transient states of high electron temperatures can be realized
with intense photoexcitation.

%%%%%%%%%%%%%%%%%%%%%%%%%%%%%%%%%%%%%%%%%%%%%%%%%%%%%%%%%%%%%%%%%%%%%%%%%%%%%
%\section{Relativistic Boltzmann equation}
%The relativistic Boltzmann equation assumes the following form:
%%%
%\begin{align}
%\left( P^\mu \partial_\mu + F_{\mu\nu} P^\mu \partial_{P_\nu} \right) f_R (X,P) =I[f_R] \,
%\label{eqn:rela_boltzmann}
%\end{align}
%%%
%where $f_R (X,P)$ is the relativistic distribution function. Note that the four-momentum $P$ contains both momentum and energy. Thus $f_R (X,P)$ is a scalar function defined on the tangent bundle of the $D+1$ dimensional space-time. If we focus on one species of particle with a fixed mass $m$, then $f_R (X,P)$ is related to the ordinary distribution function through
%%%
%\begin{align}
%f_R (X,P) = f(t,\vec{r},\vec{p}) \delta(E^2 + p^2 - m^2) \,,
%\label{eqn:relation_ordinary}
%\end{align}
%%%
%where $c$ has been set to $1$.
%Integrate Eq.~\eqref{eqn:rela_boltzmann} over $P$, we get the charge continuity equation
%%%
%\begin{align}
%&\partial_\mu J^\mu =0 \,, \notag\\
%& j^\mu = \int d P P^\mu f_R (X,P) =\int d\vec{p} \; (1, \vec{v}) f(t,\vec{r},\vec{p})  \,.
%\label{eqn:conti}
%\end{align}
%%%
%Multiply Eq.~\eqref{eqn:rela_boltzmann} by $P^\alpha$  and integrate over $P$, we get the conservation equation for the stress tensor
%%%
%\begin{align}
%&\partial_\mu T^{\mu\nu} = F_\mu^\nu J^\mu \,, \notag\\
%& T^{\mu\nu} = \int d P P^\mu P^\nu f_R (X,P) =\int d\vec{p} \; P^\mu P^\nu \frac{1}{E} f(t,\vec{r},\vec{p})  \,.
%\label{eqn:stress}
%\end{align}
%%%

%%%%%%%%%%%%%%%%%%%%%%%%%%%%%%%%%%%%%%%%%%%%%%%%%%%%%%%%%%%%%%%%%%%%%%%%%%%%
\subsection{Summary
	tables for the second-order conductivity}

As we pointed out earlier,
$\sigma^{(2)}_{i l m}$ is fully characterized by three functions $G_1$, $G_2$, and $G_3$.
Shown in Table~\ref{tbl:general}
are $G_1$, $G_2$, $G_3$ and $T_1$, $T_2$, $\tilde{T}_1$
in different regimes and for different band dispersions.
In addition, the same formulas in the clean limit are
summarized in Table~\ref{tbl:clean}.

Electron systems with parabolic dispersion is an interesting case.
In such systems $\sigma^{(2)}_{i l m}$ has the same form in the kinetic and hydrodynamic regimes
but only in the absence of momentum dissipation.
As mentioned in the main text, in this system the random-phase approximation (RPA) also gives the same
$\sigma^{(2)}_{i l m}$ to the linear order in $q$
in the absence of dissipation.
In the diagrammatic derivation~\cite{Stolz1967} of this RPA result
only the ``diamagnetic'' terms contribute to $\sigma^{(2)}_{i l m}$. Those diamagnetic terms are all determined by the linear-response Drude weight.
The ``paramagnetic'' term, that is, a single-loop diagram with three current vertices vanishes to the first order in $q$. This is superficially similar yet apparently unrelated to Furry's theorem in quantum electrodynamics, which says that fermion loops with odd number of
photon vertices vanish because of electron-positron symmetry.
For Dirac electrons in graphene
one may invoke Furry's theorem to explain vanishing of
the spectral weight at $\mu = 0$
[see Eq.~\eqref{eqn:graphene_boltzmann_finite_T}].
However, in the case of interest, $\mu \neq 0$,
the three-point current correlation function~\cite{Cheng2016, Wang2016, Rostami2016} is finite.
In fact, it is the diamagnetic contribution that vanishes, so that $\sigma^{(2)}_{i l m}$ is determined solely by this paramagnetic term.

\begin{table*}
	\begin{tabular}{ccc}
		\hline\hline
		Regime/Dispersion    &  Parabolic band (2D) &  graphene \\
		\\
		\hline
		Hydrodynamic &  &
		\\
		$G_1$ & 
		$\dfrac{D^{(2)}_p}{\omega_1^+ \omega_2^+ \omega_3^+}
		\left(-\dfrac{i\Gamma_d}{\omega_1}\right)$        
		& $\dfrac{D^{(2)}_h}{\omega_1^+ \omega_2^+ \omega_3^+}
		\left(-2 \dfrac{i\Gamma_d}{\omega_1}\right)$
		\\
		$G_2$ & 
		$\dfrac{D^{(2)}_p}{\omega_1^+ \omega_2^+ \omega_3^+}
		\left( \dfrac{\omega_3^+}{\omega_1}\right)$
		& $\dfrac{D^{(2)}_h}{\omega_1^+ \omega_2^+ \omega_3^+}
		\left(\dfrac{\omega_3^+}{\omega_1}\right)$
		\\
		$G_3$ & 
		$\dfrac{D^{(2)}_p}{\omega_1^+ \omega_2^+ \omega_3^+}
		\left( 1+\dfrac{i\Gamma_d}{\omega_1} +\dfrac{\partial P}{\partial {n_E}} \dfrac{2i\Gamma_d}{\omega+i\Gamma_E} \right)$
		& $\dfrac{D^{(2)}_h}{\omega_1^+ \omega_2^+\omega_3^+}
		\left[1+ 2\left(\dfrac{i\Gamma_d}{\omega_1} +\dfrac{\partial P}{\partial {n_E}} \dfrac{2i\Gamma_d}{\omega+i\Gamma_E} \right) \right]$
		\\
		$T_1$ & 
		$-\dfrac{8 D^{(2)}_p}{\omega \left(\omega^2 + \Gamma_d^2\right)}$ 
		&  $-\dfrac{12 D^{(2)}_h}{\omega \left(\omega^2 + \Gamma_d^2 \right)}$
		\\
		$T_2$ & 
		$0$        &  $\dfrac{4 D^{(2)}_h}{\omega \left(\omega^2 + \Gamma_d^2\right)}$
		\\
		$\tilde{T}_1$ & 
		$0$        &  $0$
		\\
		\hline
		Kinetic nonconserving &  &  \\
		$G_1$ & 
		$\dfrac{D^{(2)}_p}{\omega_1^+ \omega_2^+ \omega_3^+}
		\left(\dfrac{\omega_1^+ + \omega_2^+}{\omega_3^+}-\dfrac{\omega_1^+}{\omega_1}\right)$        
		& $ \dfrac{2D_0^{(2)}}{\omega_1^+ \omega_2^+ \omega_3^+} \left(-1- \dfrac{\omega_2^+}{2\omega_1^+} + \dfrac{i\Gamma}{\omega_3^+} - \dfrac{2i\Gamma}{\omega_1}\right)$
		\\
		$G_2$ & 
		$\dfrac{D^{(2)}_p}{\omega_1^+ \omega_2^+ \omega_3^+}
		\left(\dfrac{\omega_1^+ + \omega_2^+}{\omega_3^+}  + \dfrac{\omega_2^+}{\omega_1^+}\right) $
		&$\dfrac{2D_0^{(2)}}{\omega_1^+ \omega_2^+ \omega_3^+}
		\left(1 + \dfrac{3}{2}\dfrac{\omega_2^+}{\omega_1^+}  + \dfrac{i\Gamma}{\omega_3^+}\right)$
		\\
		$G_3$ & 
		$\dfrac{D^{(2)}_p}{\omega_1^+ \omega_2^+ \omega_3^+}
		\dfrac{\omega_1^+}{\omega_1}$
		& $\dfrac{2D_0^{(2)}}{\omega_1^+ \omega_2^+ \omega_3^+}
		\left(1 - \dfrac{\omega_2^+}{2\omega_1^+} - \dfrac{i\Gamma}{\omega_3^+}+\dfrac{2i\Gamma}{\omega_1}\right)$
		\\
		$T_1$ & 
		$ 4 D^{(2)}_p \dfrac{-3\omega^2 -\Gamma^2}{\omega(\omega^2 + \Gamma^2 )^2}  $
		& $16D_0^{(2)} 
		\dfrac{-\omega^2 -\Gamma^2}{\omega(\omega^2 + \Gamma^2 )^2}$
		\\
		$T_2$ & 
		$ 4 D^{(2)}_p \dfrac{-\omega^2 + \Gamma^2}{\omega(\omega^2 + \Gamma^2 )^2}$
		&
		$16D_0^{(2)} \dfrac{\Gamma^2}{\omega(\omega^2 + \Gamma^2 )^2} $
		\\
		$\tilde{T}_1$ & 
		$ 8 D^{(2)}_p \dfrac{\Gamma}{(\omega^2 + \Gamma^2 )^2}$
		& $ 32D_0^{(2)} \dfrac{\Gamma}{(\omega^2 + \Gamma^2 )^2}$
		\\
		\\
		\hline
		Kinetic &  &
		\\
		$G_1$ & 
		& $\dfrac{2D_0^{(2)}}{\omega_1^{(1)} \omega_2^{(1)}\omega_3^{(1)}} \bigg\{ -\dfrac{\omega_2^{(1)}}{2 \omega_1^{(2)}} \bigg(1 + \varepsilon_F \dfrac{\partial_\varepsilon i\Gamma_1}{\omega_3^{(1)}}\bigg) - \dfrac{2 \omega_1^{(1)}}{ \omega_1}  $
		\\
		&  & $ + \dfrac{\omega_1^{(1)} + \omega_2^{(1)}}{\omega_3^{(2)} } \bigg[1-\dfrac{1}{2} \varepsilon_F \bigg(\dfrac{\partial_\varepsilon i\Gamma_1}{\omega_3^{(1)}} +\dfrac{\partial_\varepsilon i\Gamma_2}{\omega_3^{(2)}} \bigg) \bigg] \bigg\}$
		\\
		$G_2$ & 
		& $\dfrac{2D_0^{(2)}}{\omega_1^{(1)} \omega_2^{(1)} \omega_3^{(1)}} \bigg\{ \dfrac{\omega_2^{(1)}}{ \omega_1} \bigg(1 - \varepsilon_F \dfrac{\partial_\varepsilon i\Gamma_1}{\omega_3^{(1)}} \bigg) +$
		\\
		&  & $ \dfrac{ \omega_2^{(1)}}{2 \omega_1^{(2)}} \bigg(1 + \varepsilon_F \dfrac{\partial_\varepsilon i\Gamma_1}{\omega_3^{(1)}}\bigg) + $
		\\
		&  & $ \dfrac{\omega_1^{(1)} + \omega_2^{(1)}}{\omega_3^{(2)} }
		\bigg[1 - \dfrac{1}{2} \varepsilon_F \bigg(\dfrac{\partial_\varepsilon i\Gamma_1}{\omega_3^{(1)}} +\dfrac{\partial_\varepsilon i\Gamma_2}{\omega_3^{(2)}} \bigg) \bigg] \bigg\}$
		\\
		$G_3$ & 
		& $\dfrac{2D_0^{(2)}}{\omega_1^{(1)} \omega_2^{(1)}\omega_3^{(1)}} \bigg( 2\dfrac{\omega_1^{(1)}}{\omega_1} - \dfrac{\omega_1^{(1)} + \omega_2^{(1)}}{\omega_3^{(0)}} \varepsilon_F \dfrac{\partial_\varepsilon i\Gamma_1}{\omega_3^{(1)}} $
		\\
		&  & $ -\dfrac{ \omega_2^{(1)}}{2 \omega_1^{(2)}} \bigg(1 + \varepsilon_F \dfrac{\partial_\varepsilon i\Gamma_1}{\omega_3^{(1)}} \bigg)  $
		\\
		&  & $ -\dfrac{\omega_1^{(1)} + \omega_2^{(1)}}{\omega_3^{(2)} } \bigg[1-\dfrac{1}{2} \varepsilon_F \bigg(\dfrac{\partial_\varepsilon i\Gamma_1}{\omega_3^{(1)}} +\dfrac{\partial_\varepsilon i\Gamma_2}{\omega_3^{(2)}} \bigg) \bigg] \bigg\}$
		\\
		$T_1$ &	& $8 D^{(2)}_0 
		\dfrac{(\omega^2 + \Gamma_2^2 ) (-3 + \varepsilon_F \partial_\varepsilon \ln \Gamma_1 ) + \omega^2 (1+ \Gamma_2 / \Gamma_1)(1 + \varepsilon_F \partial_\varepsilon \ln \Gamma_1 ) }{\omega(\omega^2 + \Gamma_1^2 )(\omega^2 + \Gamma_2^2 )}$
		\\
		$T_2$ &	&
		$8 D^{(2)}_0 \dfrac{1 + \varepsilon_F \partial_\varepsilon \ln \Gamma_1}{\omega(\omega^2 + \Gamma_1^2 )} $
		\\
		$\tilde{T}_1$ &	&
		$ -4 D^{(2)}_0 \dfrac{(1+ \Gamma_2 / \Gamma_1) \Gamma_2}{(\omega^2 + \Gamma_1^2 )(\omega^2 + \Gamma_2^2 )} (1 + \varepsilon_F \partial_\varepsilon \ln \Gamma_1 ) $
		\\
		\hline
		Quantum  & Ref.~\onlinecite{Stolz1967}   & Refs.~\onlinecite{Cheng2016, Wang2016}     \\
		\hline\hline
	\end{tabular}
	\caption{Summary for the general case. Notations:
		$D^{(2)}_h = -\dfrac{1}{2} \dfrac{n^3 v^4}{W^2} (1 - C_{\mathrm{ise}})$, $D^{(2)}_p = - \dfrac{e^3 n}{2 m^2}$, and $D_0^{(2)}= - \dfrac{g e^3 v^2}{32 \pi \hbar^2}$.
	}
	\label{tbl:general}
\end{table*}
\begin{table*}
	\begin{tabular}{cccc}
		\hline\hline
		Regime/Dispersion &	General    &  Parabolic band (2D) &  graphene \\[9pt]
		\hline
		Hydrodynamic &	&  &  
		\\[9pt]
		$G_1$ & $0$	& $0$        & $0$ 
		\\[9pt]
		$G_2$ & $\dfrac{D^{(2)}_h}{\omega_1 \omega_2 \omega_3}
		\left(\dfrac{\omega_3}{\omega_1}\right)$
		& $\dfrac{D^{(2)}_p}{\omega_1 \omega_2 \omega_3}
		\left( \dfrac{\omega_3}{\omega_1}\right)$
		& $\dfrac{D^{(2)}_h}{\omega_1 \omega_2 \omega_3}
		\left(\dfrac{\omega_3}{\omega_1}\right)$
		\\[9pt]
		$G_3$ & $\dfrac{D^{(2)}_h}{\omega_1 \omega_2 \omega_3}$
		& $\dfrac{D^{(2)}_p}{\omega_1 \omega_2 \omega_3}$
		& $\dfrac{D^{(2)}_h}{\omega_1 \omega_2 \omega_3} $
		\\[9pt]
		$T_1$ & $ 4D^{(2)}_h \left(-\dfrac{1}{1 - C_{\mathrm{ise}}} - 1\right)
		\dfrac{1}{\omega^3}$
		& $-8 D^{(2)}_p \dfrac{1}{\omega^3}$
		&  $-12 D^{(2)}_h \dfrac{1}{\omega^3}$
		\\[9pt]
		$T_2$ &  $ 4 D^{(2)}_h
		\left(\dfrac{1}{1 - C_{\mathrm{ise}}} - 1\right)
		\dfrac{1}{\omega^3}$
		& $0$        &  $ 4 D^{(2)}_h \dfrac{1}{\omega^3}$
		\\[9pt]
		$\tilde{T}_1$ &  $ 0$	& 
		$0$        &  $0$
		\\[9pt]
		\hline
		Kinetic &	&  &
		\\[9pt]
		$G_1$ &  	& $0$       & $\dfrac{D_0^{(2)}}{\omega_1 \omega_2 \omega_3}
		\left(-2 - \dfrac{\omega_2}{\omega_1} \right)$
		\\[9pt]
		$G_2$ &  	& 
		$\dfrac{D^{(2)}_p}{\omega_1 \omega_2 \omega_3}
		\dfrac{\omega_3}{\omega_1}$
		& $\dfrac{D_0^{(2)}}{\omega_1 \omega_2 \omega_3}
		\left(2 + 3\,\dfrac{\omega_2}{\omega_1}\right)$
		\\[9pt]
		$G_3$ &  
		& $\dfrac{D^{(2)}_p}{\omega_1 \omega_2 \omega_3} $
		& $\dfrac{D_0^{(2)}}{\omega_1 \omega_2 \omega_3}
		\left(2 - \dfrac{\omega_2}{\omega_1} \right)$
		\\[9pt]
		$T_1$ &  & 
		$ -12 D^{(2)}_p \dfrac{1}{\omega^3}  $        & $ 32 D^{(2)}_0 
		\dfrac{1}{\omega^3}$
		\\[9pt]
		$T_2$ &  & 
		$ -4 D^{(2)}_p \dfrac{1}{\omega^3}$        
		& $ 16 D^{(2)}_0 \dfrac{1}{\omega^3}$
		\\[9pt]
		$\tilde{T}_1$ &  $ 0$	& 
		$0$        &  $0$
		\\[9pt]
		\\[9pt]
		\hline
		Quantum &   &     & Refs.~\onlinecite{Cheng2016, Wang2016}   \\[9pt]
		\hline\hline
	\end{tabular}
	\caption{Clean limit:
		$\Gamma_d = \Gamma_1 \to 0$, $\Gamma_2 \to 0$,
		$\Gamma_2 / \Gamma_1 = 2 = \mathrm{const}$.
	}
	\label{tbl:clean}
\end{table*}
%%

%%%%%%%%%%%%%%%%%%%%%%%%%%%%%%%%%%%%%%%%%%%%%
%\bibliographystyle{apsrev4-1}
%\bibliography{./Library_hydrodynamics}

\end{document}